\documentclass[reprint,amsmath,amssymb,aps,prd]{revtex4-2}

\usepackage{graphicx}% Include figure files
\usepackage{dcolumn}% Align table columns on decimal point
\usepackage{bm}% bold math
\usepackage{xcolor} 
\usepackage{tabularx} % increase table row widths
\usepackage{amsmath}

\newcommand{\comment}[1]{} %does nothing with the input, i.e. effectively comments it out

\begin{document}

\preprint{APS/123-QED}

\title{Inference of the sound speed and related properties of neutron stars}

\author{Len Brandes}
 \email{len.brandes@tum.de}
\author{Wolfram Weise}%
\email{weise@tum.de}
\author{Norbert Kaiser}%
\email{nkaiser@tum.de}
\affiliation{%
 Physics Department, Technical University of Munich, 85748 Garching, Germany
}%

\date{\today}% It is always \today, today,

\begin{abstract}

	Information on the phase structure of strongly interacting matter at high baryon densities can be gained from observations of neutron stars and their detailed analysis.  In the present work Bayesian inference methods are used to set constraints on the speed of sound in the interior of neutron stars,  based on recent multi-messenger data in combination with limiting conditions from nuclear physics at low densities. Two general parametric representations are introduced for the sound speed $c_s$ in order to examine the independence with respect to choices for the parametrisation of Priors.  Credible regions for neutron star properties are analysed,  in particular with reference to the quest for possible phase transitions in cold dense matter.  The evaluation of Bayes factors implies extreme evidence for a violation of the conformal bound, $c_s^2 \leq 1/3$,  inside neutron stars.  Given the presently existing data base,  it can be concluded that the occurrence of a first-order phase transition in the core of even a two-solar-mass neutron star is unlikely,  while a continuous crossover cannot be ruled out.  At the same time it is pointed out that the discovery of a superheavy neutron star with a mass $ M \sim 2.3 - 2.4\, M_\odot$ would strengthen evidence for a phase change in the deep interior of the star.
	
\end{abstract}

\maketitle

%\tableofcontents

\section{\label{sec:Introduction}Introduction}

Neutron stars are among the most extreme objects in the universe.  The densities inside their cores far exceed those reachable in terrestrial experiments, so that they provide a unique window into the physics of strongly interacting matter under extreme conditions \cite{Oezel2016a,Baym2018}.  

For a long time the composition of neutron star interiors has been subject to speculation.  The situation changed drastically when,  via pulsar timing of (general relativistic) Shapiro delays,  several heavy neutron stars with masses $M \sim 2\, M_\odot$ were discovered \cite{Demorest2010,Fonseca2016,Arzoumanian2018,Antoniadis2013,Cromartie2020,Fonseca2021}.  These observations set severe constraints on the equation of state (EoS) in the neutron star interior,  since the pressure inside must be sufficiently high to stabilise such heavy objects against gravitational collapse.  Further important information came from the gravitational wave signals of binary neutron star mergers observed by the LIGO and Virgo Collaborations \cite{Abbott2019,Abbott2020}, with resulting constraints on their tidal deformabilities.  Moreover,  the NICER telescope on board the International Space Station measured the first pulse profiles of hot spots on the surfaces of rapidly rotating pulsars.  From those,  the combined mass and radius of the corresponding two neutron stars could be inferred \cite{Miller2019,Miller2021,Riley2019,Riley2021}.  Many more multi-messenger data samples are expected to become available in the future.

In principle the physics of dense matter in the core of neutron stars is governed by Quantum Chromodynamics (QCD).  At asymptotically high baryon densities far beyond those encountered in neutron stars,  with Fermi momenta in the multi GeV range,  weakly interacting quarks and gluons are the relevant active degrees of freedom,  and perturbative QCD computations become feasible \cite{Fukushima2010,Alford1999,Gorda2021}.  The opposite,  low-density limit is accessible by Chiral Effective Field Theory (ChEFT) as the low-energy realisation of QCD.  ChEFT provides a systematic framework with controllable uncertainties for dealing with nuclear many-body systems \cite{Holt2016,Drischler2021b}.  This approach gives satisfactory descriptions of nuclear and neutron matter \cite{Wellenhofer2015} in a range of validity up to densities $n\lesssim 2\,n_0$, with $n_0\simeq 0.16\,$fm$^{-3}$ the equilibrium density of nuclear matter.

Between these extremes,  the nature and location of the transition from cold dense nuclear or neutron matter to quark and gluon degrees of freedom is still largely unknown.  The ab-initio method of numerically solving QCD on a lattice has been successful in analysing the phase structure at high temperatures and vanishing baryon chemical potential \cite{Bazavov2014,Borsanyi2014}.  But at non-zero baryon densities this approach is severely hindered by the sign problem of the fermionic determinant \cite{Aarts2016}.  Consequently,  calculations extending from $n\sim 2\,n_0$ to high densities are mostly based on models. Various hypotheses have been discussed in the literature,  ranging from a first-order (chiral) phase transition to a continuous hadron-quark crossover \cite{Fukushima2010,Buballa2005,McLerran2007,McLerran2019,Holt2016,Logoteta2021,Kojo2021a}.  Examples are Nambu - Jona-Lasinio type models which, when treated in mean-field approximation,  commonly suggest the existence of a first-order chiral phase transition well within the range of densities encountered in neutron stars.  Fluctuations beyond mean-field,  on the other hand,  tend to convert such a transition into a crossover shifted to much higher baryon densities \cite{Brandes2021}.  The idea of a smooth hadron-quark crossover at high density and zero temperature has indeed gained much significance in recent years \cite{Baym2018,Baym2019,Fukushima2020,Kojo2021,Kojo2021b}.

All possible scenarios for phase transitions or crossovers in dense matter would leave their characteristic signatures in the speed of sound, $c_s(n)$,  as a function of baryon density $n$.  For example,  a first-order phase transition would indicate itself by $c_s$ rapidly dropping to zero.  A crossover would show up less dramatically,  but still visibly in terms of a more or less pronounced change of slope in $c_s(n)$.  These are the distinctive features that motivate our investigation of constraints on the behaviour of the sound speed in neutrons stars,  based on the presently available observational data. 

In this work we translate the recently collected exterior neutron star observables, together with state-of-the-art ChEFT results \cite{Drischler2021,Drischler2022},  into restrictions for interior neutron star properties using Bayesian inference,  an approach which has been used extensively in the literature \cite{Greif2019,Raaijmakers2019,Raaijmakers2020,Raaijmakers2021,Landry2020,Legred2021,Steiner2010,Raithel2016,Raithel2017,Raithel2021,Dietrich2020,Pang2021,Lim2019,Lim2022,Biswas2021,Biswas2021a,AlMamun2021,Gorda2022,Huth2022}. There are notable differences between these analyses.  Here, we largely follow the procedure outlined in Refs.\,\,\cite{Riley2018,Miller2019a}.  Alternative approaches use neural networks for the inference procedure \cite{Fujimoto2020,Fujimoto2021,Morawski2020,Ferreira2019,Traversi2020,Krastev2021,Soma2022} or remove equations-of-state that do not reproduce neutron star properties within the credible intervals of the astrophysical observables \cite{Annala2020,Annala2022,Altiparmak2022,Somasundaram2021,Somasundaram2022}.

 In our analysis we pay particular attention to the speed of sound inside neutron stars.  It is modelled using two general representations that were introduced in previous studies,  namely a Gaussian parametrisation \cite{Greif2019,Tews2019} and a parametrisation based on segment-wise linear interpolations \cite{Annala2020,Annala2022}. A comparison of inference results with these two forms as input gives an impression of the model dependence caused by possible biases in the choices of parametrisations. In contrast to previous works we suggest a new implementation of the ChEFT constraint as a Likelihood instead of employing it a priori.  This has the advantage of dealing with the low-density constraint in a way consistent with the treatment of the astrophysical data.

From the inferred behaviour of the sound speed and a detailed assessment in terms of Bayes factors we deduce implications regarding the likelihood of phase transitions inside neutron stars.  Such an extensive investigation of Bayes factors has not been performed in previous work.  Part of this discussion also concerns the possible range of validity for a description of neutron star matter in terms of conventional baryonic degrees of freedom. Recent studies have examined the potential impact of asymptotic pQCD on neutron star properties \cite{Somasundaram2022,Komoltsev2022,Gorda2022}.  In this context we investigate the role of different asymptotic behaviours,  particularly with regard to phase transitions.

This paper is organised as follows: In Section \ref{sec:Theory}, following a quick introduction of the TOV equations, the EoS and the speed of sound,  we give a brief survey of possible phases at high densities and introduce the two parametrisations to model the speed of sound inside neutron stars.  In Section \ref{sec:BayesianInference} the statistical procedure is explained, which we use to infer constraints for neutron star properties based on current empirical data and theoretical low-density conditions. The results for the sound speed and related properties are presented and discussed in Section \ref{sec:Results}. Based on these findings, implications for the phase structure inside neutron stars are examined.  A summary and conclusions follow in Section \ref{sec:Summary}. 

\section{\label{sec:Theory} Speed of sound in neutron stars}

\subsection{TOV equations and EoS}

A description of neutron star matter as a general relativistic ideal fluid with spherical symmetry leads to a coupled system of differential equations,  the Tolman-Oppenheimer-Volkoff (TOV) equations:
\begin{align}
	\frac{\partial P(r)}{\partial r}=& -\frac{G_N}{r^2} \left[\varepsilon(r) + P(r) \right] \left[m(r) + 4\pi r^3 P(r) \right] \nonumber \\ &\times  \left( 1- \frac{2 G_N\,m(r)}{r} \right)^{-1}~, \label{eq:TOV1} \\
	\frac{\partial m(r)}{\partial r}=\,& 4\pi r^2\varepsilon(r)~,
	\label{eq:TOV2}
\end{align}
where $G_N$ is the gravitational constant.  Given an equation of state (EoS) $P(\varepsilon)$, i.e. pressure as a function of energy density $\varepsilon$,  this system can be solved with the boundary condition $m(r=0) = 0$ and a central pressure $P(r=0) = P_c$.  The mass of the star is given as $M = m(R) = 4\pi \int_0^R dr\,r^2\varepsilon(r)$,  while the star radius $R$ is determined as the point at which the pressure vanishes, $P(R) = 0$. The TOV equations describe non-rotating neutron stars.  The effect of the rotation on $R$ is expected to become only relevant for very high pulsar spin frequencies \cite{Lattimer2004,Haensel2009}.

Matter in the interior of a neutron star can be described in terms of the squared speed of sound, 
\begin{eqnarray}
	c_s^2(\varepsilon) = \frac{\partial P(\varepsilon)}{\partial \varepsilon} \geq 0~,
	\label{eq:soundspeed}
\end{eqnarray}
from which the EoS is determined as 
\begin{eqnarray}
	P(\varepsilon) = \int^\varepsilon_0 d\varepsilon' \, c_s^2(\varepsilon')~.
	\label{eq:EoSfromCs}
\end{eqnarray} 
Causality demands that the speed of sound must always remain smaller than or equal to the speed of light (c =1 in our units), i.e. $c_s \leq 1$.  In addition, thermodynamic stability of the star dictates that the derivative $\partial P/\partial \varepsilon$ is non-negative.  At extremely high densities perturbative QCD calculations become feasible in terms of quark and gluon degrees of freedom.  They suggest that at densities $n \sim 50 \, n_0$ the squared speed of sound approaches the conformal bound, 
\begin{eqnarray}
	c_s^2 \rightarrow 1/3~,
\end{eqnarray}
from below \cite{Fraga2014}.
This limit can be derived from naive dimensional analysis and asymptotic freedom \cite{Hippert2021}.  In fact it is expected that this bound holds in all conformal theories, i.e.  field theories in which the trace of the energy-momentum tensor vanishes \cite{Bedaque2015,Tews2019}.  However,  recent analyses based on astrophysical observables suggest that this conformal bound can be violated inside neutron stars \cite{Landry2019,Landry2020,Legred2021,Leonhardt2020,Altiparmak2022,Gorda2022}.  Squared sound velocities with $c_s^2 > 1/3$ were also found in recent $N_c = 2$ lattice QCD computations \cite{Iida2022}.  A possible mechanism for the violation of this bound, based on the trace anomaly in strongly coupled matter, is discussed in Ref.\,\,\cite{Fujimoto2022a}.  In that context it is interesting that Hard Dense Loop resummation methods (going beyond basic perturbative QCD) indicate that the conformal limit may be approached asymptotically from above,  with $c_s^2 >1/3$ \cite{Fujimoto2022}.  In a later section we shall examine whether this changed asymptotic behaviour has an impact on the sound speed at neutron star core densities.

For neutron stars in binary systems the gravitational field of the companion induces a quadrupole moment in the star,  which can be modelled by a linearised metric perturbation \cite{Flanagan2008,Hinderer2008}. The resulting two coupled differential equations are solved simultaneously with the TOV equations \footnote{We use the LALSuite library \cite{LIGO2018} for a fast numerical solution of the full system of coupled differential equations.}. From the numerical solution the dimensionless tidal deformability $\Lambda$ can be inferred.  When the TOV equations and the equations for the tidal deformability are solved together for a variety of central pressures, $P_c$, the mass-radius relation and the tidal deformabilities $(M_i, R_i, \Lambda_i)$ are obtained for each given EoS, or equivalently, for given $c_s^2(\varepsilon)$.

\subsection{Phases of strongly interacting matter}
\label{sec:phases}

At high temperatures and small baryon chemical potentials the phase structure of the strong interaction is well understood from lattice QCD \cite{Bazavov2014,Borsanyi2014} and from high-energy heavy ion collisions \cite{BraunMunzinger2016,Andronic2018}.  At vanishing baryon chemical potential a continuous crossover proceeds from the hadronic to the quark-gluon phase around a pseudocritical temperature of about $155\,$MeV.  This behaviour is reflected in the speed of sound which grows rapidly with increasing temperature in the hadronic phase and then decreases along the crossover transition.  At asymptotically high temperatures $c_s$ increases again to reach the asymptotic value of the conformal limit from below \cite{Romatschke2019}. 

At low baryon densities it is quite well established that the thermodynamics of (isospin symmetric) nuclear matter features a first-order liquid-gas phase transition,  with a critical point located empirically \cite{Elliott2013} at a temperature $T_{crit}\simeq 18$ MeV and density $n_{crit}\simeq n_0/3$.  When viewed in a $(T,\mu)$ phase diagram,  the first-order liquid-gas transition line starting at the critical point reaches the $T=0$ axis at a baryon chemical potential $\mu = m_N - B \simeq 923 \,$MeV corresponding to the binding energy per particle $B \simeq 16\,$MeV of symmetric nuclear matter.  With an empirical symmetry energy $S \simeq 32\,$MeV, this phase transition is absent in pure neutron matter.  

At asymptotically high densities, $n \gtrsim 50\, n_0$, quark and gluon degrees of freedom take over in a colour superconducting phase \cite{Alford1999,Fukushima2010,Schaefer1999,Alford2008}.  Still unknown remains the detailed nature and density range of the transition from nuclear to quark matter.  Many models have been designed,  with a variety of hypotheses predicting different active degrees of freedom in this intermediate region.  With their core densities of up to $n \sim 6\, n_0$ \cite{Legred2021} and low temperatures,  neutron stars are the objects of choice to gain information about this speculative region of the phase diagram. 

As mentioned in the introduction,  studies based on Nambu - Jona-Lasinio type models in mean-field approximation have commonly found a first-order chiral phase transition at quite moderate baryon densities for $T = 0$ \cite{Buballa2005}.  A first-order phase transition would manifest itself in the speed of sound rapidly decreasing to zero.  It could lead to mass-radius relations with a disconnected third-family branch of compact stars containing exotic matter \cite{Benic2015}.  On the other hand,  investigations using non-perturbative Functional Renormalization Group techniques \cite{Drews2015,Drews2017,Brandes2021} found that fluctuations beyond mean-field tend to convert the first-order chiral transition into a crossover shifted to much higher baryon densities,  even beyond those realised in neutron star cores.  In any case,  to support the observed heavy neutron stars with masses $M \sim 2\, M_\odot$,  a transition to quark matter in neutron stars at relatively low densities is possible only if the quark EoS is extremely stiff,  or otherwise the transition has to take place at high densities leading to small quark cores \cite{RaneaSandoval2016,Kojo2021a}.

Models proposing a continuous crossover from hadronic to quark matter are often referred to under the key word {\it quark-hadron continuity}.  Such models describe the low-density part of the EoS in agreement with ChEFT calculations but still provide the necessary stiffness to support heavy neutron stars,  usually by introducing strongly repulsive correlations in the quark sector \cite{Baym2018,Baym2019,Fukushima2020,Kojo2021,Kojo2021b}.  A continuous crossover may be visible as a maximum in the speed of sound as a function of density and might be realized through an intermediate phase of quarkyonic matter \cite{McLerran2007,McLerran2019}, a combined phase of quarks and nucleons derived from large $N_c$ considerations \cite{Fukushima2010,Zhao2020}.  

At sufficiently high densities in neutron stars,  the formation of hyperons through electroweak processes may become energetically favourable.  It was frequently argued,  however,  that the additional degrees of freedom introduced via the hyperons lead to a softening of the equation of state such that heavy neutron stars with $M \sim 2\, M_\odot$ cannot be supported against gravitational collapse \cite{Lonardoni2015, Logoteta2021}.  Introducing repulsive hyperon-nuclear three-body forces is a possible way to inhibit the appearance of hyperons in neutron stars altogether \cite{Haidenbauer2017,Gerstung2020}.  An alternative picture \cite{Motta2021} couples baryons (including hyperons) to a density-dependent non-linear scalar field that effectively represents repulsive many-body correlations,  such that the required stiffness of the EoS can be maintained even in the presence of hyperons in the neutron star core.  A characteristic feature of this model is a sharply dropping speed of sound at the onset density for the appearance of hyperons.

\subsection{Parametrisations}
\label{sec:parametrisations}
A variety of parametrisations has been introduced to represent the equation of state in neutron stars,  among the most prominent ones are piecewise polytropes \cite{Read2009} or spectral representations \cite{Lindblom2010}. As discussed in the previous Section, various theories predict different phase structures at high densities including phase transitions or crossovers, which are reflected in the behaviour of the speed of sound \cite{Somasundaram2021}.  In the present analysis we employ two different parametrisations for $c_s^2(\varepsilon)$ inside neutron stars: a skewed Gaussian function and piecewise segmented linear interpolations. We prefer not to choose parametrisations of $P(\varepsilon)$ such as piecewise polytropes. The reason is that such representations can cause unphysical discontinuous effects in the speed of sound.  In contrast, the parametrisations employed here are continuous in $c_s^2(\varepsilon)$.  They dependent on sets $\theta$ of either six or eight parameters.  A comparative study using these two different forms will give an impression of possible systematic uncertainties induced by the choice of parametrisation.  At very low densities, $n \leq 0.5 \, n_0$, the speed of sound is matched to the neutron star crust modelled by the time-honoured Baym-Pethick-Sutherland (BPS) parametrisation \cite{Baym1971}. The effect of the neutron star crust on observables studied in this work is expected to be small.

\subsubsection{Gaussian}

Based on Refs.\,\,\cite{Greif2019,Tews2019} we represent the squared speed of sound of neutron star matter at zero temperature as a function of energy density by a skewed Gaussian.  A logistic function is added such that the parametrisation reaches the conformal limit $c_s^2 \rightarrow 1/3$ at asymptotically high energy densities.  With $x = \varepsilon/(m_N n_0)$ where $m_N$ is the free nucleon mass,  the squared speed of sound is written as:  
\begin{eqnarray}
c_s^2 (x, \theta) &\!\!\!=\!\!\!&
	a_1 \text{exp}\left[ -{\frac12}\frac{(x - a_2)^2}{a_3^2} \right] \left(1 + \text{erf}\left[ \frac{a_6}{\sqrt{2}} \frac{x - a_2}{a_3}\right] \right) \nonumber \\&&+ \frac{1/3 - a_7}{1 + \text{exp}\left[ - a_5 (x - a_4) \right]}  + a_7 ~,
	\label{eq:GaussianParametrisation}
\end{eqnarray}
with $\text{erf}(z) = \frac{2}{\sqrt{\pi}}\int_0^zdt ~ e^{-t^2}$ the conventional error function. 
The parameter $a_7$ is determined such that the transition to the neutron star crust is continuous.  Hence,  six free parameters $\theta = (a_1, \dots, a_6)$ remain. 
When $c_s^2 (x, \theta)$ becomes negative, violating thermodynamic stability of the star,  we set $c_s^2 = 0$.  In this way the Gaussian parametrisation can describe arbitrarily strong phase transitions.  The combination of the Gaussian and logistic function can also account for variable crossovers.  As argued in Section \ref{sec:phases},  a local maximum in the speed of sound can indicate a transition from baryonic to quark dynamics \cite{Baym2018,Kojo2021,Hippert2021} or the onset of hyperonic degrees of freedom \cite{Motta2021}.

\subsubsection{Segments}

The Gaussian parametrisation assumes a specific functional form of the sound speed inside neutron stars.  At the present stage the empirical data base is still limited,  so that inference procedures can depend sensitively on Prior choices including the functional form of the parametrisation \cite{Raithel2017,Raaijmakers2019}.   For an alternative test,  results of broader generality can be produced using a more universal parametrisation of the speed of sound based on segment-wise linear interpolations,  similar to Refs.\,\,\cite{Annala2020,Annala2022}. 
The parameters of the model are $N$ points $\theta = (c_{s,i}^2, \varepsilon_i)$.  The squared speed of sound $c_{s}^2(\varepsilon, \theta)$ is modelled as a linear interpolation between these $N$ points, i.e. for $\varepsilon \in [\varepsilon_{i}, \varepsilon_{i+1}]$ with $ i = 0, \dots, N$:
\begin{eqnarray}
	c_{s}^2(\varepsilon, \theta) = \frac{(\varepsilon_{i+1} - \varepsilon) c_{s,i}^2 + (\varepsilon - \varepsilon_{i})c_{s,i+1}^2}{\varepsilon_{i+1} - \varepsilon_{i}}~.
\end{eqnarray}
Here we choose $N = 5$. The $i=0$ point is the transition point to the neutron star crust and the last point is chosen such that the conformal limit is reached at very high energy densities $(c_{s,5}^2, \varepsilon_5) = (1/3, 10\,\text{GeV\,fm}^{-3})$. We have checked that the results do not depend on the specific choice of $\varepsilon_5$ as long as its value is large enough. The asymptotic end point at $\varepsilon_5 = 10\,\text{GeV\,fm}^{-3}$ corresponds to a baryon chemical potential of $\mu \sim 2.4\,$GeV in the pQCD results from Ref.\,\,\cite{Gorda2021}. Ref.\,\,\cite{Altiparmak2022} uses a similar parametrisation based on piecewise segments. There it is found that five segments are sufficient to avoid numerical artefacts,  namely that for a larger number of segments the results do not change significantly any more. An equivalent or smaller number of segments is used to interpolate over the full range between ChEFT and pQCD constraints in Refs.\,\,\cite{Annala2020,Annala2022}. The parametrisation in terms of segments can also incorporate a variety of phase transitions or crossovers.  In contrast to the Gaussian parametrisation it can also accommodate possible steep rises as well as plateaus in the speed of sound.

\section{Bayesian inference}
\label{sec:BayesianInference}

\subsection{Basics}
\label{sec:BayesianInferenceBasics}

Making use of a set of neutron star observables, we aim now to find credible regions for the free parameters $\theta$ of the two parametrisations described in Sec.\,\ref{sec:parametrisations}.  For that purpose we use Bayesian inference, similar to Refs.\,\,\cite{Greif2019,Raaijmakers2019,Raaijmakers2020,Raaijmakers2021,Landry2020,Legred2021,Steiner2010,Raithel2016,Raithel2017,Raithel2021,Dietrich2020,Pang2021,Lim2019,Lim2022,Biswas2021,Biswas2021a,AlMamun2021,Gorda2022,Huth2022}, 
and follow  Refs.\,\,\cite{Riley2018,Miller2019a}.  For given data $\mathcal{D}$ and a model $\mathcal{M}$ which includes all assumptions such as the choice of parametrisation,  the {\it Posterior} probability distribution for the parameters $\theta$ can be computed using Bayes' theorem:
\begin{align}
	\text{Pr} &(\theta|\mathcal{D}, \mathcal{M}) \nonumber \\
	&= \int dP_c ~ \frac{ \text{Pr} (\mathcal{D}|\theta, P_c, \mathcal{M}) \, \text{Pr} (P_c|\theta, \mathcal{M}) \, \text{Pr} (\theta| \mathcal{M})}{\text{Pr} (\mathcal{D}| \mathcal{M}) }~,
	\label{eq:BayesTheorem}
\end{align} 
where the Posterior probability distribution is marginalised over the central pressures $P_c$.
The probability distribution $\text{Pr} (\theta| \mathcal{M})$ for the parameters given the model $\mathcal{M}$ is denoted the {\it Prior} and given by the chosen parameter distributions. The Prior for the central pressures, $ \text{Pr} (P_c|\theta, \mathcal{M})$, depends on $\theta$ because the maximum central pressure leading to a stable solution is different for each set of parameters. The probability $\text{Pr} (\mathcal{D}|\theta, P_c, \mathcal{M})$ for the data $\mathcal{D}$ to occur, given the parameters $\theta$, the central pressures $P_c$ and the model $\mathcal{M}$, is usually referred to as the {\it Likelihood}. By numerically solving the coupled system of TOV equations and the equations for the tidal deformability, a set of parameters $\theta$ and central pressures $P_c$ is deterministically linked to a mass-radius relation $M$, $R$ and tidal deformabilities $\Lambda$. Therefore we can write
\begin{eqnarray}
	\text{Pr} ( \mathcal{D}|\theta, P_c, \mathcal{M}) = \text{Pr} ( \mathcal{D}|M, R, \Lambda, \mathcal{M})~.
\end{eqnarray}
For computational feasibility we assume that we can use the Posterior distributions from the analyses of neutron star observables as Likelihoods for external neutron star parameters, similar to the treatments in Refs.\,\,\cite{Riley2018,Greif2019}:
\begin{eqnarray}
	\text{Pr} ( \mathcal{D}|M, R, \Lambda, \mathcal{M}) \propto \text{Pr} ( M, R, \Lambda |\mathcal{D}, \mathcal{M})~.
\end{eqnarray}
This is valid if the Prior of $(M, R, \Lambda)$ used in the analyses of the observational data is sufficiently flat, which is the case for the observables analysed in this work \cite{Raaijmakers2021}. The Likelihood $\text{Pr} ( M, R, \Lambda |\mathcal{D}, \mathcal{M})$ can then be evaluated for a given set of parameters based on the observables as explained in Sec.\,\ref{sec:measurements}. The probability distribution $\text{Pr} (\mathcal{D}| \mathcal{M})$ in the denominator of Eq.\,\,(\ref{eq:BayesTheorem}) is usually referred to as the {\it Evidence} or marginal Likelihood. It is determined by the normalization of the Posterior:
\begin{align}
	\text{Pr} &(\mathcal{D}|\mathcal{M}) \nonumber \\
	&=  \int d \theta \int dP_c \,~  \text{Pr} (\mathcal{D}|\theta, P_c, \mathcal{M}) \, \text{Pr} (P_c|\theta, \mathcal{M}) \, \text{Pr} (\theta| \mathcal{M})  ~.
\end{align} 
Depending on the number of parameters this may be a high-dimensional integral which can be difficult to solve numerically.  In Bayesian inference,  sampling algorithms such as Markov Chain Monte Carlo or nested sampling are commonly used.  For a sufficiently low-dimensional parameter space,  samples from the Prior $\text{Pr} (\theta| \mathcal{M})$ weighted with the Likelihood $\text{Pr} ( \mathcal{D}|\theta, P_c, \mathcal{M})$ marginalised over the central pressures yield the Posterior probability distribution up to a multiplicative constant.  In this case it needs to be checked whether the number of samples is large enough such that sufficient probability mass of the Posterior has been covered.  From this Posterior probability distribution credible regions for the parameters $\theta$ can be inferred. 

To transform these credible regions to the EoS space,  we follow Ref.\,\,\cite{Greif2019} in discretizing energy densities on a grid $\{\varepsilon_{i}\}$.  For each Posterior sample the pressure is determined at each discrete energy density $P(\varepsilon_{i}, \theta)$, up to the maximum central energy density $\varepsilon_{c,max}$, corresponding to the endpoint  $M_{max}$ of the mass-radius relation.  In this way we obtain the Posterior distribution for the pressure $\text{Pr}\big(P \big| \varepsilon_i,  \mathcal{D}, \mathcal{M}  \big)$ at each energy density.  We can then determine the highest density credible interval $[a,b]$ at the levels $\alpha = 68\%$ or $95\%$ as 
\begin{eqnarray}
	\alpha = \int_a^b dP \,~ \text{Pr}\big(P \big| \varepsilon_i,  \mathcal{D}, \mathcal{M}  \big)~.
	\label{eq:CredInt}
\end{eqnarray}
Combining the credible intervals at each $\varepsilon_i$ gives a Posterior credible band for $P(\varepsilon)$. Similarly we can find credible bands for $c_s^2(\varepsilon)$, $R(M)$, $\Lambda(M)$, etc. In contrast, displaying neutron star properties such as the EoS $P(\varepsilon)$ via a two-dimensional credible region depends on the chosen Prior in $\varepsilon$,  so that different Prior choices can lead to different results.  Hence in the literature, with few exceptions, the procedure in terms of credible bands is favoured \cite{Greif2019,Raaijmakers2019,Raaijmakers2020,Raaijmakers2021,Landry2020,Legred2021,Huth2022}.  Note that each EoS is only used up to its respective endpoint,  i.e.  the point at which the central energy density $\varepsilon_{c,max}$ generates the maximum mass $M_{max}$  of the neutron star.  At higher energy densities (or masses),  the credible intervals computed via Eq.\,\,(\ref{eq:CredInt}) are determined on the basis of correspondingly fewer equations of state. This loss of expressive power at higher energy densities and masses is not reflected in the credible bands.

For two competing hypotheses the ratio of their marginal Likelihoods is referred to as the Bayes factor.  It permits a quantification of the evidence for one hypothesis over the other one,  based on the data.  More details on the evaluation of Bayes factors as well as a commonly used classification scheme from Refs.\,\,\cite{Jeffreys1961,Lee2016} can be found in Appendix \ref{sec:Bayesfactors}.

\subsection{Priors}

To compute the Posterior probability distribution (\ref{eq:BayesTheorem}),  Prior probability distributions for the parameters and for the central pressures must be chosen. The central pressures are taken from a uniform distribution $P_c \in [1.56\,$MeV$\,$fm$^{-3},P_{c,max}(\theta)]$, where $P_{c,max}(\theta)$ is the maximum pressure corresponding to the last stable solution with maximum mass for each parameter set $\theta$.  As noted in Ref.\,\,\cite{Miller2021}, the pressure increases rapidly towards the upper end of the mass-radius relation,  so that this chosen Prior leads to a higher weighting of larger masses. In the literature some works use instead Priors that are uniform in the individual masses.  However, with increasing maximum mass,  the EoS probability to support such masses decreases.  So, a stronger weighting of larger masses is indeed advised. %because uniform Priors in individual masses only reach the respective maximum mass 

In order to ensure maximum generality of the results we choose very broad parameter ranges for both parametrisations,  covering most of the speed-of-sound space.  However,  as in Refs.\,\,\cite{Raaijmakers2019,Raaijmakers2020,Raaijmakers2021}, we discard parameter sets that lead to multiple disconnected stable mass-radius relations.  EoS with multiple stable branches were found to be disfavoured by the data in previous analyses,  because their majority cannot support neutron star masses as high as $2\,M_\odot$ \cite{Essick2020}. The similar radii of the two pulsars measured with NICER also render a twin-star scenario unlikely at high densities.  At the same time,  ChEFT in combination with EoS constraints from heavy-ion collisions  \cite{Fevre2016}, as inferred in Ref. \cite{Huth2022},  rule out such a scenario at densities $n\lesssim 3\,n_0$.

\subsubsection{Gaussian}

The six free parameters of the Gaussian parametrisation (referred to in the following as version G) are sampled from uniform intervals listed in Table \ref{tab:GaussianParameterPrior}. These parameter ranges were chosen guided by previous studies \cite{Greif2019,Tews2019}. The resulting functions cover the speed-of-sound space sufficiently well.  Only those combinations of parameters are kept that lead to causal EoS, i.e. $c_s^2(\varepsilon) < 1$ for all energy densities.  In our default version G the asymptotic conformal limit, $c_s^2 = 1/3$, is approached from below as in standard pQCD \cite{Fraga2014},  implying that the derivative of the speed of sound must be positive,  $\partial c_s^2/\partial \varepsilon >0$,  at very high energy densities.  In practice this onset of asymptotic behaviour is imposed at three different values,  $\varepsilon = 4, \,8$ and $16\, \text{GeV}\,\text{fm}^{-3}$.  We have checked that this specific choice does not affect the inference results as long as these energy densities are sufficiently large.

\begin{table}[tp]
	\centering
	%\begin{ruledtabular}
	\begin{tabularx}{0.46\linewidth}{|X|lcl|}
		\hline \hline
		Parameter & \multicolumn{3}{l|}{Range\textcolor{white}{LANIII}} \\ \hline
		$a_1$ & 0.2 & $\,-\,$ & 3 \\ 
		$a_2$ & 0.5 & $\,-\,$ & 12 \\
		$a_3/a_2$ & 0.05 & $\,-\,$ & 10 \\
		$a_4$ & 0.1 & $\,-\,$ & 15 \\
		$a_5$ & 0.1 & $\,-\,$ & 5 \\
		$a_6$ & -15 & $\,-\,$ & 15 \\
		\hline \hline
	\end{tabularx}
	%\end{ruledtabular}
	\caption{Prior ranges for the six parameters of the Gaussian parametrisation of the speed of sound inside neutron stars given in Eq.\,\,(\ref{eq:GaussianParametrisation}).}
	\label{tab:GaussianParameterPrior}
\end{table}  

\subsubsection{Segments}

The segment-wise parametrisation (referred to in the following as version S) depends on four speeds of sound and energy densities $(c_{s,i}^2, \varepsilon_i)$. The energy densities are sampled logarithmically from $\varepsilon_i \in [\varepsilon_{\text{crust}},  4 \, \text{GeV} \, \text{fm}^{-3}]$, where $\varepsilon_{\text{crust}}$ refers to the endpoint of the neutron star crust. With this sampling the large multitude of EoS's in the Prior is represented,  on average,  by 3 - 4 segments. The speed-of-sound values are collected from logarithmic intervals $c_{s,i}^2 \in [0,1]$,  so they are causal by construction and at the same time open to the possible occurrence of phase transitions. The asymptotic conformal limit is approached from below,  which is realised by restricting the last speed-of-sound value before the end point to $c_{s,4}^2 < 1/3$. With two more parameters and a more general functional form,  the Segments parametrisation allows,  in principle,  to describe more complex structures.  As a stability test  we have checked that shifting the upper limit of the logarithmic interval downward from its value $\varepsilon_{i,max} = 4\,$GeV/fm$^3$ induces small changes in the Prior but does not affect the final Posterior results. 

\begin{figure*}[tp]
	\begin{center}
		\includegraphics[height=55mm,angle=-00]{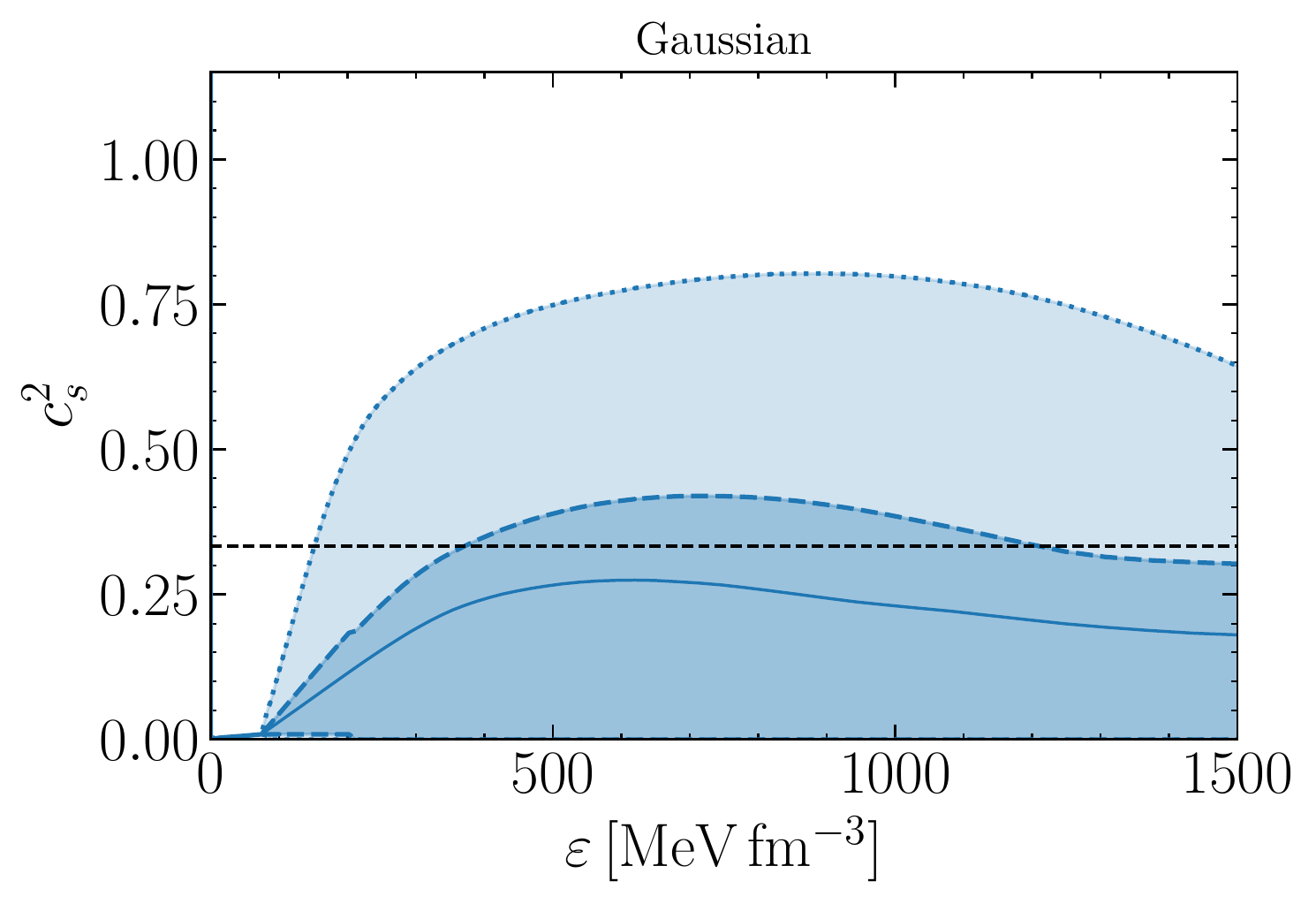} 
		\includegraphics[height=55mm,angle=-00]{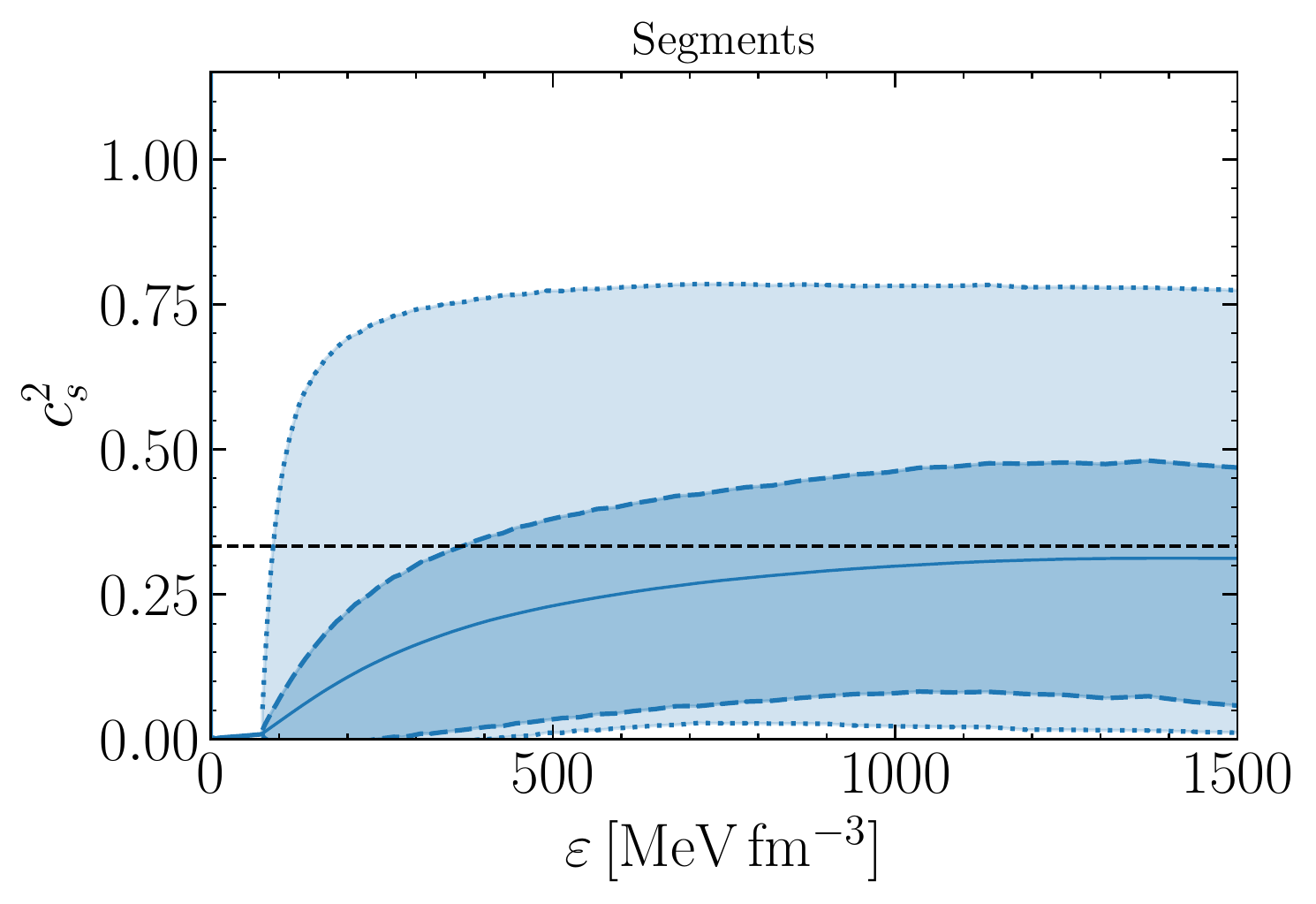} \\
		\includegraphics[height=55mm,angle=-00]{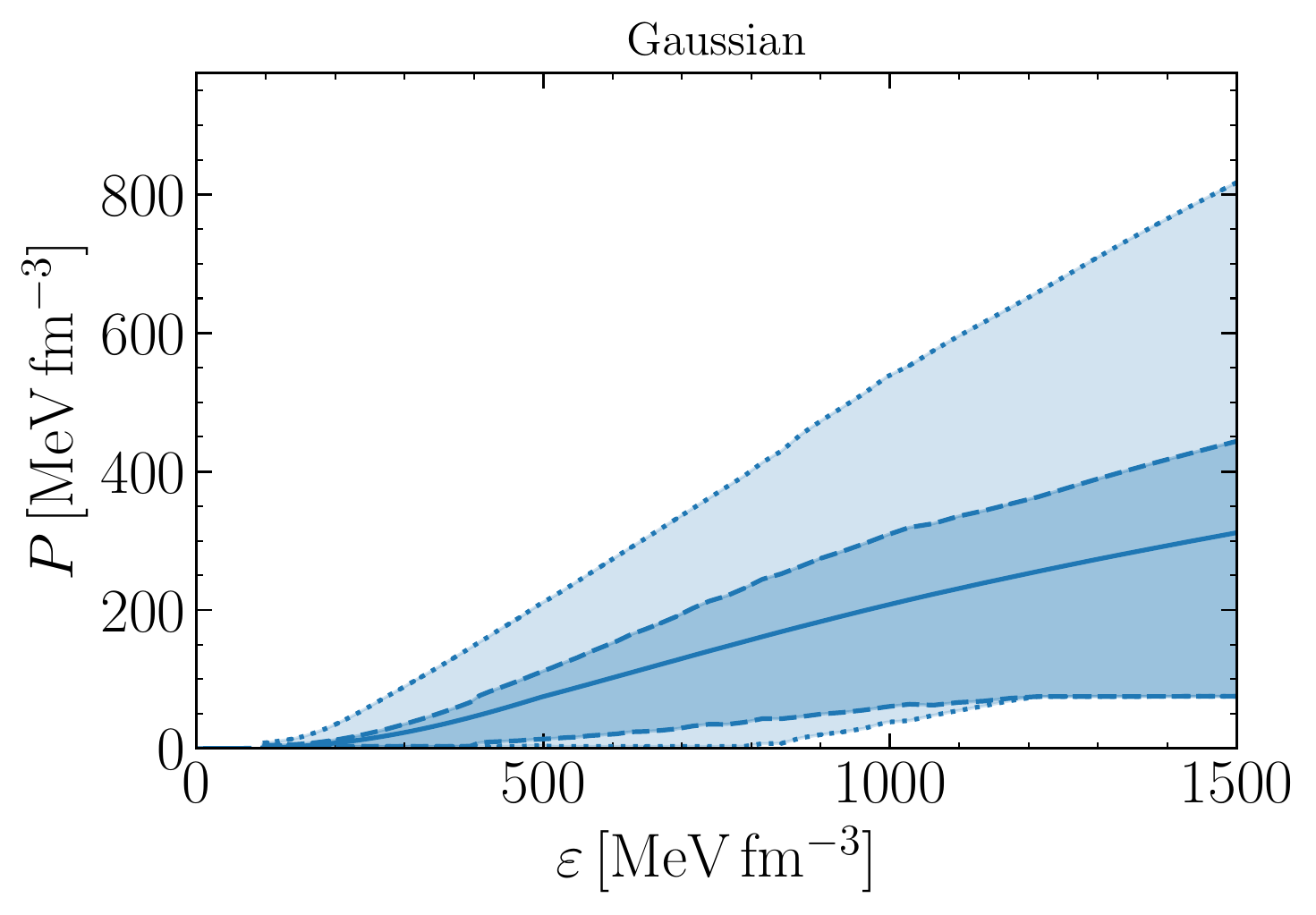}
		\includegraphics[height=55mm,angle=-00]{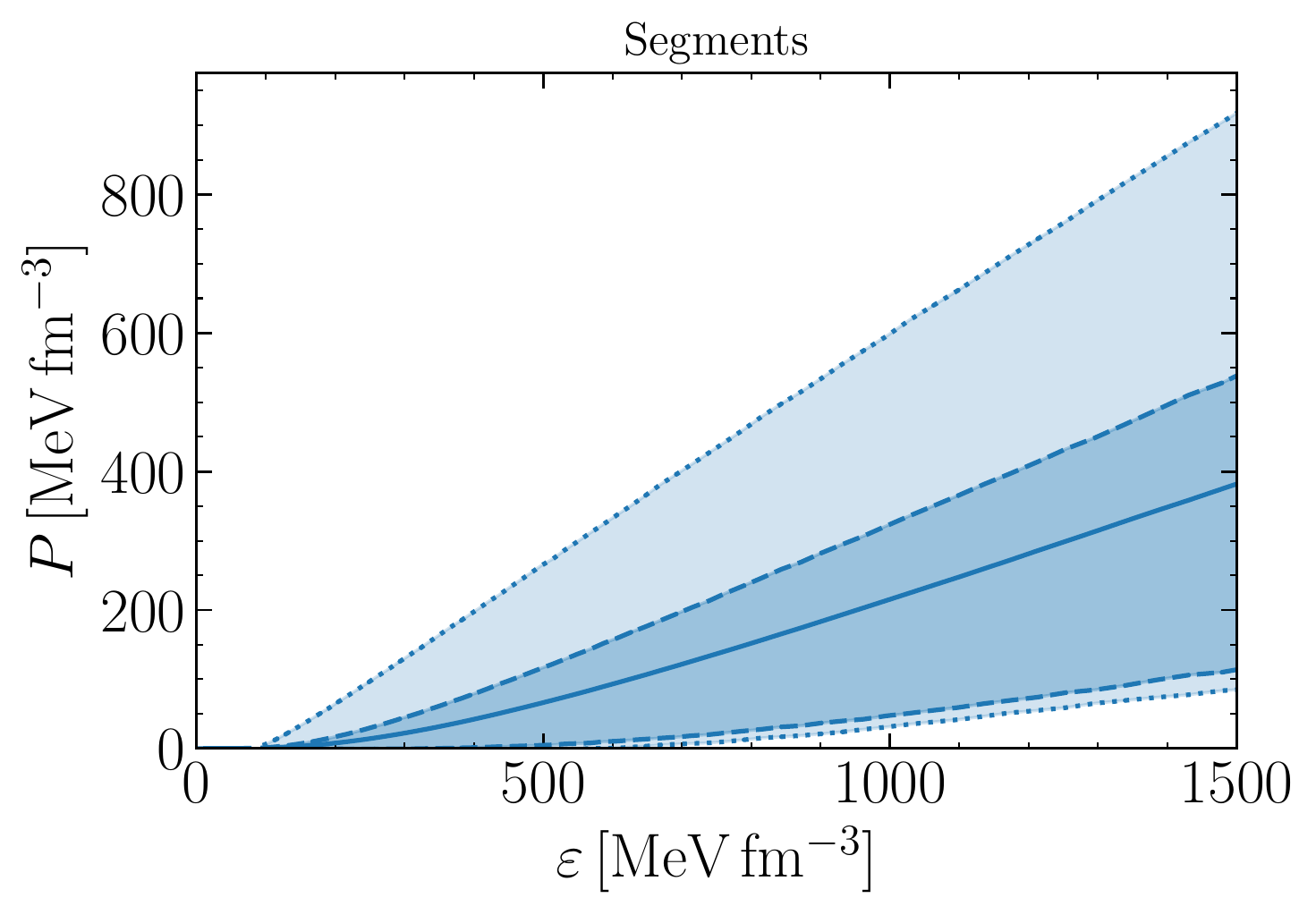} \\
		\includegraphics[height=55mm,angle=-00]{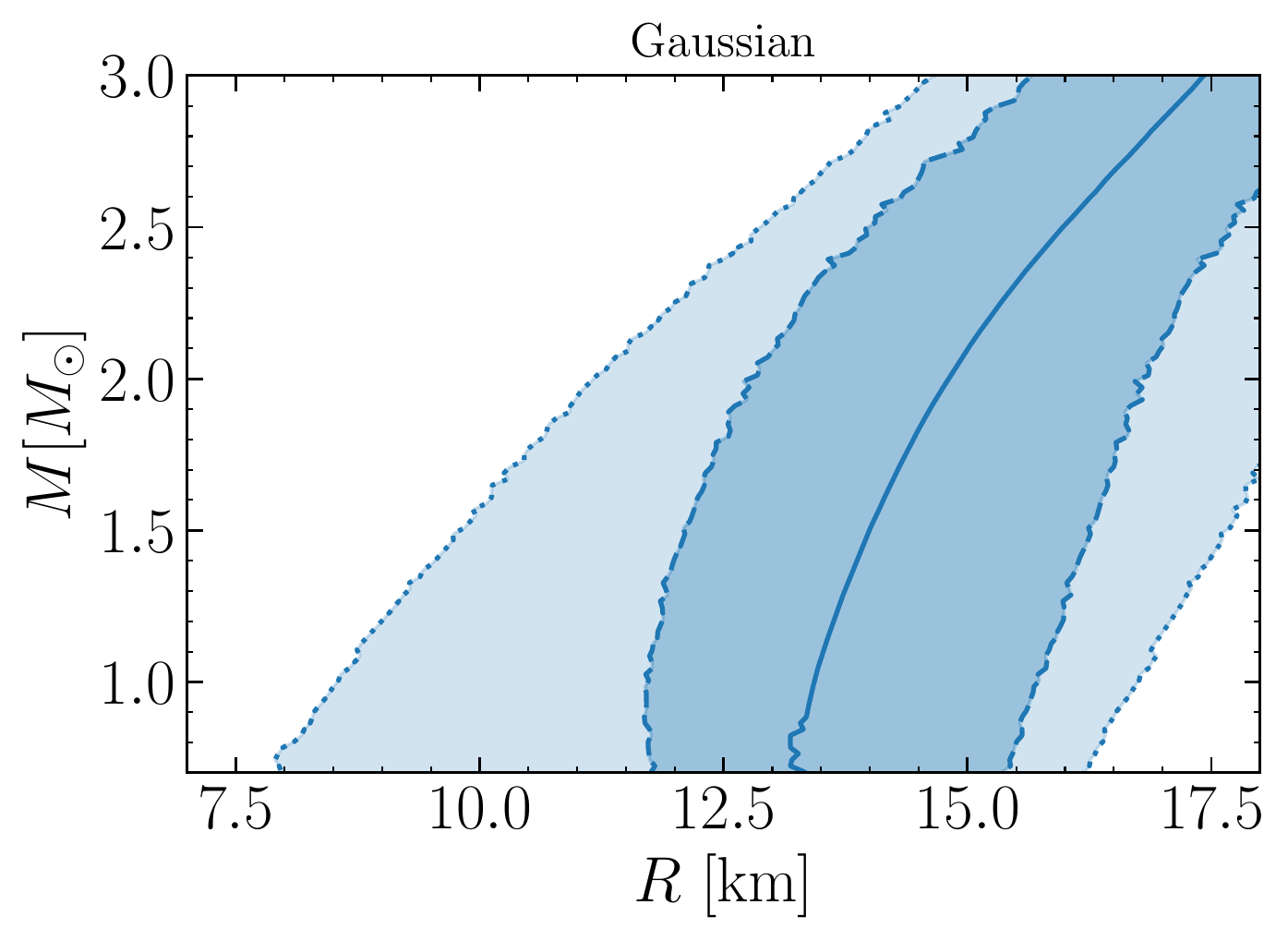}
		\includegraphics[height=55mm,angle=-00]{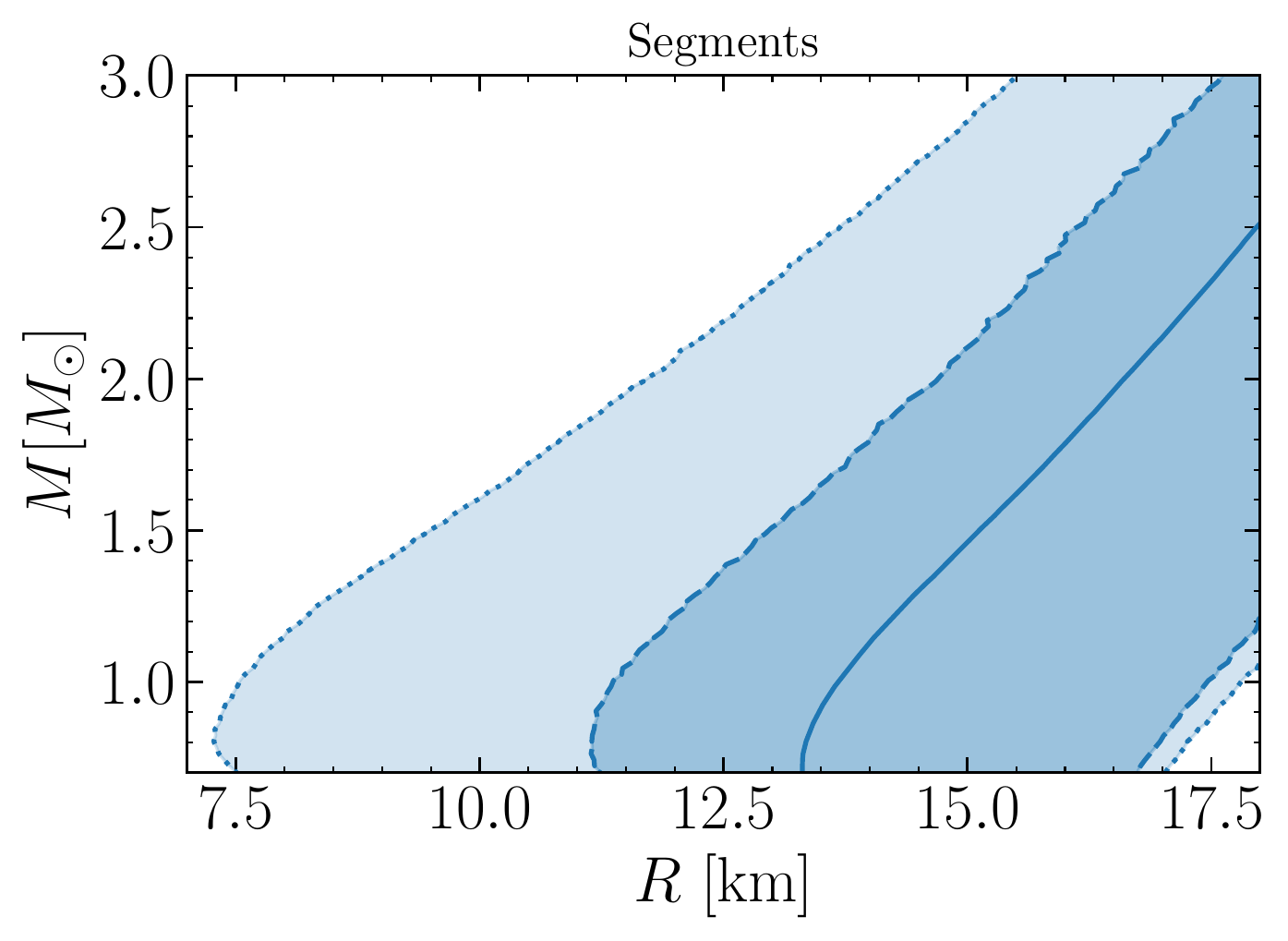} \\
		\includegraphics[height=55mm,angle=-00]{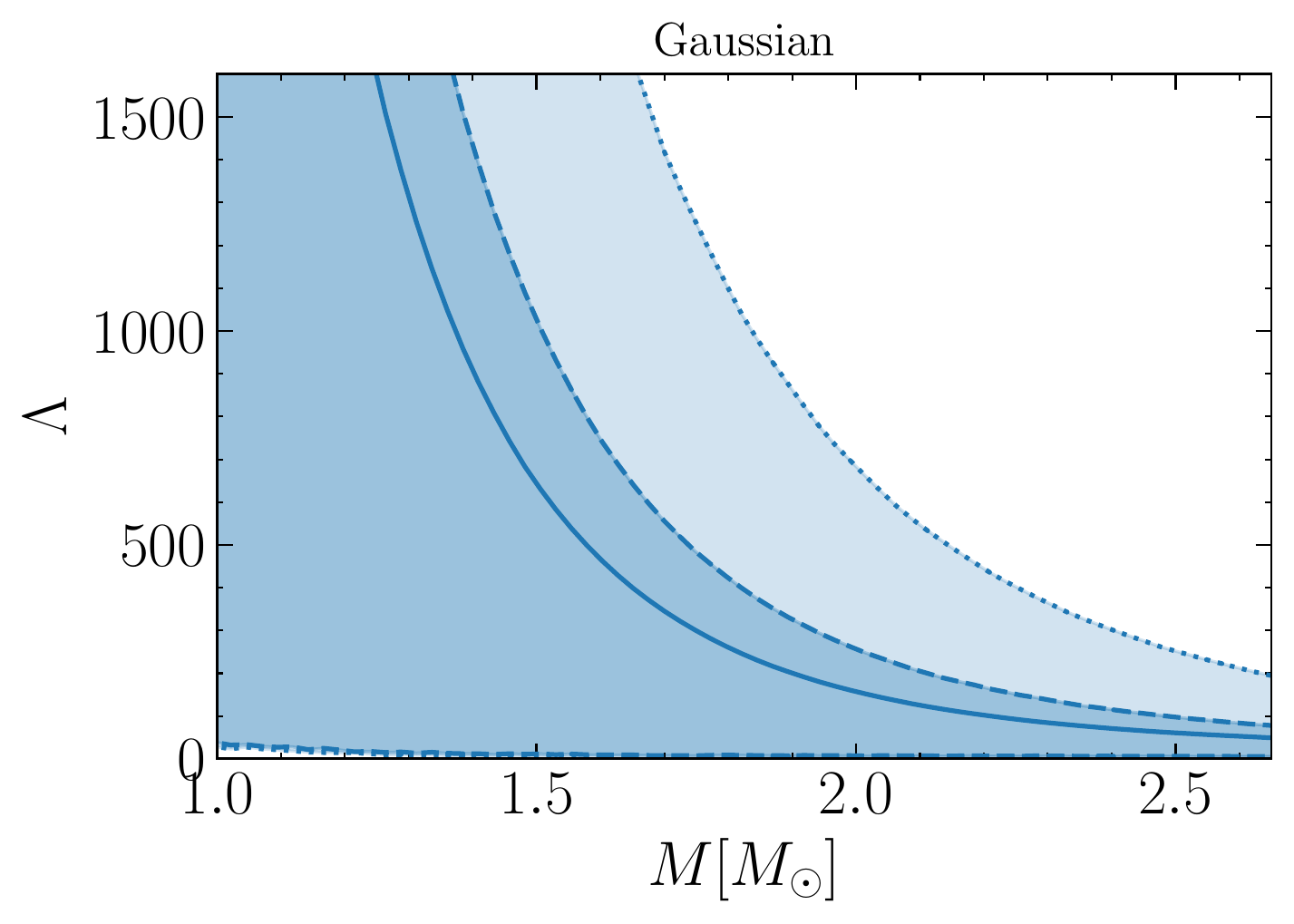}
		\includegraphics[height=55mm,angle=-00]{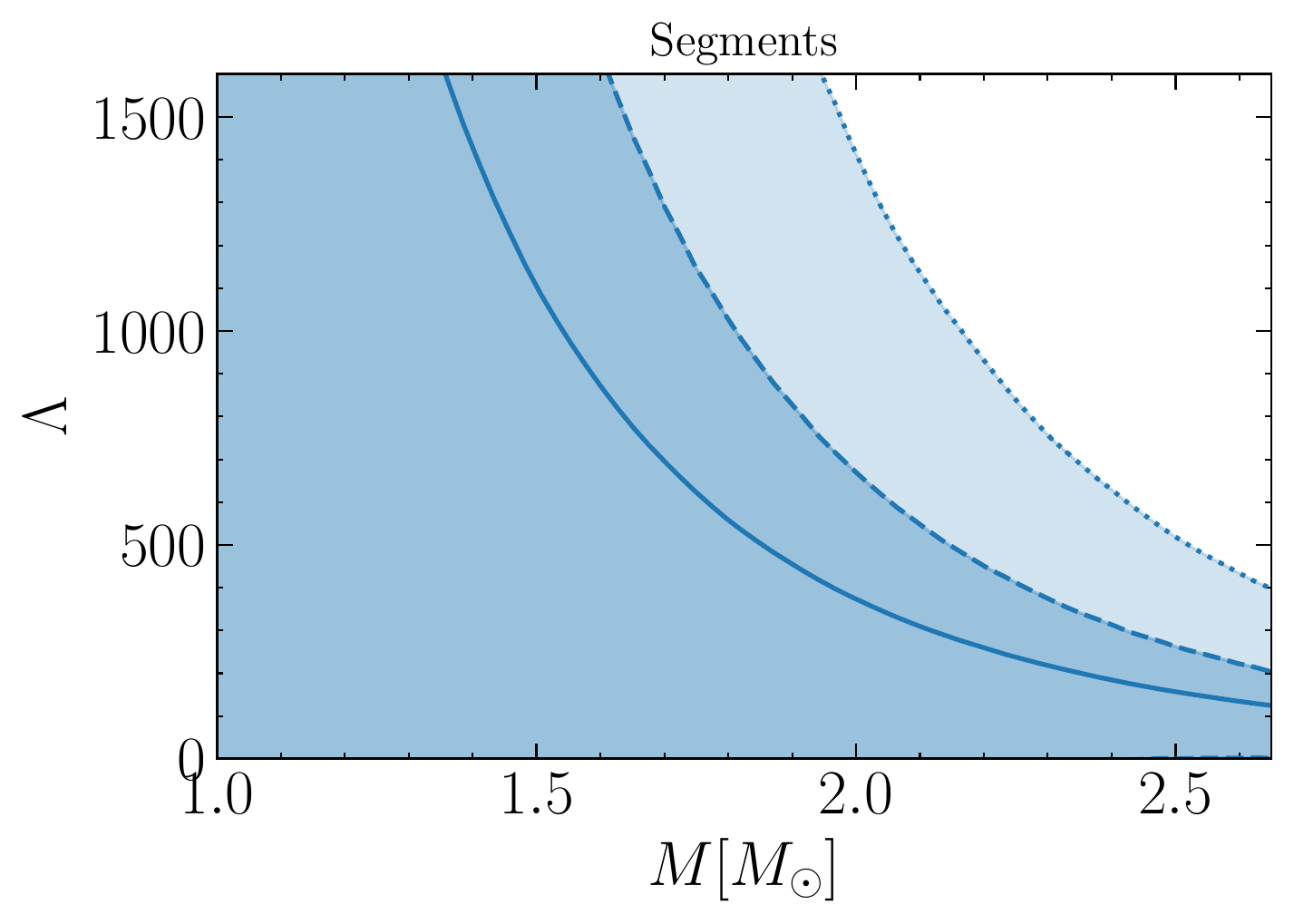} \\
		\caption{Marginal Prior probability distributions at the 95\% and 68\% level for the Gaussian (left) and Segments parametrisation (right) of the squared speed of sound $c_s^2$ and pressure $P$ as a function of energy density $\varepsilon$.  Also shown is the Prior for the mass-radius relation and the tidal deformability, $\Lambda$, as a function of neutron star mass $M$ in units of the solar mass $M_\odot$. At each $\varepsilon$ or $M$,  there exist 95\% and 68\% Prior credible intervals for $c_s^2(\varepsilon)$, $P(\varepsilon)$ or $R(M)$, $\Lambda(M)$. These intervals are connected to obtain the Prior credible bands. Similarly, the medians of the Prior probability distributions at each $\varepsilon$ or $M$ are connected (solid lines). For the speed of sound the dashed black line indicates the value of the conformal limit.}
		\label{fig:Prior}
	\end{center}
\end{figure*}

The Prior credible bands for both G and S parametrisations are depicted in Fig.\,\,\ref{fig:Prior}. The bands are very broad in both interior and exterior parameter spaces. Because the ChEFT constraint is employed as Likelihood and hence not present in the Priors,  there is Prior support for rapidly increasing speeds of sound at low densities,  leading to large neutron star radii.  Hence the Prior credible bands have strong weights both at small sound speeds and large radii. 

The parameter ranges are chosen to minimise any possible restrictions,  such that the Posterior distribution has maximum freedom to be governed by the empirical data. The Prior probability distributions of versions G and S differ because of the different functional forms and chosen parameter ranges.  This permits an assessment of the impact of different Prior choices on the inference results.  If the results for versions G and S turn out to be very similar,  we can conclude that the inference procedure is robust against variations in the functional form of the Prior. 

Both Priors at the 95\% level support very small speeds of sound, $c_s^2 \lesssim 0.05$.  In fact the 68\% credible band of the G version reaches down to $c_s^2 = 0$. Accordingly,  every fourth EoS in the Gaussian parametrisation potentially has a first-order phase transition in the sense that the minimum speed of sound becomes smaller than $c_{s,min}^2 \leq 0.1$,  whereas every fifth EoS in the Segments parametrisation features such a phase transition.  In contrast,  each EoS in the (later determined) Posterior credible bands is constrained by astrophysical data and thus limited by its emerging maximum central energy density, $\varepsilon_{c,max}$.  The mass-radius trajectory deduced from each given EoS,  with or without a phase transition, terminates at this point.  An EoS's mass-radius sequence normally ends after a first-order phase transition.  As a consequence small sound speeds appear with lower weight in the Posteriors than in the Priors. 

For both parametrisations, the pressures at asymptotic energy densities agree with the pQCD results of Ref.\,\,\cite{Gorda2021} within the uncertainty band from a variation of the renormalisation scale.	

%Notice that, according to Sec.\,\ref{sec:BayesianInferenceBasics}, each EoS is only considered up to its respective central energy density when calculating the credible bands. As a result, the lower bounds of the $c_s^2$ credible bands remain above zero, because the mass-radius sequence usually ends after a strong first-order phase transition. Instead, if each Eos is used up to arbitrarily high energy densities, the 95\% bands approach or equal zero. In fact, ca. every fifth (fourth) EoS in the Segments (Gaussian) parametrisation contains a strong first-order phase transition in the sense that the minimum speed of sound becomes less than $c_{s,min}^2 \leq 0.1$.   

\subsubsection{Monotonically rising speed of sound}
\label{sec:AdditionalPrior}

The previous,  general choice of Priors is open,  in principle,  to possible phase transitions in the EoS if the data suggest such an option.  At the same time we also, additionally, wish to investigate a more restrictive case,  namely the hypothesis that neutron star matter is composed of conventional hadronic (nucleon and meson) degrees of freedom,  with spontaneously broken chiral symmetry intact and no complex phase structure.  A successful historical example of this kind is the APR equation of state \cite{Akmal1998}.  Another example,  already mentioned previously,  is a model based on chiral nucleon-meson field theory  \cite{Drews2015,Drews2017}.  The parameters of the model are adjusted to reproduce low-density nuclear phenomenology including the liquid-gas phase transition.  Fluctuations beyond mean-field approximation  are treated non-perturbatively using Functional Renormalization Group methods.  The effect of these fluctuations is to stabilize dense matter against a first-order chiral restoration phase transition and to convert it into a crossover at high densities \cite{Brandes2021}.  In such a scenario matter behaves as a strongly correlated,  relativistic Fermi liquid \cite{Friman2019}. The speed of sound rises monotonically with increasing density and exceeds the conformal boundary at densities $n\sim 4\,n_0$.  

With the aim of studying whether such a picture is compatible with the empirical data base,  the additional Prior assumption is implemented that neutron star matter displays no phase transition or crossover up to a given transition density $n_{tr}$.  This is equivalent to the speed of sound rising monotonically up to $n_{tr}$:
\begin{eqnarray}
	\frac{\partial c_s^2}{\partial \varepsilon} > 0 \quad \text{for} \quad n < n_{tr}~.
\end{eqnarray}
For densities $n > n_{tr}$,  the system is allowed any freedom to undergo transitions or changes of degrees freedom.  In practice, based on the findings in \cite{Brandes2021}, we vary the transition density in the range $n_{tr} = 3 - 6 \, n_{0}$.

\subsection{Constraints from observations and theory}
\label{sec:measurements}

In this Section we give a summary of the data base used to compute the Likelihoods that are needed, in turn,  to determine Posterior distributions. This includes neutron star masses, radii and tidal deformabilities,  and low-density constraints from nuclear theory.

\subsubsection{Shapiro time delay}
\label{sec:Shapiro}

If a pulsating neutron star is in a binary system with a white dwarf companion,  the gravitational field of the companion changes the pulsar signal frequency,  an effect referred to as the Shapiro time delay.  Via a General Relativistic modelling of this delay the pulsar mass can be extracted with high precision.  The most interesting measurements are those of the heaviest neutron stars as this sets a lower limit on the maximum mass that the neutron star EoS has to support.  Several neutron stars with masses as high as two times the solar mass $M_\odot$ were measured in this way,  namely PSR J1614–2230 \cite{Demorest2010,Fonseca2016,Arzoumanian2018}, PSR J0348+0432 \cite{Antoniadis2013} and PSR J0740+6620 \cite{Cromartie2020,Fonseca2021}, with masses evaluated at the 68\% level:
\begin{eqnarray}
	&\text{PSR J1614–2230} \qquad &M = 1.908 \pm 0.016 \, M_\odot ~, ~ \label{eq:ShapiroMass1}\\
	&\text{PSR J0348+0432} \qquad &M = 2.01 \pm 0.04 \, M_\odot ~, \label{eq:ShapiroMass2}\\
	&\text{PSR J0740+6620} \qquad &M = 2.08 \pm 0.07 \, M_\odot ~. \label{eq:ShapiroMass3}
\end{eqnarray}
To compute the respective Likelihoods we follow previous analyses \cite{Raaijmakers2020,Raaijmakers2021,Dietrich2020} and assume that the mass measurements based on the Shapiro time delay are distributed as Gaussians, $\mathcal{N}\left(M, \mu, \sigma\right) = 1/\sqrt{2\pi \sigma^2} \,\text{exp}[-1/2 \, (M - \mu)^2/\sigma^2]$ with mean values $\mu$ and standard deviations $\sigma$.  For a given set of parameters $\theta$ the solution of the TOV Eqs.\,\,(\ref{eq:TOV1} - \ref{eq:TOV2}) yields the maximum supported mass $M_{max} (\theta)$ for this respective EoS. Then the Likelihood for each measurement is computed as 
\begin{align}
	\text{Pr} \big(M(\theta)&\big|\mathcal{D}_{\text{Shapiro}}, \mathcal{M}\big) = \nonumber \\
	& \int_{0}^{M_\text{max}(\theta)} \text{d}M \,~ \mathcal{N}\left(M, \mu, \sigma\right) \, \text{Pr}(M(\theta)) ~.
	\label{eq:LikelihoodShapiro}
\end{align}
The mass Prior $\text{Pr}(M(\theta))$ is implicitly defined via the chosen Prior distribution of central pressures.  Note that this implies a general mass population of neutron stars,  similar to the one used in the analyses in \cite{Landry2020,Legred2021}.  Once the number of available data increases by future measurements,  the resulting Posterior distributions may be affected \cite{Mandel2019},  such that the neutron star population may have to be inferred together with the Posterior \cite{Wysocki2020}. The total Likelihood for all Shapiro time delay measurements is given by the product of the individual Likelihoods.

\subsubsection{Pulse profile modelling}

The emission of soft X-rays from hot spots on the magnetic polar caps of rapidly rotating neutron stars is modulated by the gravitational field of the star.  The pulse amplitude and shape depend on the compactness, $M/R$,  of the star and on its mass $M$.  Using a model of the hot spots and the neutron star atmosphere,  Bayesian Posterior distributions for the mass and radius can be inferred from X-ray profiles measured by the Neutron Star Interior Composition ExploreR (NICER).  So far two neutron stars were measured and subsequently analysed by two independent groups.  Here we use the results of Miller et al. \cite{Miller2019,Miller2021} for masses and radii at the 68\% level:
\begin{eqnarray}
	\text{PSR J0030+0451} \qquad &R&=13.02^{+1.24}_{-1.06}\,\text{km} ~, \nonumber \\ &M&=1.44^{+0.15}_{-0.14} \, M_\odot ~, \\
	\text{PSR J0740+6620} \qquad &R& = 13.7^{+2.6}_{-1.5}\,\text{km} ~, \nonumber \\ &M& = 2.08 \pm 0.07 \, M_\odot ~.
\end{eqnarray}
to be seen in comparison with an alternative analysis by Riley et al. \cite{Riley2019,Riley2021}:
\begin{eqnarray}
	\text{PSR J0030+0451} \qquad &R&=12.71^{+1.14}_{-1.19}\,\text{km} ~, \nonumber \\ &M&=1.34^{+0.15}_{-0.16} \, M_\odot ~, \\
	\text{PSR J0740+6620} \qquad &R& = 12.39^{+1.30}_{-0.98}\,\text{km} ~, \nonumber \\ &M& = 2.072^{+0.067}_{-0.066} \, M_\odot ~.
\end{eqnarray}

The results of these analyses are publicly available as samples from the mass-radius Posterior. To approximate the underlying probability distribution we use the Kernel Density Estimation (KDE) technique (for a brief introduction see Appendix \ref{sec:KDE}). The TOV equations are solved for a given set of parameters $\theta$ and for $N$ central pressures $P_c$, resulting in $N$ points of the mass-radius relation $(M_i, R_i)(\theta)$. The Likelihood for each measurement is given by the KDE computed at these points:
\begin{equation}
	\text{Pr} \big( (M,R)(\theta) \big| \mathcal{D}_{\text{NICER}}, \mathcal{M}\big) = \frac{1}{N}\sum_i^N \text{KDE} \big( (M_i, R_i) (\theta) \big) ~.
\end{equation}
For large $N$ this is equivalent to the line integral along the mass-radius curve $C\big((M, R)(\theta)\big)$ weighted with the mass-radius Prior $\text{Pr}\big((M,R)(\theta)\big)$ that is implicitly defined by the Prior distribution $ \text{Pr} (P_c|\theta, \mathcal{M})$ of the central pressures:
\begin{align}
	\lim_{N\rightarrow\infty} & \text{Pr} \big( (M,R)(\theta) \big| \mathcal{D}_{\text{NICER}}, \mathcal{M}\big) = \nonumber\\  &\int_{C\left((M,R)(\theta)\right)} \text{d}s \,~ \text{KDE}\big((M,R)(\theta)\big) \text{Pr}\big((M,R)(\theta)\big) ~. 
\end{align}
Here, $\text{d}s$ denotes the line element along the mass-radius curve $C\big((M, R)(\theta)\big)$.
Because the mass measurement in Eq.\,\,(\ref{eq:ShapiroMass3}) has been used in the NICER analysis of PSR J0740+6620, we do not include this mass measurement in the total Likelihood to avoid double counting.

\subsubsection{Neutron star mergers}

The merger of two neutron stars in a binary produces gravitational waves that are detectable in earth-based detectors by the LIGO and Virgo Scientific Collaborations. A detected merger signal can be compared to theoretical waveform models, which depend on the mass ratio of the two neutron stars, $M_2/M_1$, and a mass-weighted combination of their tidal deformabilities
\begin{equation}
	\bar{\Lambda} = \frac{16}{13}\frac{(M_1 + 12M_2)M_1^4\Lambda_1 + (M_2 + 12M_1)M_2^4\Lambda_2}{(M_1+M_2)^5}~.
\end{equation} 
Hence from the gravitational wave measurement a Bayesian Posterior for the masses $(M_1, M_2)$ and tidal deformabilities $(\Lambda_1, \Lambda_2)$ can be inferred. So far, two binary neutron star merger events, GW170817 \cite{Abbott2019} and GW190425 \cite{Abbott2020},  were detected, yielding the following constraints at the 90\% level
\begin{align}
	\text{GW170817} \qquad & \bar{\Lambda}=320^{+420}_{-230} ~,\nonumber\\
	\text{GW190425} \qquad & \bar{\Lambda} \leq 600 ~.
\end{align}
Notice that different analyses of the gravitational wave data produced slightly changed results \cite{De2018}. The first one of these events (GW170817) was further evaluated together with electromagnetic signals \cite{Raaijmakers2021,Capano2020,Fasano2019}. The following masses and tidal deformabilities of the individual neutron stars in the binary were reported in Ref.\,\,\cite{Fasano2019}:
\begin{align}
	M_1 = 1.46^{+0.13}_{-0.09}\,M_\odot \qquad & \Lambda_1 =255^{+416}_{-171} ~,\nonumber\\
         M_2 = 1.26^{+0.09}_{-0.12}\,M_\odot \qquad & \Lambda_2 =661^{+858}_{-375} ~.
\end{align}
Other detected events are under discussion as possibly being neutron star-black hole mergers and so we do not include them in the present analysis. 

For a given set of parameters $\theta$, we numerically solve the differential equations to obtain $(\Lambda_i, M_i)$. From this we can interpolate the function $\Lambda (M)$. We approximate the Posterior $(\Lambda_1, \Lambda_2)$ for each measurement using again Kernel Density Estimation. To compute the Likelihood we insert the $N$ mass Posterior samples $(M_{1j}, M_{2j})$ into the function $\Lambda(M)$ which is in turn inserted into the $(\Lambda_1, \Lambda_2)$ KDE  
\begin{align}
	\text{Pr} &\big( (\Lambda, M)(\theta)\big| \mathcal{D}_{\text{GW}}, \mathcal{M}\big) = \nonumber \\
	&\frac{1}{N}\sum_j^N \text{KDE} \big(\Lambda (M_{1j}, \theta) , \Lambda(M_{2j}, \theta)\big)~.
\end{align}
Using the above procedure, we do not have to assume that the chirp mass $M_{chirp} = (M_1M_2)^{3/5}(M_1 + M_2)^{-1/5}$ of the event is fixed as was done in previous analyses \cite{Miller2019a,Raaijmakers2019,Raaijmakers2020,Raaijmakers2021}. Note that we again implicitly assumed the population of neutron stars is given by our central pressure Prior.

\subsubsection{Low densities: Chiral effective field theory}

In addition to the measurements listed above we can use constraints from theory to determine the neutron star EoS at low densities. Chiral effective field theory involves a systematic expansion of nuclear forces at low energies with controlled uncertainties.  It can be extended to finite densities using many-body methods and gives a good description of nuclear phenomenology \cite{Wellenhofer2015}.  At its current state of development ChEFT is believed to be valid up to densities of $n \sim 1 - 2 \, n_0$.  The nuclear and neutron matter results have been extended to neutron star matter by including beta equilibrium conditions \cite{Hebeler2013}.  

Previous analyses \cite{Greif2019,Raaijmakers2019,Raaijmakers2020,Raaijmakers2021,Dietrich2020,Pang2021} have incorporated this low-density constraint by only considering those EoS which fall inside the ChEFT uncertainty band.  We take a different approach here,   using a treatment similar to that of the previously discussed observables which is considered to be statistically more meaningful.  It allows to take possible structure within the uncertainty band into account and gives a finite (but low) probability to EoS outside that band.  There is no obvious reason to trust the uncertainty estimates of ChEFT calculations more than those of the observables.  A statistically well-defined treatment similar to that of the empirical data base allows a balancing between different constraints. 

Recently a Bayesian framework was introduced to compute the combined uncertainty resulting from many-body approximations and convergence errors in ChEFT \cite{Melendez2019,Drischler2020,Drischler2020a}.  Based on this approach,  constraints for the sound velocity in neutron star matter can be derived \cite{Drischler2021}.  In their work the authors trust the next-to-next-to-next-to-leading-order (N3LO) results up to a density of $n = 2\, n_0$.  However,  the differences between N2LO and N3LO results may hint towards possible convergence issues. Therefore we take a more conservative choice,  using the uncertainty estimate at $n = 1.3 \, n_0$,  similar to the one derived in Ref.\,\,\cite{Keller2022}.  Also in Ref.\,\,\cite{Essick2020},  the authors find that ChEFT results are preferred by astrophysical data up to $n \sim 1.3\, n_0$. 
In addition, we set the Likelihood to zero for all EoS with speed of sound larger than the 99.7\% credible interval at $n = 2.0 \, n_0$. This prohibits EoS which rise very quickly beyond the ChEFT constraint at $n = 1.3 \, n_0$. Because the uncertainty estimate in Refs.\,\,\cite{Drischler2021,Drischler2022} is based on a Gaussian process, the Likelihood for the ChEFT constraint can be computed in terms of a Gaussian distribution.  Accordingly,  the upper limit of the 99.7\% credible interval can be determined as the mean value $\mu$ plus three times the standard deviation $\sigma$. This upper limit at $2\, n_0$ is implemented via a Heaviside step function multiplied to the Gaussian normal distribution:
\begin{align}
	\text{Pr} \big( c_s^2(n, \theta) &\big|\mathcal{D}_{\text{ChEFT}}, \mathcal{M}  \big) = \nonumber \\ & \mathcal{N}\left(c_s^2\big(1.3 \,n_0, \theta\big), \mu_{1.3n_0}, \sigma_{1.3n_0}\right) \nonumber \\
	&\times
	\theta(\mu_{2n_0} + 3 \, \sigma_{2n_0} - c_s^2\big(2.0 \,n_0, \theta\big)) ~ .
\end{align}
It turns out that a statistically well-defined incorporation of the ChEFT constraint leads to more freedom at small and intermediate densities compared to previous approaches.  In addition the incorporation of the ChEFT constraint into the Likelihood in contrast to using it as a Prior assumption avoids an unphysical discontinuity in the speed of sound at low densities.

In several previous studies measurements from quiescent low-mass X-ray binaries and thermonuclear bursters were also used \cite{Steiner2010,Oezel2016a,Oezel2016,Bogdanov2016}.  However, these data involve lots of specific model features and are therefore neglected in modern analyses.  Recent measurements of the neutron skin thickness of $^{208}$Pb suggest a stiff EoS for densities close to the nuclear saturation density $n \sim n_0$ \cite{Adhikari2021}.  Uncertainties are still large and there is potential tension with other laboratory probes \cite{Reed2021}. Therefore, we do not include this measurement in the total Likelihood. Together with the neutron star merger event GW170817 the short gamma ray burst GRB170817A and the kilonova AT2017gfo were detected \cite{Abbott2017}.  Some recent Bayesian analyses include information about this kilonova \cite{Raaijmakers2021,Dietrich2020,Pang2021},  which however introduces a series of model assumptions and consequently raises the systematic uncertainties.\\

Summarizing the preceding Sections, the full procedure to obtain credible bands for neutron star properties consists of the following steps: first a set of parameters $\theta$ is sampled from the Prior $\text{Pr}(\theta | \mathcal{M})$.  We need in total more than  $300,000 - 600,000$ samples for each parametrisation in order to generate statistically solid results that cover enough probability mass such that they remain stable after a further increase in the number of samples.  For this sampled set of parameters we compute the speed of sound for the respective parametrisation and then the EoS, $P(\varepsilon, \theta)$, using Eq.\,\,(\ref{eq:EoSfromCs}). Given the equation of state we can numerically solve the coupled system of differential equations for $\big(M, R, \Lambda\big) (\theta)$. The total Likelihood,  already marginalised over the central pressures $P_c$,  can then be determined as the product of the individual marginalised Likelihoods for the different measurements and constraints:
\begin{align}
	\text{Pr} (\mathcal{D} |\theta, \mathcal{M})  \propto  & \, \text{Pr} \big(M(\theta)\big|\mathcal{D}_{\text{Shapiro}}, \mathcal{M}\big) \nonumber \\ & \times \text{Pr} \big( (M,R)(\theta) \big| \mathcal{D}_{\text{NICER}}, \mathcal{M}\big) \nonumber \\ & \times \text{Pr} \big( (\Lambda, M)(\theta)\big| \mathcal{D}_{\text{GW}}, \mathcal{M}\big) \nonumber \\ & \times \text{Pr} \big( c_s^2(n, \theta) \big|\mathcal{D}_{\text{ChEFT}}, \mathcal{M}  \big) ~.
\end{align} 
The Prior probability distribution weighted with the above Likelihood yields the Posterior probability distribution,  $\text{Pr}\big(\theta \big| \mathcal{D}, \mathcal{M} \big)$, for the parameters.  We can then marginalize over this Posterior probability distribution to compute the median as well as the highest density credible intervals at the $68\%$ and $95\%$ level for different neutron star properties,  as well as the credible bands at different levels as explained in Sec.\,\ref{sec:BayesianInferenceBasics}.

\comment{
\begin{enumerate}
	\item Sample a set of parameters $\theta$ from the Prior $\text{Pr}(\theta | \mathcal{M})$. Hereby, we need in total $\sim 300,000$ samples for each parametrisation for statistically solid results.
	\item For this set of parameters determine the EoS $P(\varepsilon, \theta)$ using Eq.\,\,(\ref{eq:EoSfromCs}).
	\item Then, numerically solve the coupled system of differential equations for $\big(M, R, \Lambda\big) (\theta)$.
	\item Determine the total Likelihood as the product of the Likelihoods for the different constraints
	\begin{align}
		\text{Pr} (\mathcal{D} |\theta, \mathcal{M})  \propto  & \, \text{Pr} \big(M(\theta)\big|\mathcal{D}_{\text{Shapiro}}, \mathcal{M}\big) \nonumber \\ & \times \text{Pr} \big( (M,R)(\theta) \big| \mathcal{D}_{\text{NICER}}, \mathcal{M}\big) \nonumber \\ & \times \text{Pr} \big( (\Lambda, M)(\theta)\big| \mathcal{D}_{\text{GW}}, \mathcal{M}\big) \nonumber \\ & \times \text{Pr} \big( c_s^2(n, \theta) \big|\mathcal{D}_{\text{ChEFT}}, \mathcal{M}  \big) ~.
	\end{align}  
	The Prior samples weighted with the above Likelihood yields the Posterior probability distribution for the parameters $\text{Pr}\big(\theta \big| \mathcal{D}, \mathcal{M} \big)$.
	\item Marginalize over the Posterior probability distribution to compute the credible intervals at the $[68\%, 95\%]$ level.
\end{enumerate}
}

\section{Results}
\label{sec:Results}
\subsection{General Priors}

\begin{figure*}[tp]
	\begin{center}
		\includegraphics[height=55mm,angle=-00]{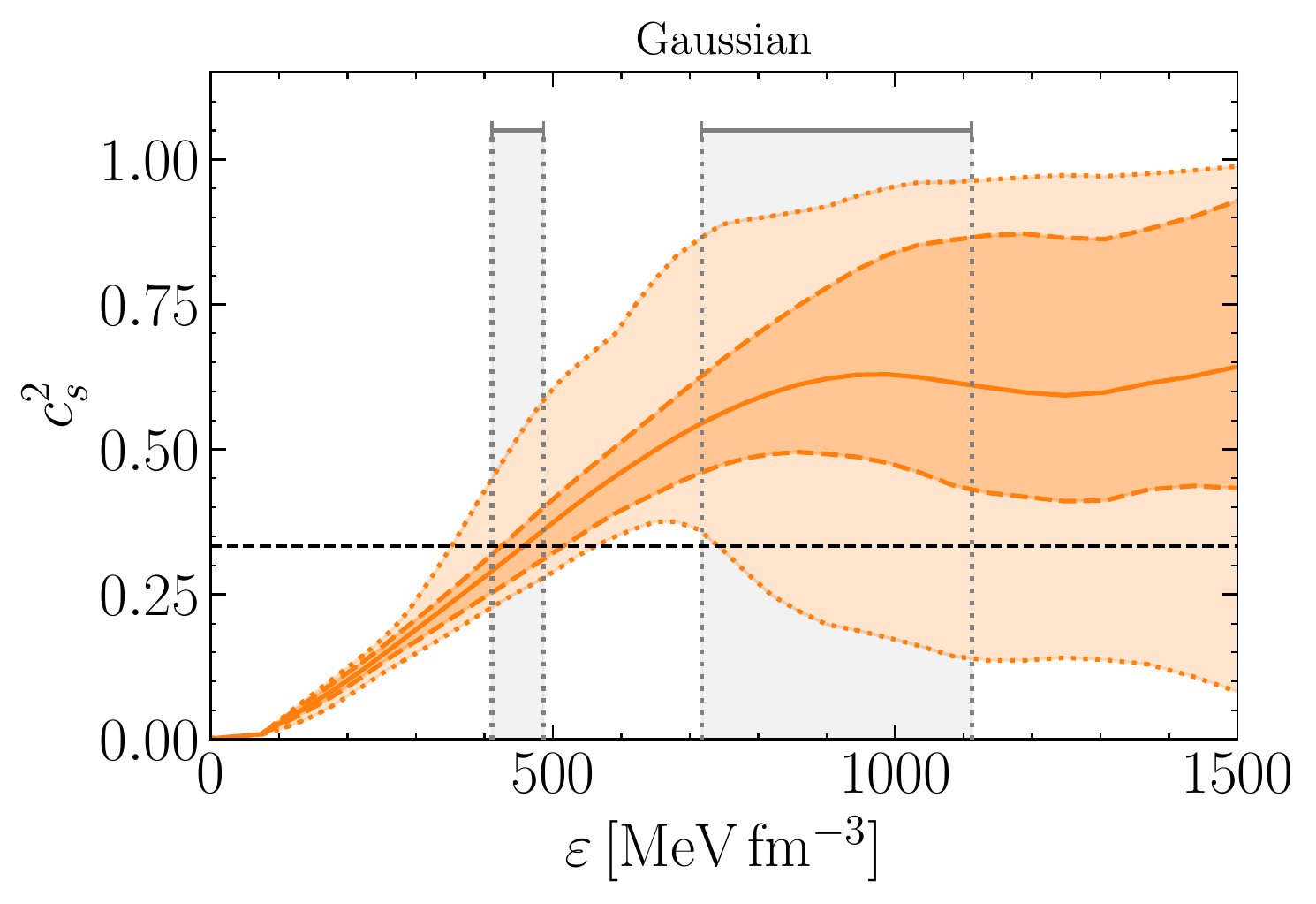} 
		\includegraphics[height=55mm,angle=-00]{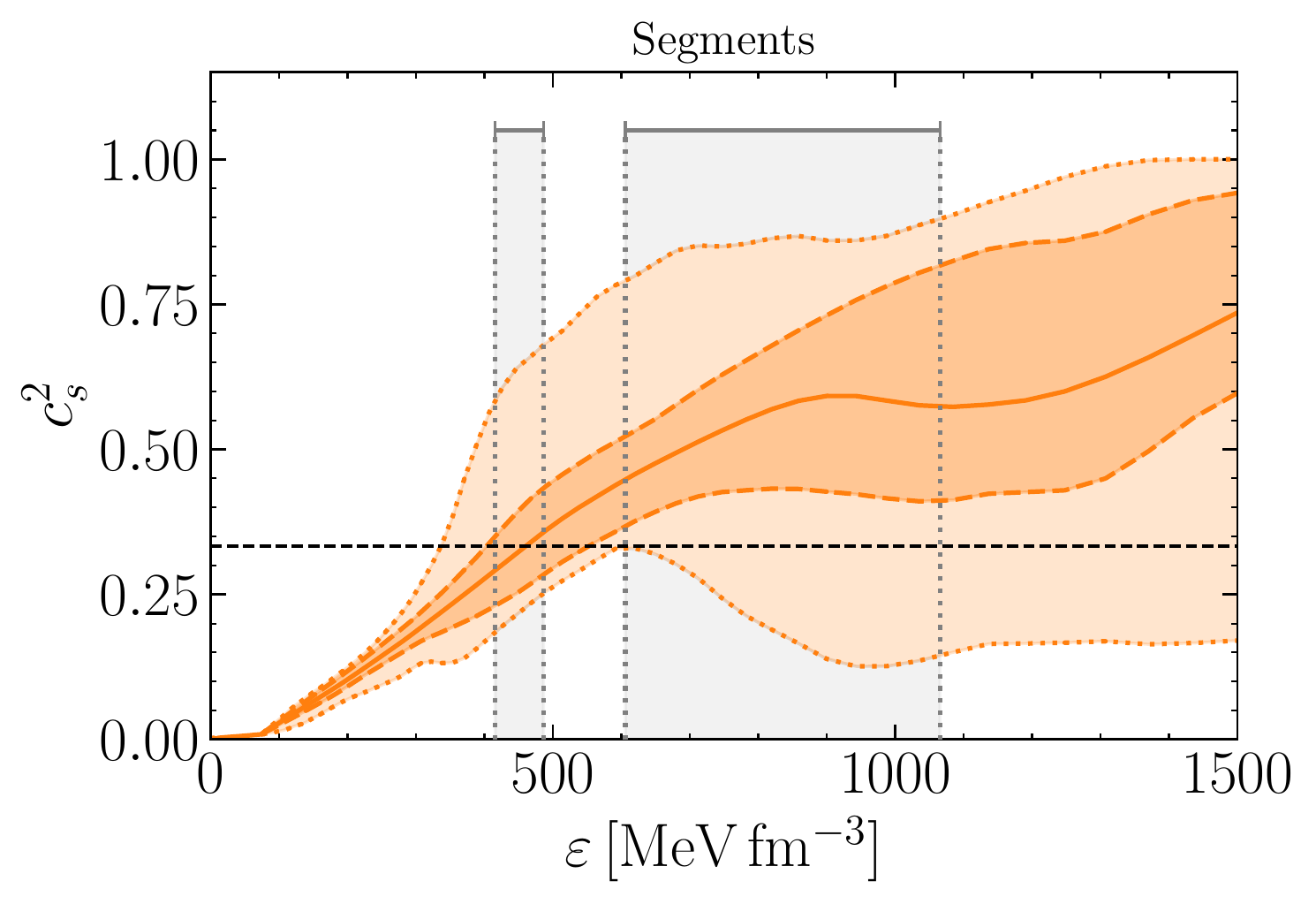} \\
		\includegraphics[height=55mm,angle=-00]{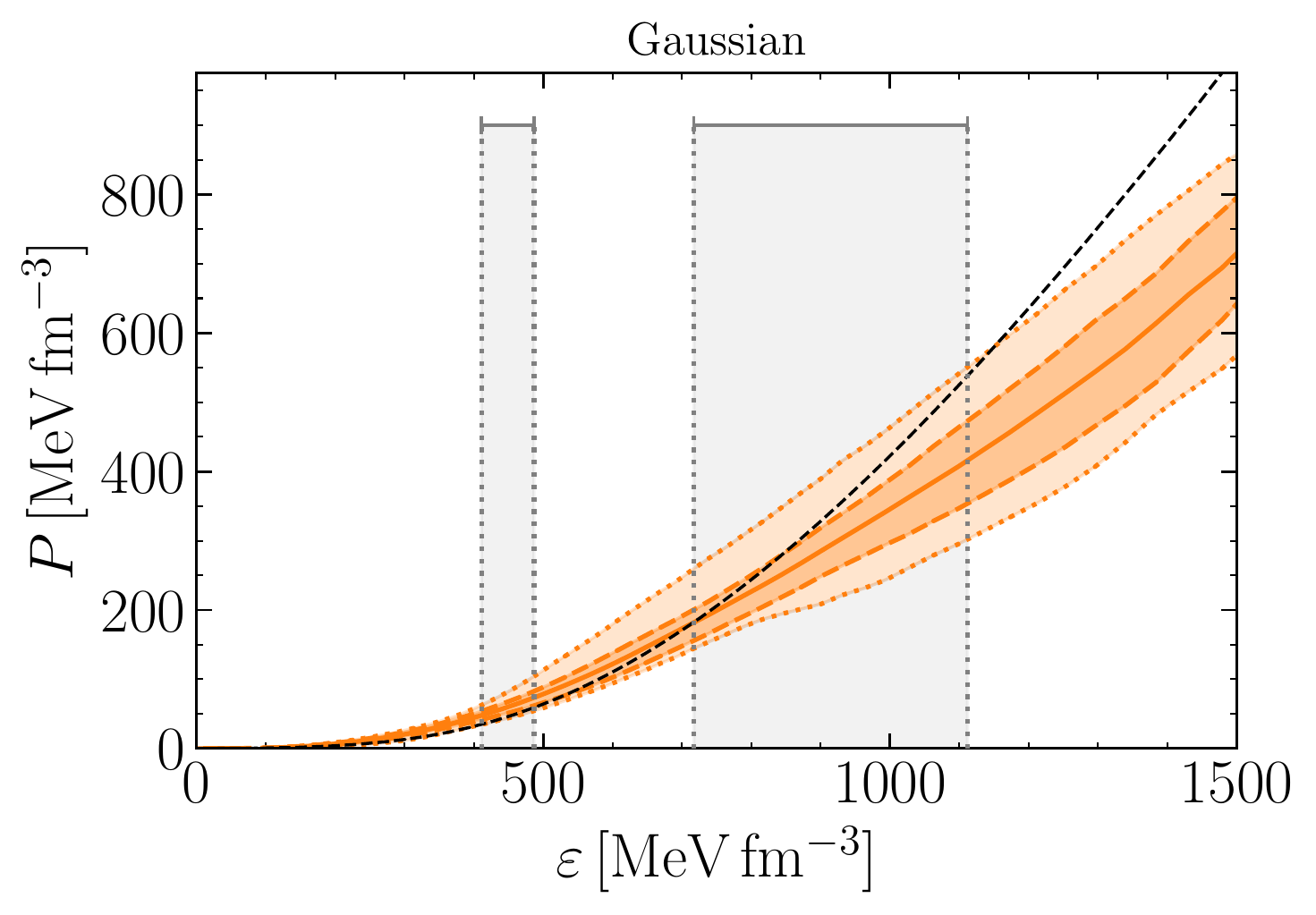}
		\includegraphics[height=55mm,angle=-00]{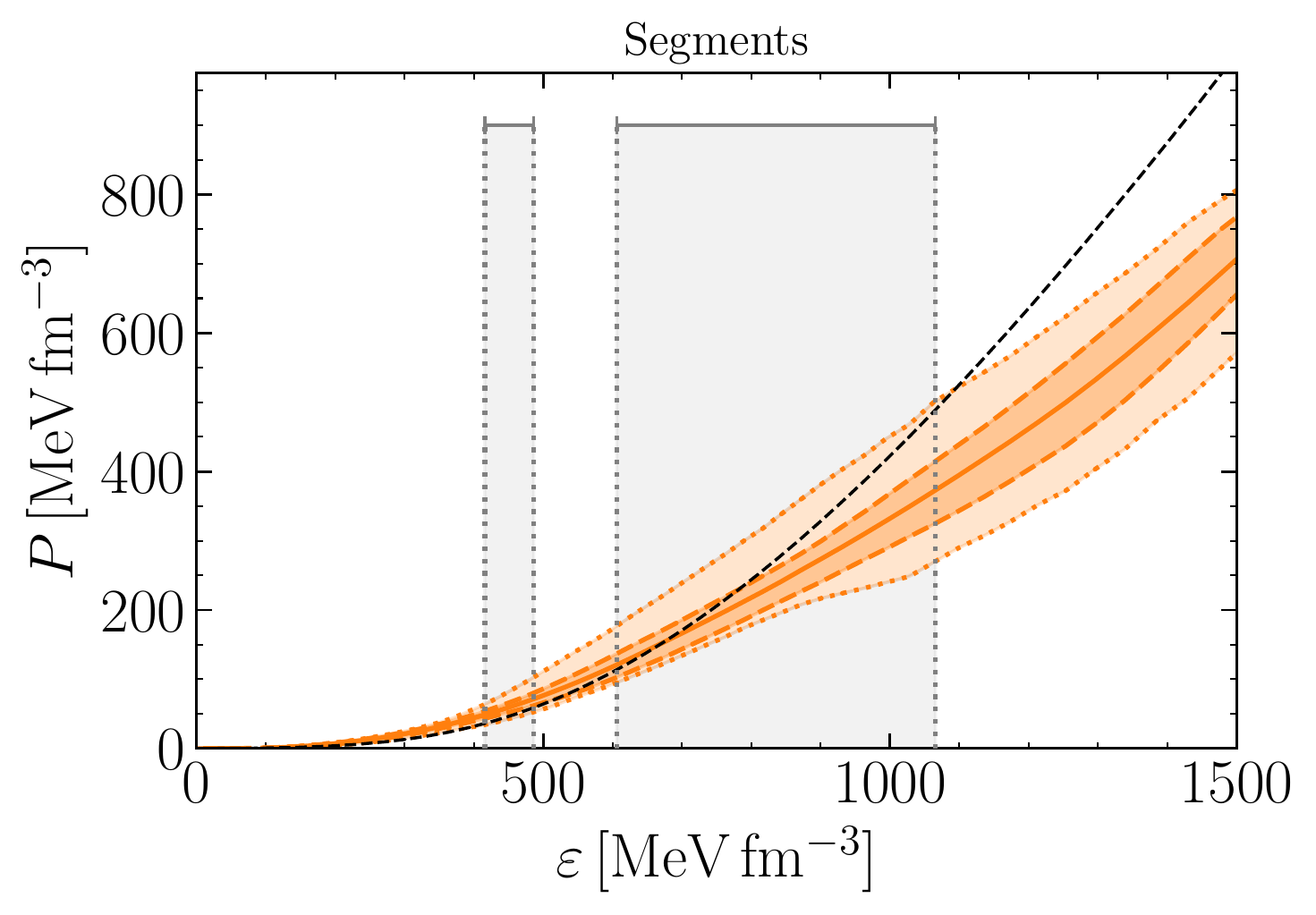} \\
		\caption{Marginal Posterior probability distributions at the 95\% and 68\% level for the Gaussian (left) and Segments parametrisation (right) for the squared speed of sound $c_s^2$ and pressure $P$ as a function of energy density $\varepsilon$. At each $\varepsilon$, there exist 95\% and 68\% Posterior credible intervals for $c_s^2(\varepsilon)$ and $P(\varepsilon)$. These intervals are connected to obtain the Prior credible bands. Similarly, the medians of the marginal Posterior probability distributions at each $\varepsilon$ are connected (solid lines).  Grey areas mark the 68\% credible intervals of the central energy densities of neutron stars with masses $M = 1.4 \, M_\odot$ (left columns) and $2.1 \, M_\odot$ (right columns) in each figure. The dashed black line indicates the value of the conformal limit for the speed of sound and the APR EoS \cite{Akmal1998} for the pressure.}
		\label{fig:PosteriorBands1}
	\end{center}
\end{figure*}

Following the procedures outlined in the preceding Sections, the resulting marginal Posterior credible bands for the squared speed of sound, $c_s^2(\varepsilon)$,  and for the pressure $P(\varepsilon)$ are displayed in Fig.\,\,\ref{fig:PosteriorBands1} for both parametrisations. Note that the general Prior used here is free of assumptions about monotonically rising sound speeds as introduced in Sec.\,\ref{sec:AdditionalPrior}. Compared to the Prior credible bands in Fig.\,\,\ref{fig:Prior}, we can see how the Posterior bands have become much narrower because of the constraints implied by the observational data. The credible bands for both parametrisations agree rather well, especially at energy densities $\varepsilon \lesssim 700\,$MeV$\,$fm$^{-3}$. In this regime the squared speed of sound rises until it exceeds the conformal limit,  $c_s^2 = 1/3$,  around $\varepsilon \sim 600\,$MeV$\,$fm$^{-3}$ at the 95\% level.  As in Refs.\,\,\cite{Greif2019,Raaijmakers2019,Raaijmakers2020,Raaijmakers2021,Gorda2022} we implemented the transition to the neutron star crust discontinuously. This is visible in the speed-of-sound credible bands at very low energy densities but of no quantitative significance.

According to Fig.\,\,2,  there are only small differences at the 68\% level between the two parametrisations at low energy densities. The 95\% band of the Segments parametrisation extends to smaller sound speeds at low $\varepsilon$ and then rises to higher speeds of sound,  as this parametrisation allows for steeper slopes. The conservative upper limit at $n = 2\, n_0$ based on the ChEFT calculation in Refs.\,\,\cite{Drischler2021,Drischler2022} prohibits extremely steep rises of $c_s^2$ in the low-density region that were seen in some previous works \cite{Altiparmak2022,Gorda2022,Han2022}. Even though there remain uncertainties about the convergence of ChEFT at higher densities,  there is no indication of a steep rise in the sound speed at small densities in different ChEFT analyses \cite{Lynn2016,Tews2018,Lonardoni2020}.    

In the FOPI heavy-ion experiment Au nuclei were collided at energies from 0.4 to 1.5$\,$GeV/A.  The EoS of symmetric nuclear matter deduced from these data \cite{Fevre2016, Huth2022} indicates that smaller pressures are allowed for densities $n \leq 2.1 \, n_0$ as compared to recent ChEFT results \cite{Drischler2019}. This further contests a strong increase in the speed of sound at low densities.

Going to higher energy densities, the 68\% credible band stays above the conformal limit whereas the 95\% band allows for sound speeds below this bound from $\varepsilon \gtrsim 700\,$MeV$\,$fm$^{-3}$ onward,  but with very low probability. The 68\% credible range for the S version appears to continue increasing while it tends towards a plateau for the G version.

The region beyond $\varepsilon \gtrsim 1.2\,$GeV$\,$fm$^{-3}$ already exceeds the central credible densities for neutron stars with mass $M \sim 2.1 \, M_\odot$.  This region is therefore much less restricted by the observational data collected in Sec.\,\ref{sec:measurements}.  So this extrapolated behaviour is more sensitive to  the Prior of each parametrisation.  The 95\% credible bands for both parametrisations remain systematically larger than $c_s^2 > 0.1$ up to the central energy density of a $2.1 \, M_\odot$ neutron star, $\varepsilon_c = 0.9\pm 0.4\,$GeV$\,$fm$^{-3}$ \footnote{If not stated otherwise, credible intervals for neutron star properties correspond to the median and 95\% intervals for the Segments parametrisation.}, making a first-order phase transition unlikely. This behaviour is reflected in the credible bands for the equation of state $P(\varepsilon)$ in Fig.\,\, \ref{fig:PosteriorBands1}.  

For the 68\% bands the squared sound velocity always stays above the conformal limit from energy densities $\varepsilon > 500\,$MeV$\,$fm$^{-3}$ upward.  Due to the integrated nature of the pressure,  Eq.\,\,(\ref{eq:EoSfromCs}),  the differences between the two parametrisations are less prominently visible compared to those in the speed of sound.  At energy densities $\varepsilon \lesssim 800\,$MeV$\,$fm$^{-3}$ the median of the Posterior distributions shows a close correspondence to the APR EoS \cite{Akmal1998} \footnote{available via the CompOSE library \cite{Typel2015,Oertel2017}},  but the inference results suggest a lower pressure at high energy densities.  It is a known feature that the APR EoS becomes too stiff and even violates causality at the highest energy densities. The credible bands of the pressure at an energy density of $\varepsilon = 1\,$GeV$\,$fm$^{-3}$ agree within uncertainties with the softer EoS extrapolated from pQCD in Ref. \cite{Gorda2022}.

Using the method described in Appendix \ref{sec:Bayesfactors}, we can compute the Bayes factor comparing the evidence for the Gaussian and the Segments parametrisation,  with the result:
\begin{equation}
	\mathcal{B}^{\text{Gaussian}}_{\text{Segments}} = 1.65 ~,
\end{equation}
which indicates that neither parametrisation is preferred by the data following the classification in Tab. \ref{tab:BayesFactorInterpretation}. The Bayes factor  $\mathcal{B}^{c_{s,max}^2 > 1/3}_{c_{s,max}^2 \leq 1/3}$ comparing equations of state with maximum squared speed of sound larger than $1/3$, i.e. violating the conformal limit, versus EoS with maximum squared sound speed smaller than $1/3$ is 
\begin{equation}
	\mathcal{B}^{c_{s,max}^2 > 1/3}_{c_{s,max}^2 \leq 1/3} = 26.2\times 10^2	
	\label{eq:BayesConLimit}
\end{equation}
in the Segments parametrisation and $\mathcal{B}^{c_{s,max}^2 > 1/3}_{c_{s,max}^2 \leq 1/3} = 16.2 \times 10^4$ in the Gaussian parametrisation. The Segments parametrisation can describe steeper slopes as well as plateaus,  resulting in a description more consistent with the data for EoS with maximum sound speeds smaller than $1/3$ and consequently a smaller Bayes factor. Nevertheless, in both parametrisations there is extreme evidence that the speed of sound reaches values larger than $c_s^2 = 1/3$ inside neutron stars,  exceeding the conformal limit. This is consistent with other recent studies  \cite{Landry2019,Landry2020,Legred2021,Leonhardt2020,Altiparmak2022,Gorda2022}.

\begin{figure*}[tp]
	\begin{center}
		\includegraphics[height=55mm,angle=-00]{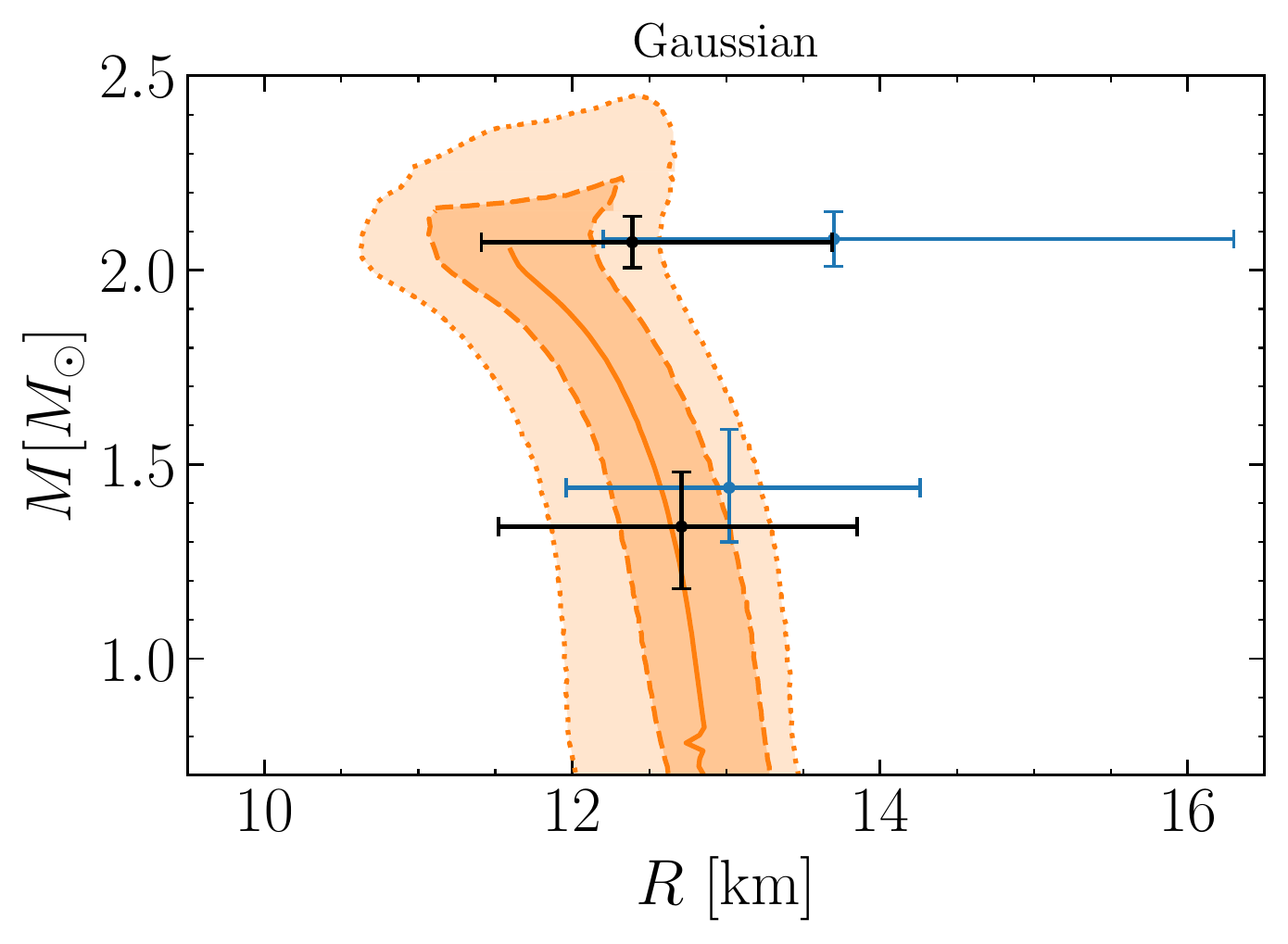} 
		\includegraphics[height=55mm,angle=-00]{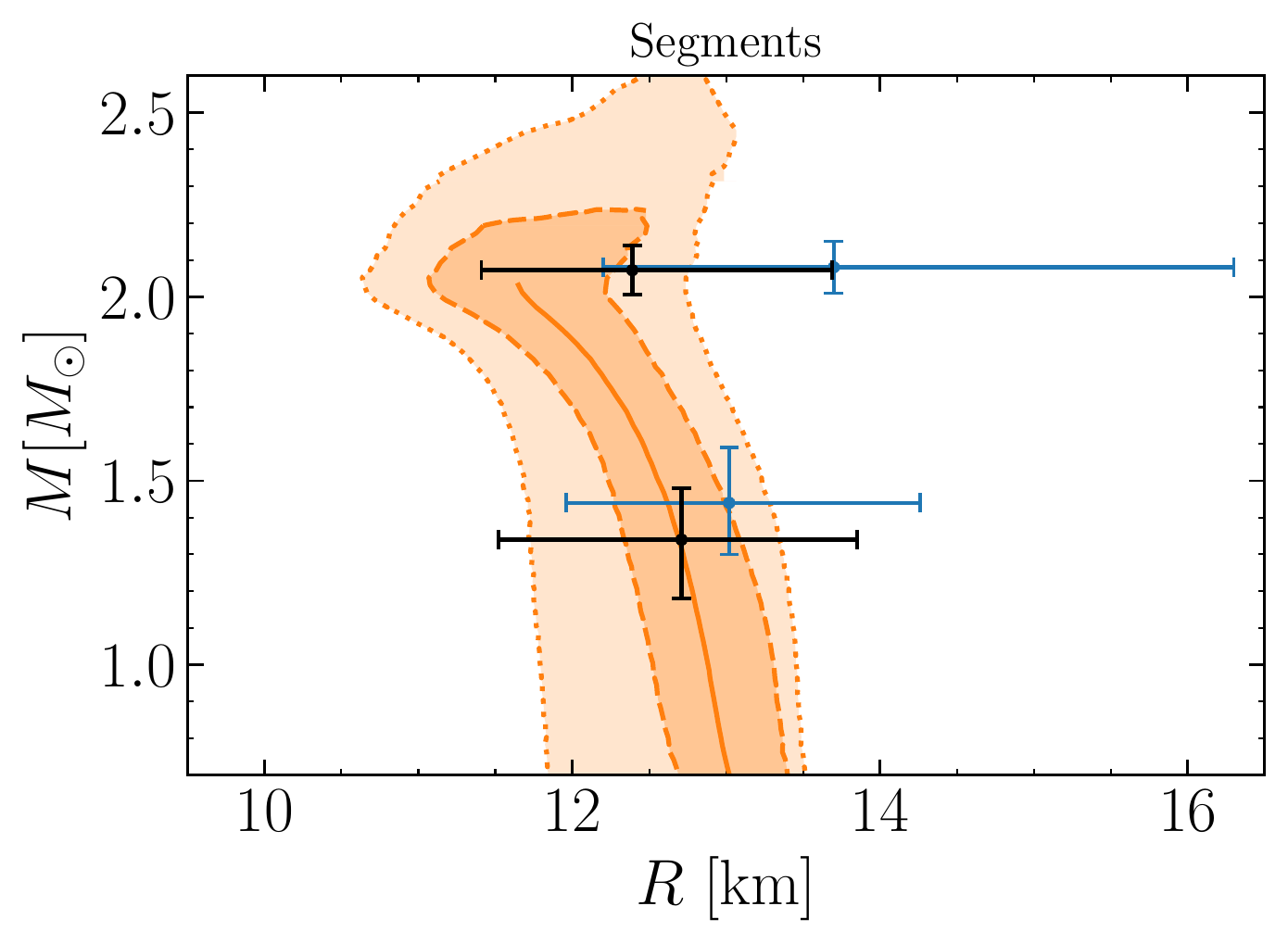} \\
		\includegraphics[height=55mm,angle=-00]{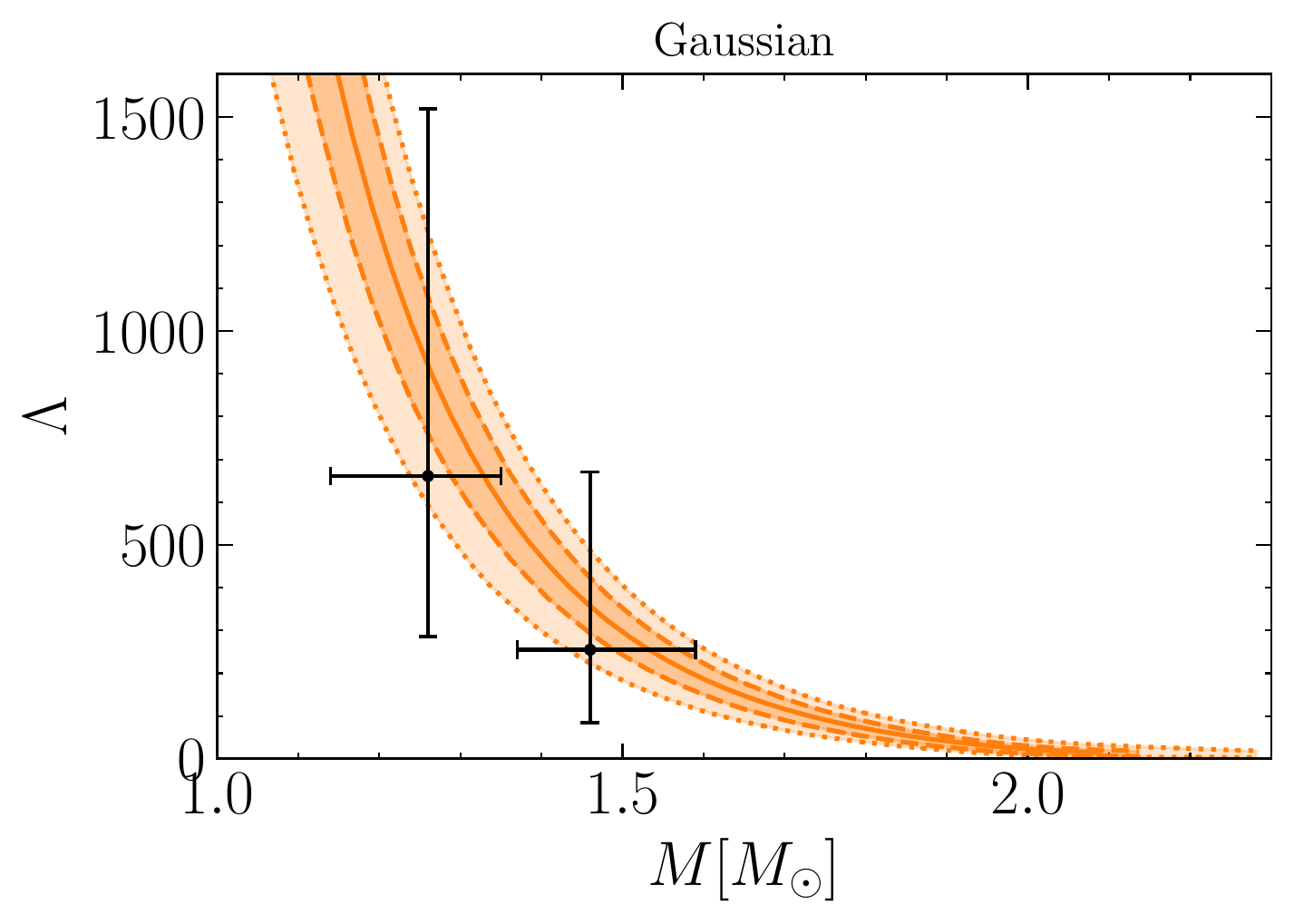}
		\includegraphics[height=55mm,angle=-00]{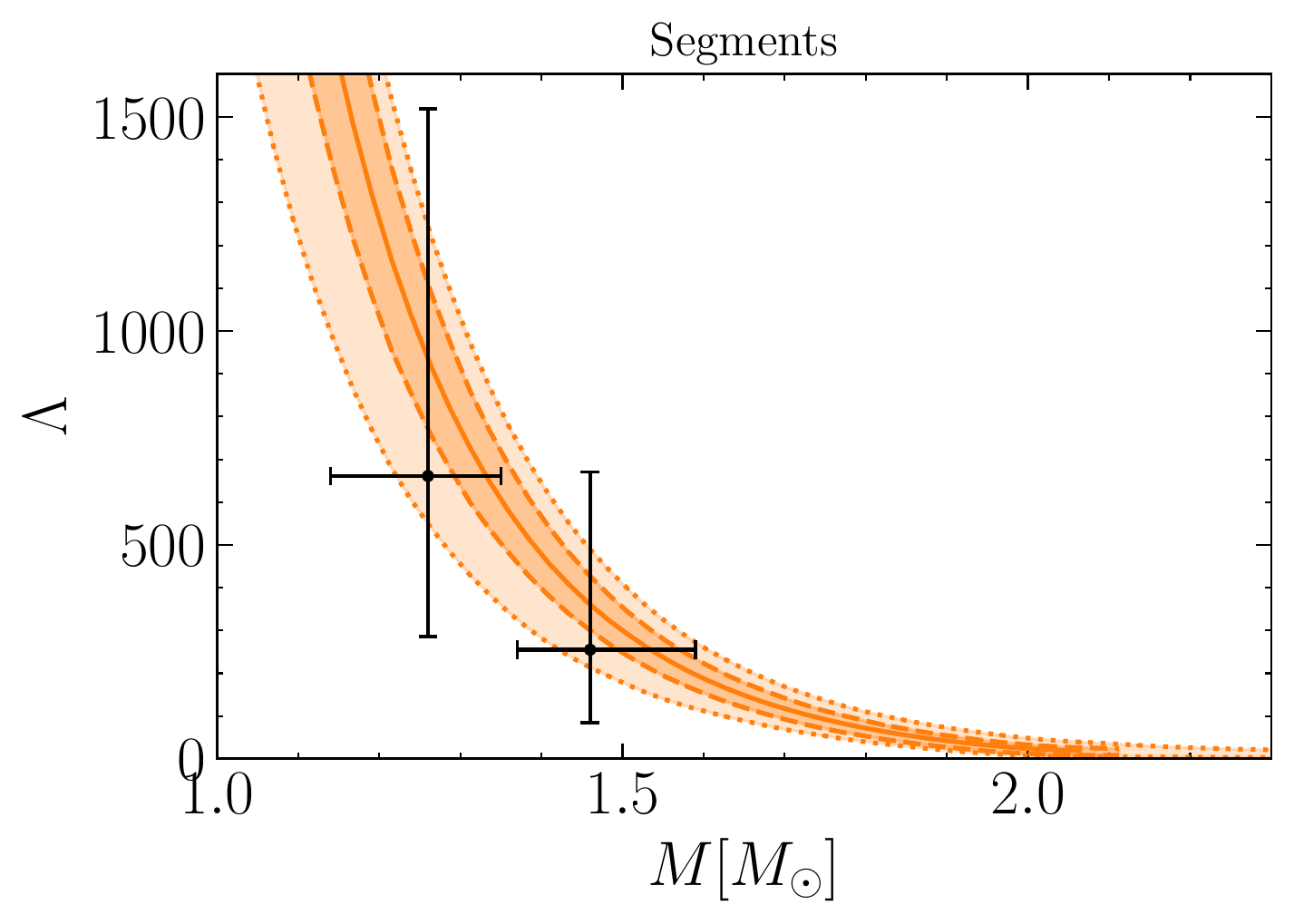} \\
		\caption{Marginal Posterior probability distributions at the 95\% and 68\% level for the Gaussian (left) and Segments parametrisation (right) for the radius $R$ and tidal deformability $\Lambda$ as a function of mass $M$. At each $M$, there exist 95\% and 68\% Posterior credible intervals for $R(M)$ and $\Lambda(M)$. These intervals are connected to obtain the Posterior credible bands. Similarly, the medians of the marginal Posterior probability distributions at each $M$ are connected (solid lines). The $R(M)$ median and credible bands are plotted until the median, upper 68\% or 95\% interval of the maximum mass at each radius. $R(M)$ is compared to the marginalised intervals at the 68\% level from the analysis of the NICER measurements of PSR J0030+0451 and PSR J0740+6620 by Miller et al. (blue) \cite{Miller2019,Miller2021} and by Riley et al. \cite{Riley2019,Riley2021} (black). $\Lambda(M)$ is compared to the masses and tidal deformabilities inferred in Ref.\,\,\cite{Fasano2019} for the two neutron stars in the merger event GW170817 at the 90\% level.}
		\label{fig:PosteriorBands2}
	\end{center}
\end{figure*}

\begin{figure*}[tp]
	\begin{center}
		\includegraphics[height=55mm,angle=-00]{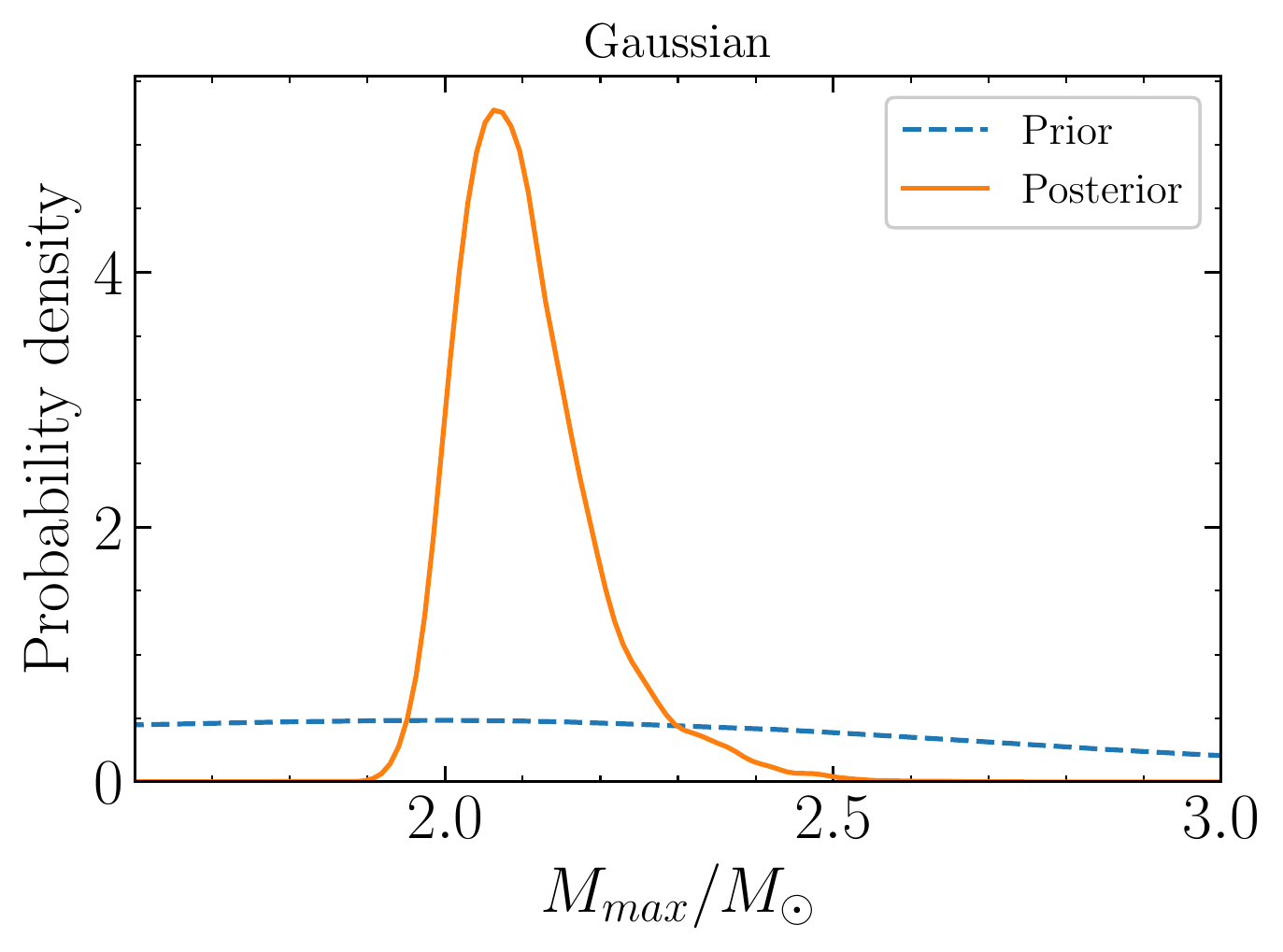} 
		\includegraphics[height=55mm,angle=-00]{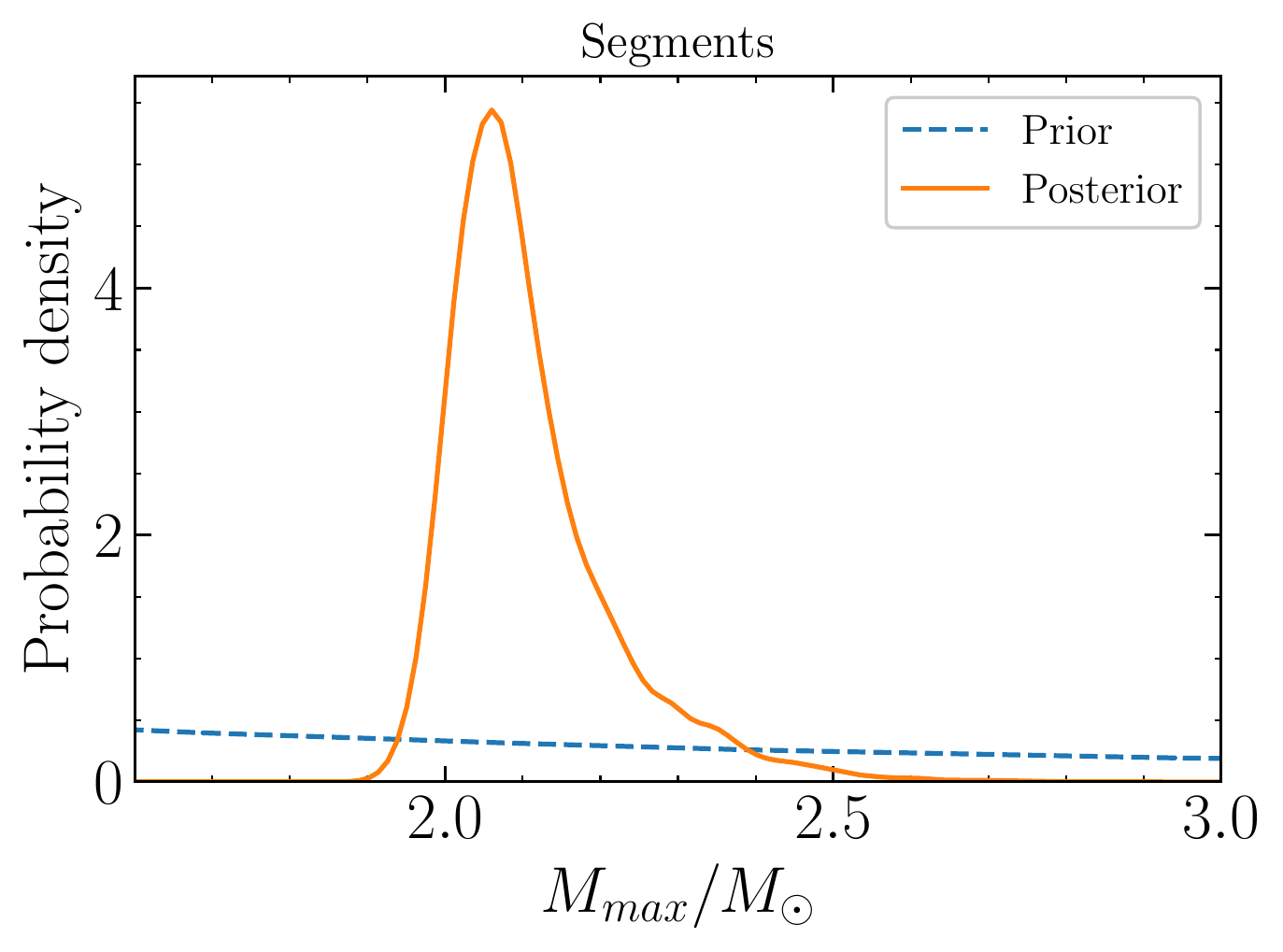} \\
		\caption{Marginal Posterior probability distributions of the maximum mass $M_{max}$ for the Gaussian (left) and Segments parametrisation (right) compared to the respective Prior.  The Prior distributions are nearly uniform over a wide mass range,  so the probability densities at individual masses are small.}
		\label{fig:MaximumMass}
	\end{center}
\end{figure*}

The Posterior credible bands for the mass-radius relation and for the tidal deformability $\Lambda(M)$ are shown in Fig.\,\,\ref{fig:PosteriorBands2}.  We choose to limit the $R(M)$ median and credible bands by the median, the upper 68\% and 95\% credible intervals of the maximum mass at each radius. As explained in Sec.\,\ref{sec:BayesianInferenceBasics}, we display credible bands instead of two-dimensional credible regions, because the former are independent of the Priors for the variables $\varepsilon$ and $M$.  However,  in the credible bands representation there is no natural ending criterion for the bands. Therefore the results for the mass-radius relation are often just cut after the upper 95\% interval of the maximum possible mass for all radii \cite{Legred2021,Riley2021,Biswas2021,Biswas2021a,Gorda2022,Huth2022}. The results for both parametrisations agree well up to the observed $\sim 2\, M_\odot$ mass range.  The extrapolated inference to larger masses is much less constrained by observations and the two parametrisations start varying,  reflecting the differences in the speed of sound in Fig.\,\,\ref{fig:PosteriorBands1}. The Segments parametrisation leads to a maximum  mass of $M_{max} = 2.08_{-0.13}^{+0.26} \,M_\odot$ and a maximum central density of $n_{c,max} = 6.5_{-1.3}^{+1.2} \, n_0$. The marginalized Posterior distributions for $M_{max}$ are compared to the Priors in Fig.\,\,\ref{fig:MaximumMass}.  Both Priors are almost uniform throughout a large range of masses,  and the Posterior distributions are nearly identical. In Fig.\,\,\ref{fig:PosteriorBands1}, there is good agreement with the marginalised 68\% error bars inferred from the NICER measurement of PSR J0030+0451.  However, the error bars inferred from the NICER measurement of PSR J0740+6620 are shifted to smaller radii compared to the $R(M)$ credible bands at $M \sim 2.08\,M_\odot$. This is because the gravitational wave event GW170817 prefers smaller radii, as noted in Ref.\,\,\cite{Raithel2018}.  Here the balancing between different observables and theoretical constraints becomes visible which requires a statistically well-defined analysis in contrast to simple cuts used e.g. in Refs.\,\,\cite{Annala2020,Annala2022,Altiparmak2022,Somasundaram2021,Somasundaram2022}.  Furthermore,  the error bars in Fig.\,\,\ref{fig:PosteriorBands2} display only the 68\% levels inferred from the NICER measurements.  The 95\% levels would extend to smaller radii.  The NICER analyses chose to use central credible intervals.  If instead,  as in the present work,  highest density intervals were used,  these intervals would reach to smaller radii. There is good agreement with the masses and tidal deformabilities derived in Ref.\,\,\cite{Fasano2019} for the two neutron stars in the merger event GW170817.  Finally,  if we use the NICER data analyses by Riley et al.  for the inference procedure instead of the one by Miller et al.,  we find very similar results.  So we can restrict ourselves to the latter. 

\setlength{\extrarowheight}{6pt}
\begin{table*}[tp]
	\centering
	\begin{tabularx}{\linewidth}{|X|X|XX|XX|}
		\hline \hline 
		\multicolumn{2}{|l|}{} & \multicolumn{2}{l|}{Gaussian} & \multicolumn{2}{l|}{Segments} \\
		\multicolumn{2}{|l|}{} & 95\% & 68 \% & 95\% & 68\% \\ \hline
		&$n_c/n_0$ & $2.8^{+0.5}_{-0.4}$ & $\pm0.2$ & $2.8_{-0.4}^{+0.5}$ & $\pm0.2$ \\
		&$\varepsilon_c  \, $[MeV$\,$fm$^{-3}$] & $451_{-68}^{+88}$ & $_{-40}^{+35}$ & $453_{-72}^{+83}$ & $_{-37}^{+34}$ \\
		$1.4\, M_\odot$& $P_c \, $[MeV$\,$fm$^{-3}$] & $60_{-13}^{+19}$ & $_{-8}^{+7}$ & $60_{-13}^{+20}$ & $_{-7}^{+6}$ \\
		&$R \, $[km] & $12.6_{-0.8}^{+0.6}$ & $\pm0.3$ & $12.7_{-0.9}^{+0.6}$ & $_{-0.3}^{+0.4}$ \\
		&$\Lambda$ & $471_{-168}^{+167}$ & $_{-76}^{+88}$ & $485_{-194}^{+157}$ & $_{-71}^{+83}$\\ \hline 
		&$n_c /n_0$ & $4.9_{-1.6}^{+1.3}$ & $_{-0.8}^{+0.9}$ & $4.8 \pm 1.6$ & $_{-0.9}^{+1.1}$\\
		&$\varepsilon_c  \, $[MeV$\,$fm$^{-3}$] & $904_{-372}^{+329}$ & $_{-187}^{+208}$ & $883_{-371}^{+390}$ & $_{-277}^{+183}$ \\
		$2.1\, M_\odot$& $P_c \, $[MeV$\,$fm$^{-3}$] & $312_{-169}^{+226}$ & $_{-134}^{+69}$ & $300_{-171}^{+257}$ & $_{-143}^{+84}$\\
		&$R \, $[km] & $11.6 \pm 0.9$ & $\pm 0.5$ & $11.6 \pm 1.0$ & $\pm 0.6$\\
		&$\Lambda$ & $15_{-9}^{+16}$ & $_{-7}^{+5}$ & $15_{-10}^{+18}$ & $_{-8}^{+6}$\\
		\hline \hline 
	\end{tabularx}
	%\end{ruledtabular}
	\caption{Median, 95\% and 68\% credible intervals for selected neutron star properties for the Gaussian and Segments parametrisations. These are computed from the one-dimensional Posterior probability distribution marginalised over all other parameters. Listed are the central density, energy density, pressure, radius and tidal deformability of neutron stars with masses $M = 1.4 \, M_\odot$ and $2.1 \, M_\odot$.}
	\label{tab:NS_properties1}
\end{table*}
\setlength{\extrarowheight}{4pt}

Tab. \ref{tab:NS_properties1} shows medians and credible intervals for selected properties of neutron stars with characteristic masses $M = 1.4\, M_\odot$ or $2.1 \, M_\odot$,  including the central density, the energy density and pressure as well as the radius and tidal deformability.  Again these numbers demonstrate agreement within uncertainties between the two parametrisations. 

At the 95\% level (version S) the inferred radius of a $1.4 \, M_\odot$ neutron star,  $R = 12.7_{-0.9}^{+0.6}\,$km,  agrees with the values found in Ref.\,\,\cite{Raaijmakers2021} for a piecewise polytrope parametrisation and a speed of sound model similar to our Gaussian parametrisation,  while the authors additionally included constraints from modelling of the kilonova AT2017gfo.  The 68\% credible intervals of the radius and tidal deformability of a $1.4 \, M_\odot$ neutron star listed in Tab. \ref{tab:NS_properties1} agree within uncertainties with the results in Ref.\,\,\cite{Lim2022} which include a theory prediction and the PREX II measurement of the $^{208}$Pb neutron skin thickness.  Our result for the $1.4 \, M_\odot$ radius also agrees with the value found in Ref.\,\,\cite{Huth2022},  where the authors additionally incorporated constraints on the EoS deduced from relativistic heavy-ion collisions. 

For a $2.1 \, M_\odot$ neutron star representative of the heaviest currently observed star,  the inferred radius is $R = 11.6 \pm 1.0\,$km,  the tidal deformability is $\Lambda = 15^{+18}_{-10}$ and the central density is $n_c = 4.8\pm 1.6 \, n_0$.  In the Bayesian analysis of Ref.\,\,\cite{Miller2021},  no ChEFT constraint was included at low densities.  Their prediction for the radius of a neutron star with mass  $M = 1.4\, M_\odot$,  based on multiple different parametrisations,  agrees nonetheless with our result at the 68\% level.  Their result for the radius of the $2.08\, M_\odot$ neutron star is larger compared to our result for the radius of a generic $2.1\, M_\odot$ neutron star.  However, within the 68\% credible intervals the two results are still consistent and the differences can be accounted for by the ChEFT constraint.  At the current state of investigations with a limited neutron star data base and correspondingly large uncertainties,  it is still justified to use parametric functional forms as long as they are sufficiently general. 

In Ref. \cite{Legred2022} the authors compare different parametrisations and argue that inferred neutron star results depend on the chosen parametrisation.  However, in their comparison the primary differences in the inferred equations of state occur at small densities,  mainly because of different implementations of the neutron star crust,  and in the high density regime not constrained by data.  In the intermediate region, $n \sim 1.5 - 6\,n_0$, the different parametrisations agree within their uncertainties.  One exception is a Gaussian parametrisation which, unlike our G version,  does not allow skewed Gaussians and is therefore not sufficiently general to reproduce the current astrophysical data,  a feature that is already visible from its Prior.  Our point regarding the stability of inference results with respect to different parametrisations is further supported by the work of Ref. \cite{Huth2022} where very similar neutron star properties are found for two qualitatively different parametrisations, namely a Segments parametrisation similar to our S version and a piecewise polytrope representation.  In the future many more data are expected and a non-parametric description of the EoS in terms of a Gaussian Process \cite{Landry2019,Landry2020,Legred2021,Legred2022} or neural network \cite{Han2021,Han2022} might be preferable.

\setlength{\extrarowheight}{6pt}
\begin{table*}[tp]
	\centering
	\begin{tabularx}{\linewidth}{|X|XX|XX|} 
		\hline \hline
		& \multicolumn{2}{l|}{Gaussian} & \multicolumn{2}{l|}{Segments} \\
		& 95\% & 68 \% & 95\% & 68\% \\ \hline
		$c_{s,max}^2$ & $0.68_{-0.24}^{+0.29}$ & $_{-0.19}^{+0.13}$ & $0.76_{-0.26}^{+0.24}$ & $_{-0.15}^{+0.17}$ \\
		$n(c_{s,max})/n_0$ & $4.8\pm1.9$ & $_{-1.1}^{+1.0}$ & $5.3_{-2.8}^{+1.9}$ & $_{-1.2}^{+1.7}$ \\
		$c_{s,min}^2$ & $0.34_{-0.34}^{+0.44}$ & $_{-0.34}^{+0.15}$ & $0.56_{-0.50}^{+0.37}$ & $_{-0.19}^{+0.32}$ \\
		$n(c_{s,min})/n_0$ & $6.3_{-2.3}^{+1.3}$ & $_{-0.7}^{+0.8}$ & $6.7_{-1.7}^{+1.0}$ & $_{-0.5}^{+0.7}$ \\
		\hline \hline 
	\end{tabularx}
	\caption{Median, 95\% and 68\% credible intervals for selected neutron star properties for the Gaussian and Segments parametrisations. These are computed from the one-dimensional Posterior probability distribution marginalised over all other parameters.  Displayed are the maximum squared speed of sound, $c_{s,max}^2$,  together with the density $n(c_{s,max})$ at which this maximum is located,  and the minimum speed of sound $c_{s,min}^2$ following the maximum at a higher density, $n(c_{s,min}) > n(c_{s,max})$.}
	\label{tab:NS_properties2}
\end{table*}
\setlength{\extrarowheight}{4pt}

The quest for a possible phase transition or crossover within the density range realised in the core of neutron stars can be addressed by assuming that the squared sound speed develops a maximum, $c_{s,max}^2$,  at some density $n(c_{s,max})$, and then a minimum,  $c_{s,min}^2$,  at a higher density $n(c_{s,min}) > n(c_{s,max})$.  At even much higher (asymptotic) densities, $c_s^2$ should approach the conformal limit from below if one follows the standard pQCD scenario.  

In Tab. \ref{tab:NS_properties2} credible intervals for $c_{s,max}^2$ are displayed for both S and G parametrisations in comparison.  For version S at 68\% level such a (shallow) maximum is reached at densities $n(c_{s,max}) \sim 5.3_{-1.2}^{+1.7}\,n_0$.  (Notably this density range is much higher than values of $n(c_{s,max})$ deduced from analyses in which the ChEFT constraint is implemented only at a density as low as $n = 1.1\,n_0$ \cite{Altiparmak2022,Gorda2022}. ) A similarly shallow minimum at higher densities follows at $n(c_{s,min}) \sim 6.7^{+0.7}_{-0.5}\,n_0$.  Such a high density could possibly be reached in a speculative super-heavy neutron star for which the speed of sound is, however, only marginally constrained by the existing data.  For version G the resulting densities $n(c_{s,max})$ and $n(c_{s,min})$ are located at slightly lower values,  indicating the principal possibility of a phase transition or crossover in the deep interior of a $2\,M_\odot$ neutron star,  but with low probability.

\begin{table}[tp]
	\centering
	\begin{tabularx}{\linewidth}{|X|XX|}
		\hline \hline 
		& \multicolumn{2}{l|}{$\mathcal{B}^{c_{s,min}^2 > 0.1}_{c_{s,min}^2 \leq 0.1}$} \\ 
		$M /M_\odot$ & Gaussian & Segments \\ \hline
		1.9 & 5.89 $\times 10^3$  & 1.75 $\times 10^4$ \\
		2.0 & 15.17 & 17.76\\
		2.1 & 2.51 & 2.01\\
		2.2 & 1.67 & 1.39\\
		\hline \hline  
	\end{tabularx}
	\caption{Bayes factors $\mathcal{B}^{c_{s,min}^2 > 0.1}_{c_{s,min}^2 \leq 0.1}$ comparing EoS samples with the following competing scenarios: a) minimum squared speed of sound  (following a maximum),  with $c^2_{s,min}$ larger than 0.1,  excluding a first-order phase transition; versus b) EoS with $c_{s,min}^2 \leq 0.1$.  The Bayes factors are calculated for a given maximum neutron star mass $M$,  i.e. the minimum speed of sound up to the corresponding mass is used. There is extreme evidence that the minimum squared sound speed, after exceeding the conformal limit,  does not drop to values smaller than $0.1$ for neutron stars with mass $M \leq 1.9 \, M_\odot$. There is strong evidence that $c_{s,min}^2$ does not become smaller than $0.1$ in neutron stars with mass $M \leq 2.0\, M_\odot$. }
	\label{tab:BayesFactorSmallCsM}
\end{table}

This discussion can be further quantified by considering the Bayes factors in Tab. \ref{tab:BayesFactorSmallCsM}. Here the evidence for EoS with $c^2_{s,min} < 0.1$ is compared to EoS with larger speeds of sound.  The Bayes factors are computed for different maximum neutron star masses, i.e. analysing the sound speed minimum up to this mass.  For neutron stars with masses up to $M \leq 1.9\, M_\odot$,  there is extreme evidence that the squared speed of sound inside the neutron star does not fall below $0.1$ after exceeding the conformal limit.  For heavier neutron stars smaller minimal sound speeds become more likely,  but there is still strong evidence for the absence of a first-order phase transition inside a $M \leq 2.0\, M_\odot$ star.  Altogether,  Tables \ref{tab:NS_properties2} and \ref{tab:BayesFactorSmallCsM} suggest that a first-order phase transition can only possibly take place in a very heavy neutron star with mass $M > 2\, M_\odot$.  In the future, observations of such very heavy neutron stars will be of prime interest. 

\begin{figure*}[tp]
	\begin{center}
		\includegraphics[height=55mm,angle=-00]{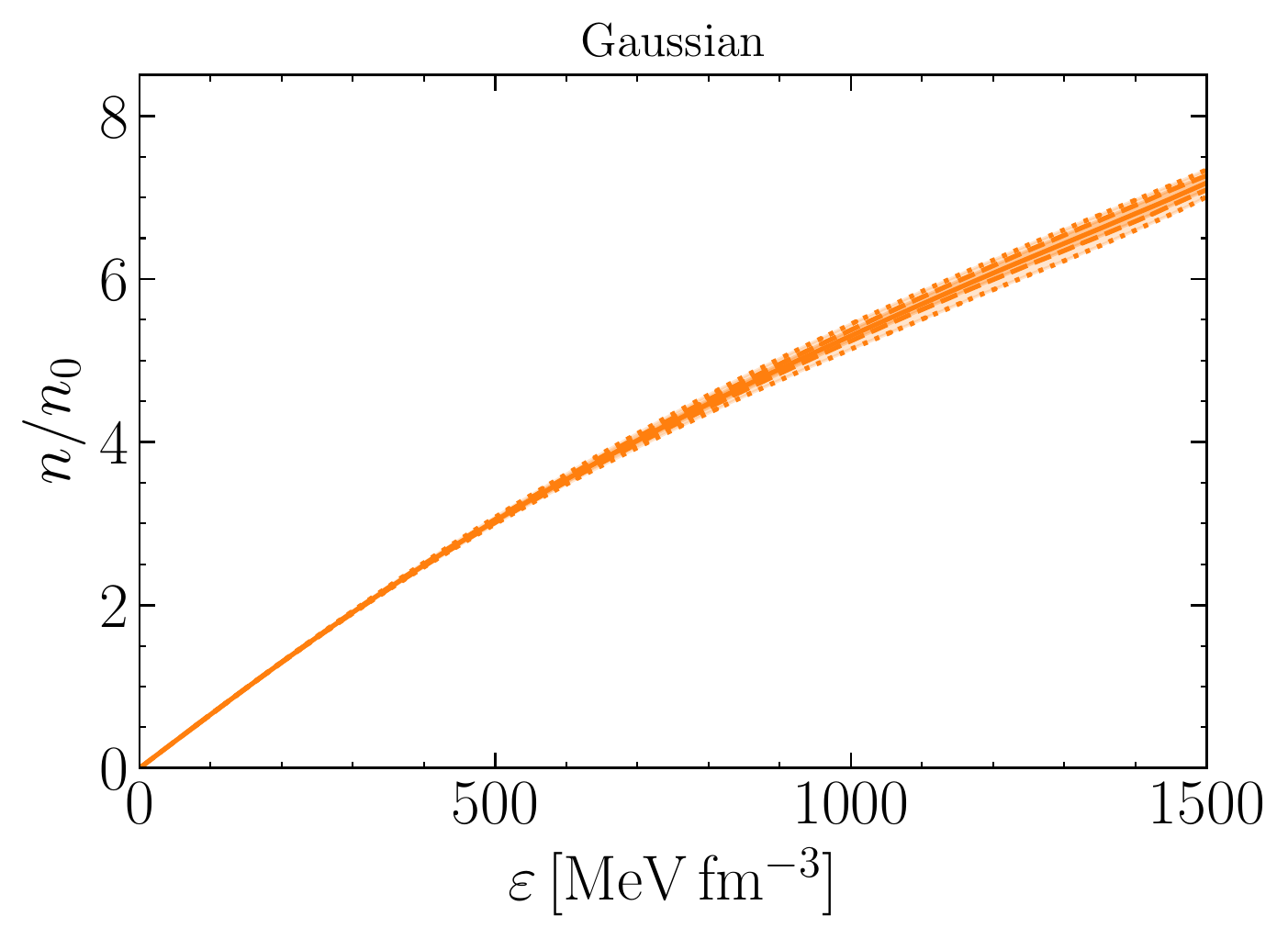}
		\includegraphics[height=55mm,angle=-00]{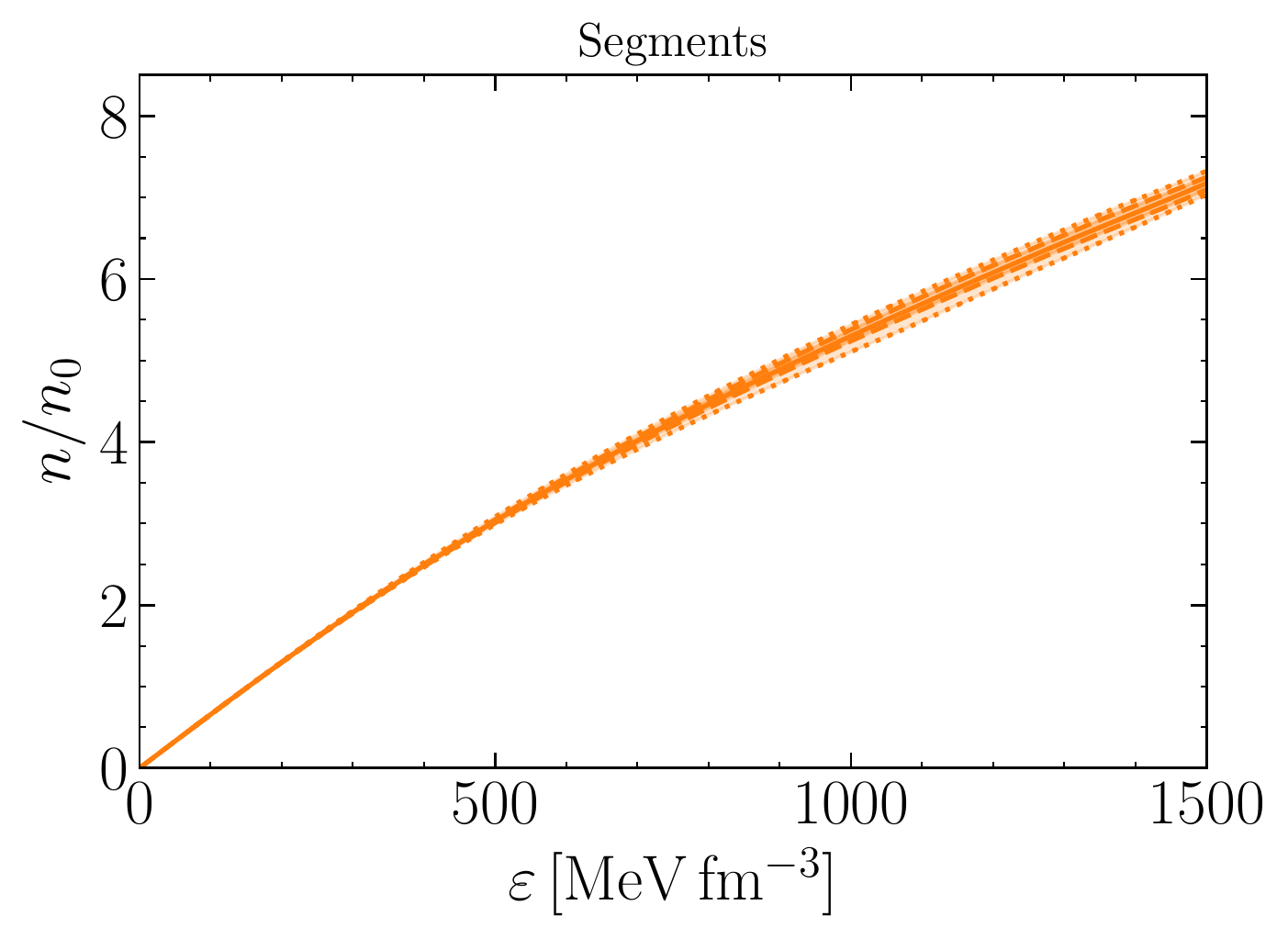} \\
		\caption{Posterior 95\% and 68\% credible bands and medians for the Gaussian (left) and Segments parametrisation (right): baryon density $n$, in units of the nuclear saturation density $n_0 = 0.16\,$fm$^{-3}$, as a function of energy density $\varepsilon$. For tabulated values of the median in the Segments parametrisation see Tab. \ref{tab:ThermoDynQuantities}.}
		\label{fig:DensityOfEnergDensity}
	\end{center}
\end{figure*}

\begin{figure*}[tp]
	\begin{center}
		\includegraphics[height=55mm,angle=-00]{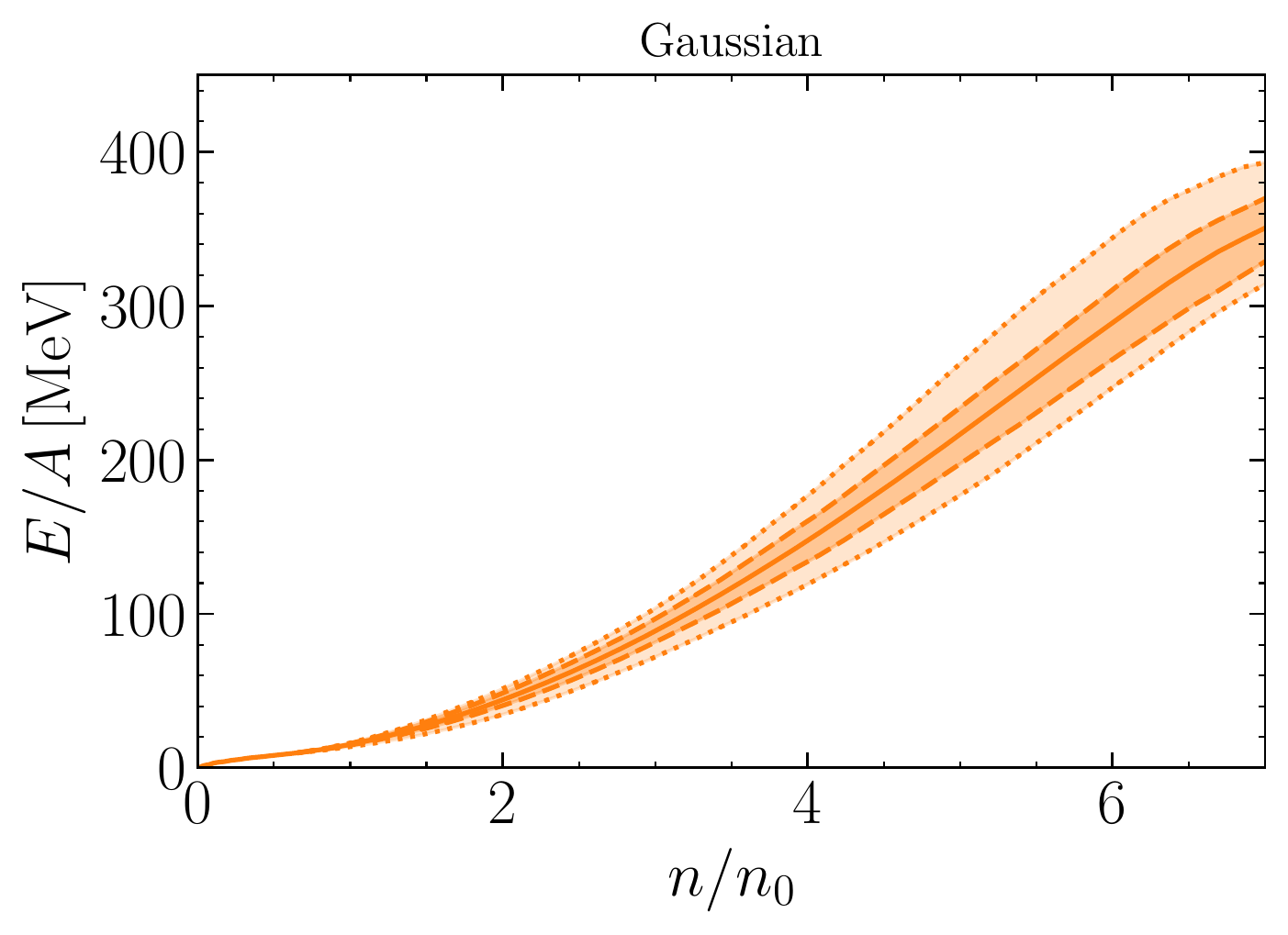}
		\includegraphics[height=55mm,angle=-00]{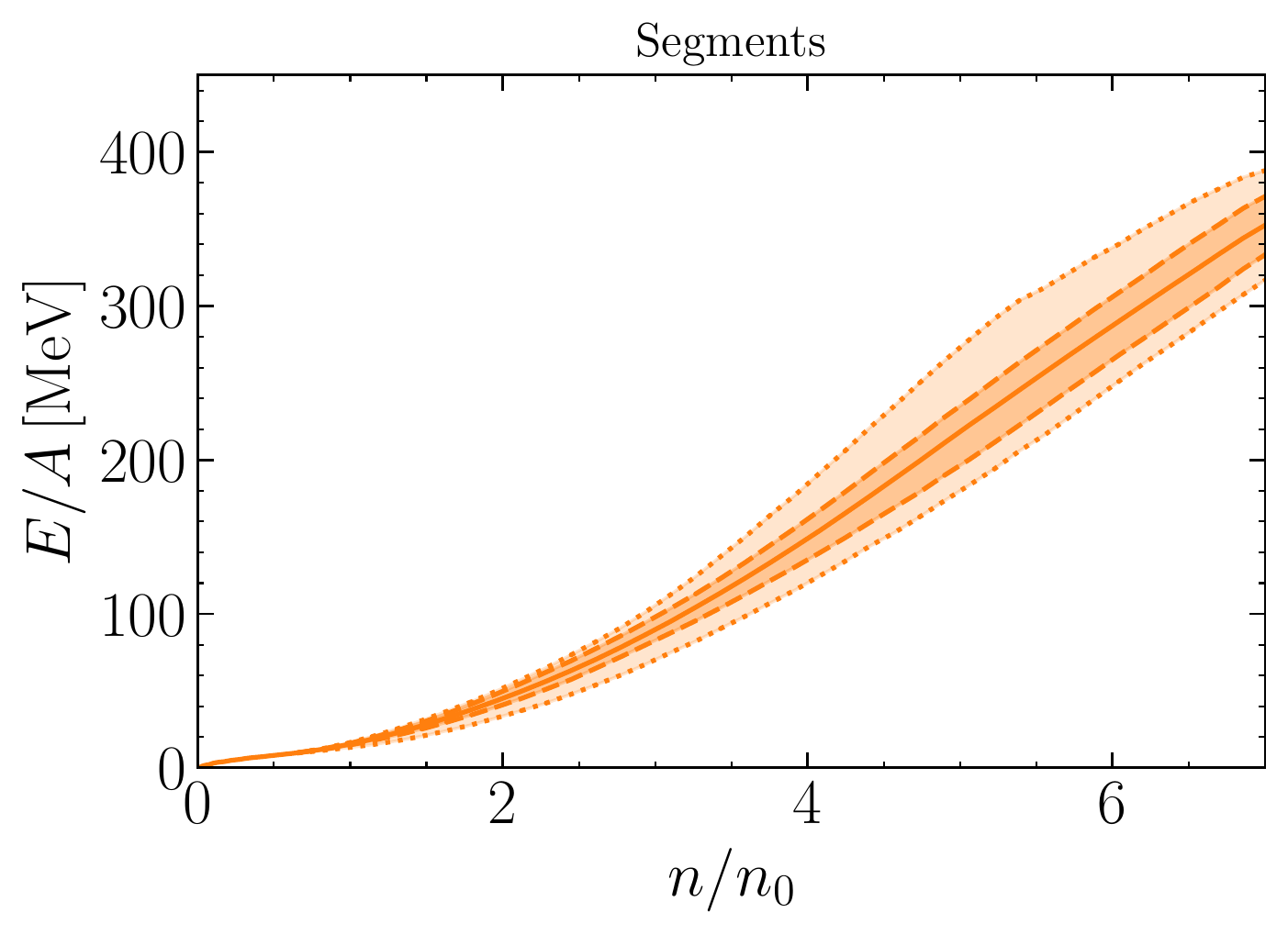} \\
		\caption{Posterior 95\% and 68\% credible bands and medians for the Gaussian (left) and Segments parametrisation (right): energy per particle $E/A = \varepsilon/n - m_N$ as function of baryon density $n$ in units of nuclear saturation density, $n_0 = 0.16\,$fm$^{-3}$.}
		\label{fig:EnergyPerParticleOfEnergDensity}
	\end{center}
\end{figure*}

In Fig.\,\,\ref{fig:DensityOfEnergDensity} Posterior credible bands for the baryon density as a function of energy density are displayed.  From this relation or its inverse, $\varepsilon(n)$,  all other quantities at $T=0$ can be deduced via the Gibbs-Duhem relation
\begin{equation}
	n \frac{\partial\varepsilon}{\partial n} = P + \varepsilon~.
	\label{eq:GibbsDuhem}
\end{equation}
For practical applications the EoS $P(\varepsilon)$,  the baryon density,  the energy per particle $E/A$ and the squared sound speed are tabulated in Tab. \ref{tab:ThermoDynQuantities} of Appendix \ref{sec:ThermoDynQuantities} using a fit to the median of $n(\varepsilon)$ in the Segments parametrisation. The energy per particle can be computed as 
\begin{equation}
	\frac{E}{A} = \frac{\varepsilon}{n} - m_N ~,
	\label{eq:EA}
\end{equation}
which is displayed in Fig.\,\,\ref{fig:EnergyPerParticleOfEnergDensity} as a function of baryon density.  Here we take $m_N \sim 939.5\,$MeV,  the neutron mass with minor adjustment for a small proton fraction of $\sim 10\%$ as in the APR EoS. 

\begin{figure*}[tp]
	\begin{center}
		\includegraphics[height=55mm,angle=-00]{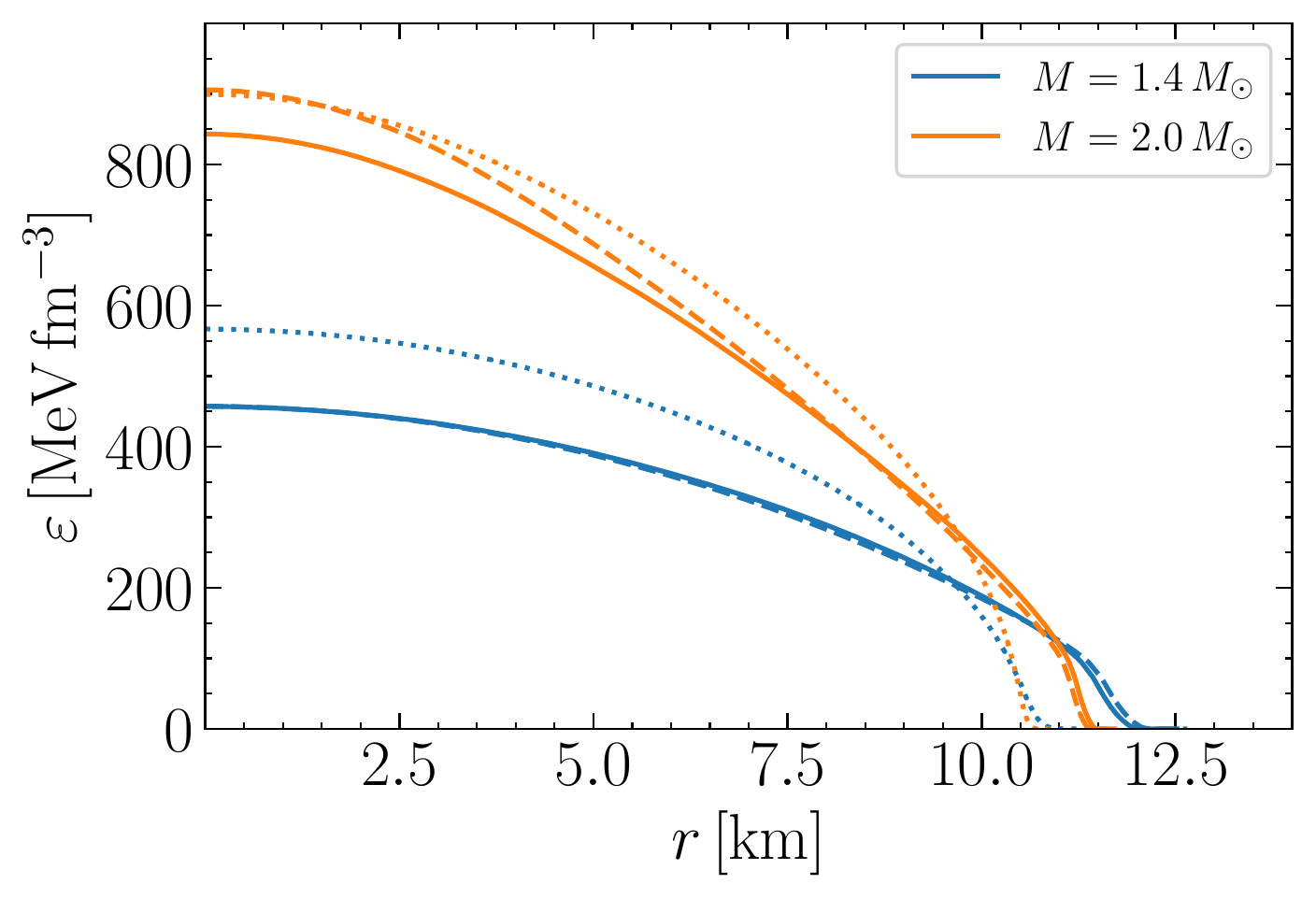} 
		\includegraphics[height=55mm,angle=-00]{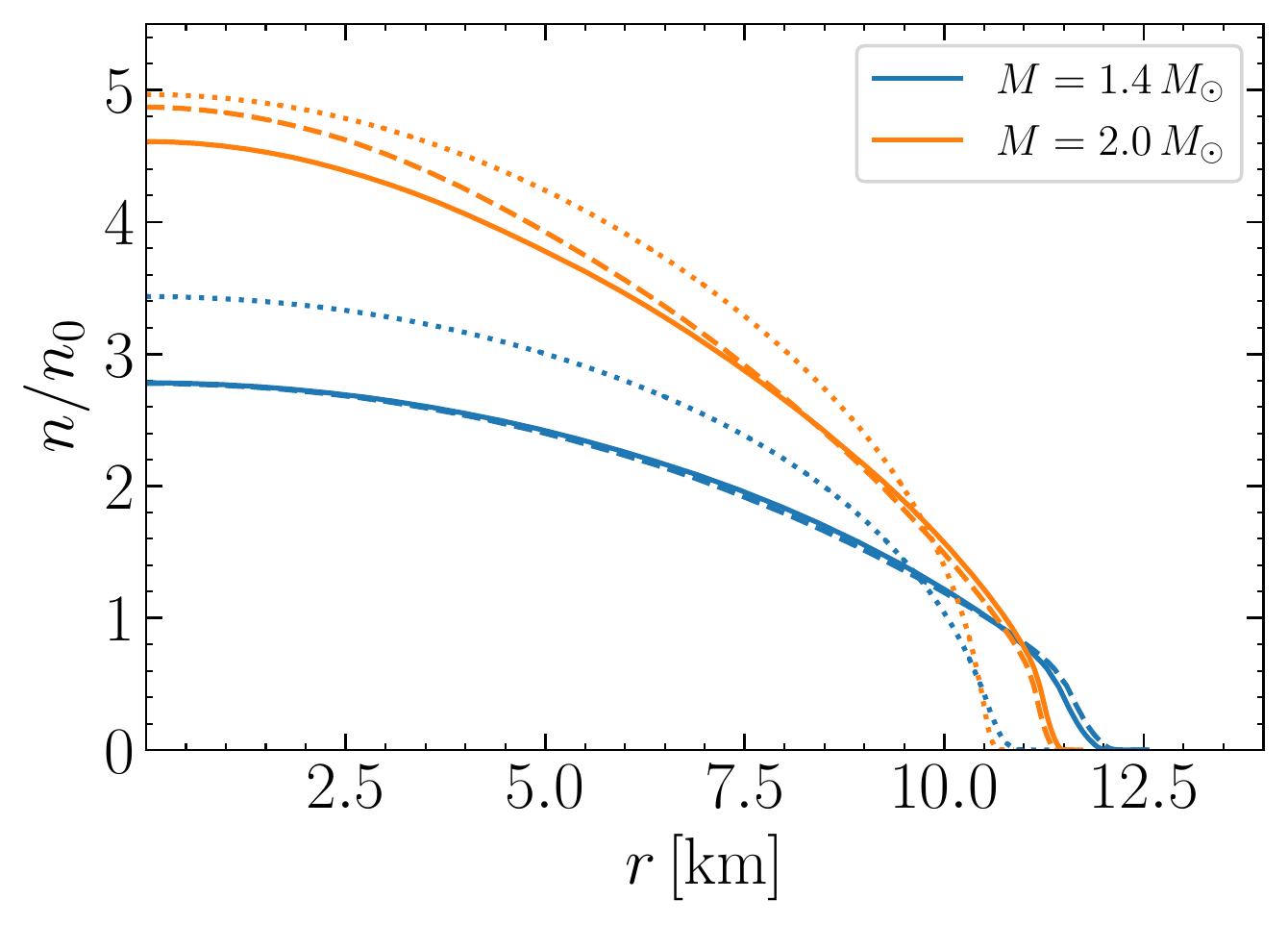} 
		\caption{Energy density profiles (left) and baryon density profiles (right) of neutron stars with masses of $M = 1.4\,M_\odot$ (blue) and $2.0\,M_\odot$ (orange).  The equations of state used correspond to the median values of the credible bands in Fig.\,\,\ref{fig:PosteriorBands1} for the Gaussian parametrisation (solid) and the Segments parametrisation (dashed), respectively.  The energy density and density profiles derived from the APR EoS are displayed for comparison (dotted lines).}
		\label{fig:DensityProfiles}
	\end{center}
\end{figure*}

Fig.\,\,\ref{fig:DensityProfiles} shows examples of energy density and baryon density profiles for neutron stars with masses $M = 1.4\, M_\odot$ and $2.0 \, M_\odot$.  For each parametrisation the median displayed in Fig.\,\,\ref{fig:PosteriorBands1} is used as the corresponding equation of state.  There is once again good agreement between both parametrisations for $M = 1.4\, M_\odot$.  Compared to the APR EoS the resulting neutron star radii are larger by about 1.3 km.  For a $2.0\, M_\odot$ neutron star,  both parametrisations lead to similar radii but the Segments parametrisation prefers a larger central energy density, $\varepsilon_c$, and central density $n_c$.

\subsection{Monotonically rising speed of sound} 

\begin{table}[tp]
	\centering
	\begin{tabularx}{\linewidth}{|X|XX|}	  
		\hline \hline 
		& \multicolumn{2}{l|}{$\mathcal{B}^{n_{-}\leq n_{tr}}_{n_{-} > n_{tr}}$} \\
		$n_{tr} /n_0$ & Gaussian & Segments \\ \hline
		3 & 0.02 & 0.32 \\
		4 & 0.17 & 0.76 \\
		5 & 0.45 & 1.50 \\
		6 & 1.71 & 3.20 \\
		\hline \hline  
	\end{tabularx}
	%\end{ruledtabular}
	\caption{Bayes factors $\mathcal{B}^{n_{-}\leq n_{tr}}_{n_{-} > n_{tr}}$ comparing EoS in which the derivative of the squared sound velocity, $\partial c_s^2/\partial \varepsilon$,  turns negative at a density $n_{-}$ below the transition density $n_{tr}$,  versus EoS with $n_{-}> n_{tr}$. There is strong evidence for the Gaussian parametrisation and moderate evidence for the Segments parametrisation that $\partial c_s^2/\partial \varepsilon > 0$ at least up to $n_{tr} = 3\, n_0$.} 
	\label{tab:BayesFactorntr}
\end{table}

With Priors prepared in broad generality and unrestricted initialization of the speed of sound,  the previous inference results pointed out some (perhaps) unexpected properties of a heavy ($2.1\,M_{\odot}$) neutron star.  In particular, the central baryon density in the core of the star is not extreme: it does not exceed $n_c \simeq (5 \pm 1)\, n_0$ (at 68\% credibility) in both parametrisations.  For an interpretation,  suppose that the neutron star centre is composed of baryons viewed as rigid spheres with a typical ``hard core" radius of $R \simeq 0.5\,$fm.  Then the critical density even for random close packing of such hard spheres, $n_{crit} \simeq 0.16/R^3 \sim 8\,n_0$ (for a packing volume fraction $\phi\simeq 0.66$ \cite{Zaccone2022}), is still significantly higher than $n_c$. In the same context, the average distance between two baryons at density $n \sim (5-6)\,n_0$ is still around $1\,$fm, considerably larger than the typical hard core range. One expects that such a scenario is characterized by a monotonically increasing $c_s^2$ as function of baryon density $n$,  with no phase transition in the neutron star core.  It is then instructive to investigate whether or not this picture is compatible with the existing empirical data. 

Following Sec.\,\ref{sec:AdditionalPrior},  related insights can be gained by examining a restrictive scenario in which $c_s^2$ is assumed to increase monotonically up to a given transition density,  $n_{tr}$.  No phase transition or crossover occurs at baryon densities $n \leq n_{tr}$,  while freedom for phase changes or the appearance of new degrees of freedom exists at densities $n > n_{tr}$,  within the constraints provided by the empirical data.  For the preparation of a corresponding Bayes factor analysis,  introduce a generic density, $n_{-}$, characterised by the slope $\partial c_s^2/\partial \varepsilon < 0$ being negative at that density,  i.e.  the counter example to a continuously increasing sound velocity. (For $n_-$ exceeding neutron star central densities, i.e. in the range not constrained by data, we assume that the corresponding equation-of-state P($\varepsilon$) continues rising.) Consider now the following two scenarios:  hypothesis $H_0$ corresponds to the case $n _-> n_{tr}$,  i.e.  the sound velocity increases monotonically at densities up to $n_{tr}$.  The sound speed may then change its slope and decrease at some higher density,  $n _-$.  The counter hypothesis,  $H_1$,  assumes that this change of slope in $c_s^2$ occurs at a lower density instead, $n _- \leq n_{tr}$,  in which case $n_{tr}$ simply acts as a density scale for comparison with the opposite hypothesis $H_0$.  The Bayes factors ${\cal B}^{n_-\leq n_{tr}}_{n_- > n_{tr}}$ then ask for the Likelihoods of the competing hypotheses and quantify the evidence of $H_1$ over $H_0$ for given values of $n_{tr}$. In Tab. \ref{tab:BayesFactorntr} these  Bayes factors are listed for different values of $n_{tr}$. There is strong evidence in the Gaussian and moderate evidence in the Segments parametrisation that $n_{tr} = 3 \, n_0$ is preferred by the data. This means that an EoS with monotonically rising sound speed, $\partial c_s^2/\partial \varepsilon > 0$ up to $n \lesssim n_{tr} = 3 \, n_0$,  is on average more likely than an EoS that develops a plateau or decreasing sound speed in this regime,  indicating that a crossover or phase transition below this transition density is unlikely.  

In Ref.\,\,\cite{Kojo2021a} the author argues that if a nuclear description is trusted up to $\sim 2\, n_0$,  then a first-order phase transition is unlikely and a quark-hadron continuity scenario may be favoured.  Such a picture would be consistent with the results of the present work which uses a conservative ChEFT constraint and assumes a description in terms of nucleonic matter to be valid up to about $2\, n_0$. The Bayes factors in Tab. \ref{tab:BayesFactorntr} actually indicate that such a description might remain valid at least up to $3\, n_0$. There is no evidence for or against larger transition densities,  which implies that a description of neutron stars in terms of nucleonic matter up to $n \sim 6\, n_0$, characterized by a monotonically rising speed of sound \cite{Friman2019},  cannot be excluded by the currently available data.

\begin{figure*}[tp]
	\begin{center}
		\includegraphics[height=55mm,angle=-00]{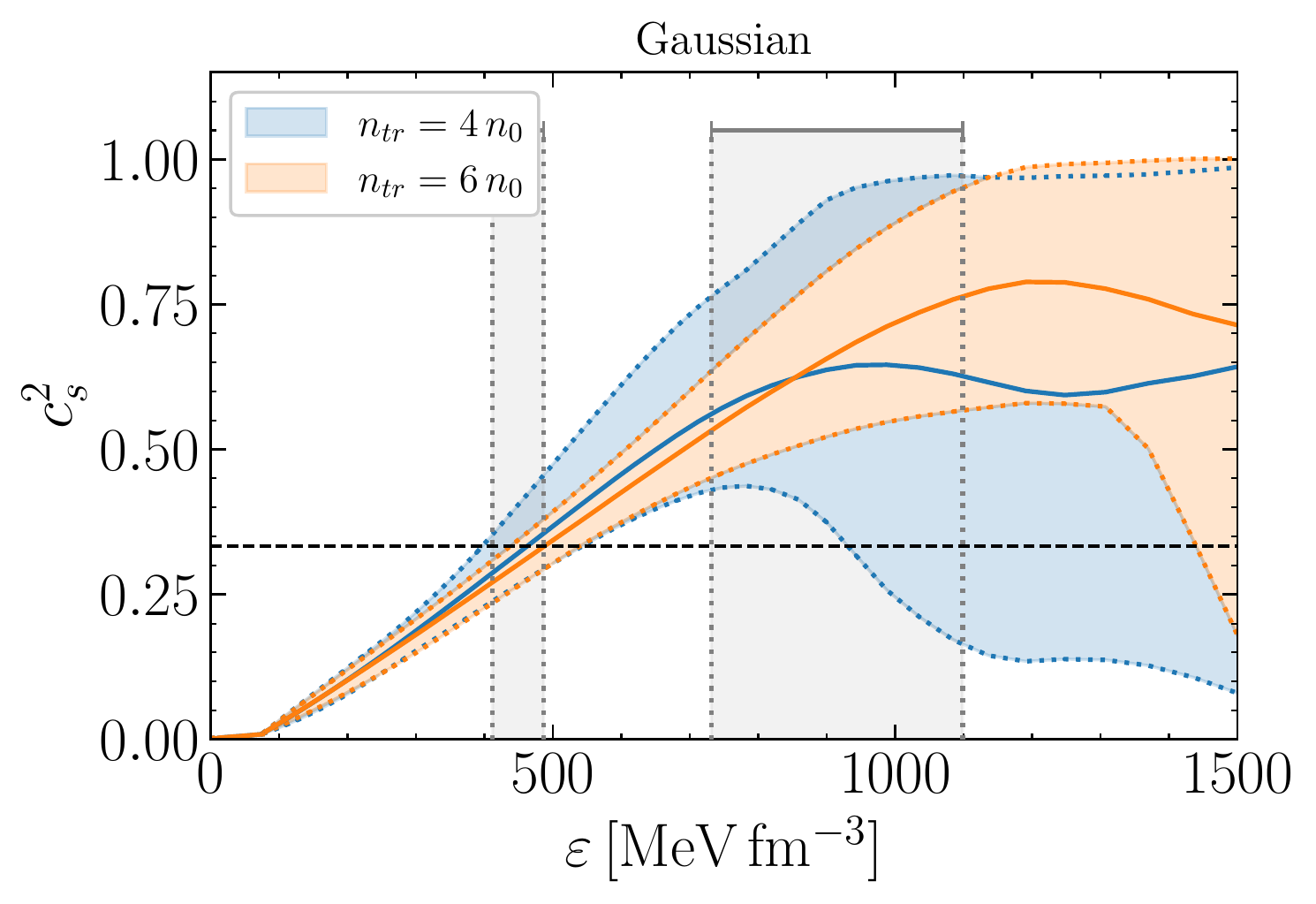} 
		\includegraphics[height=55mm,angle=-00]{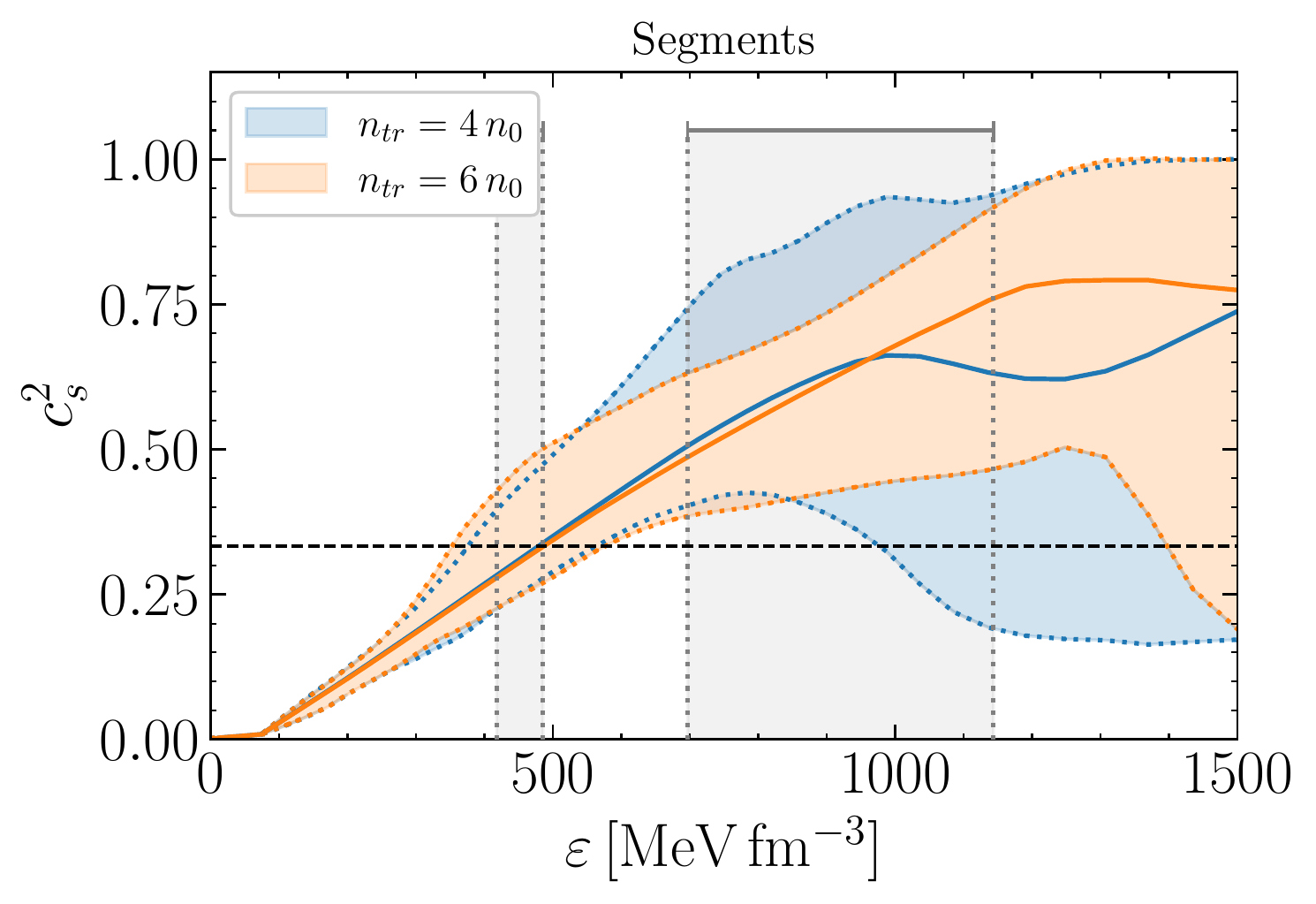} \\
		\includegraphics[height=55mm,angle=-00]{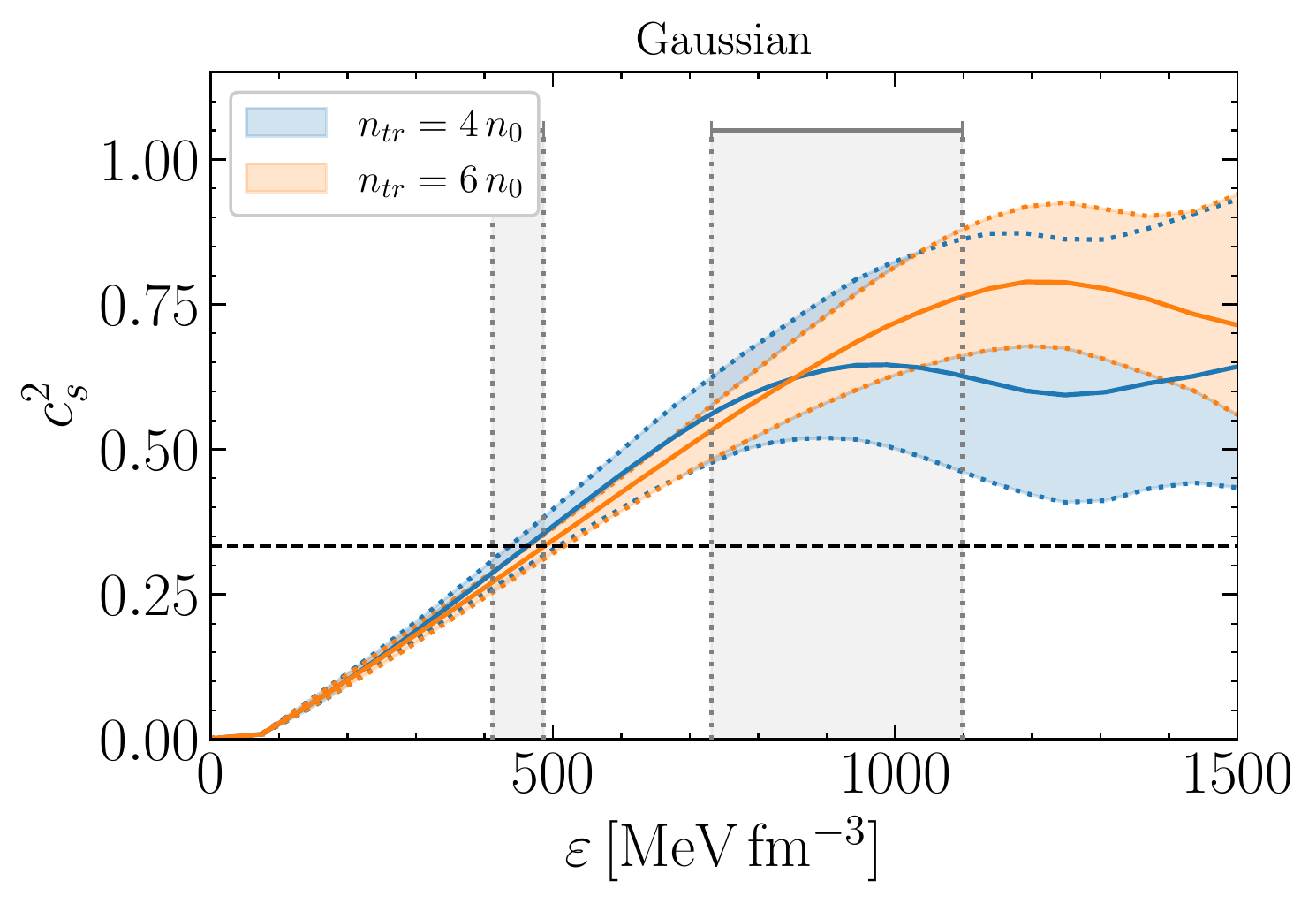} 
		\includegraphics[height=55mm,angle=-00]{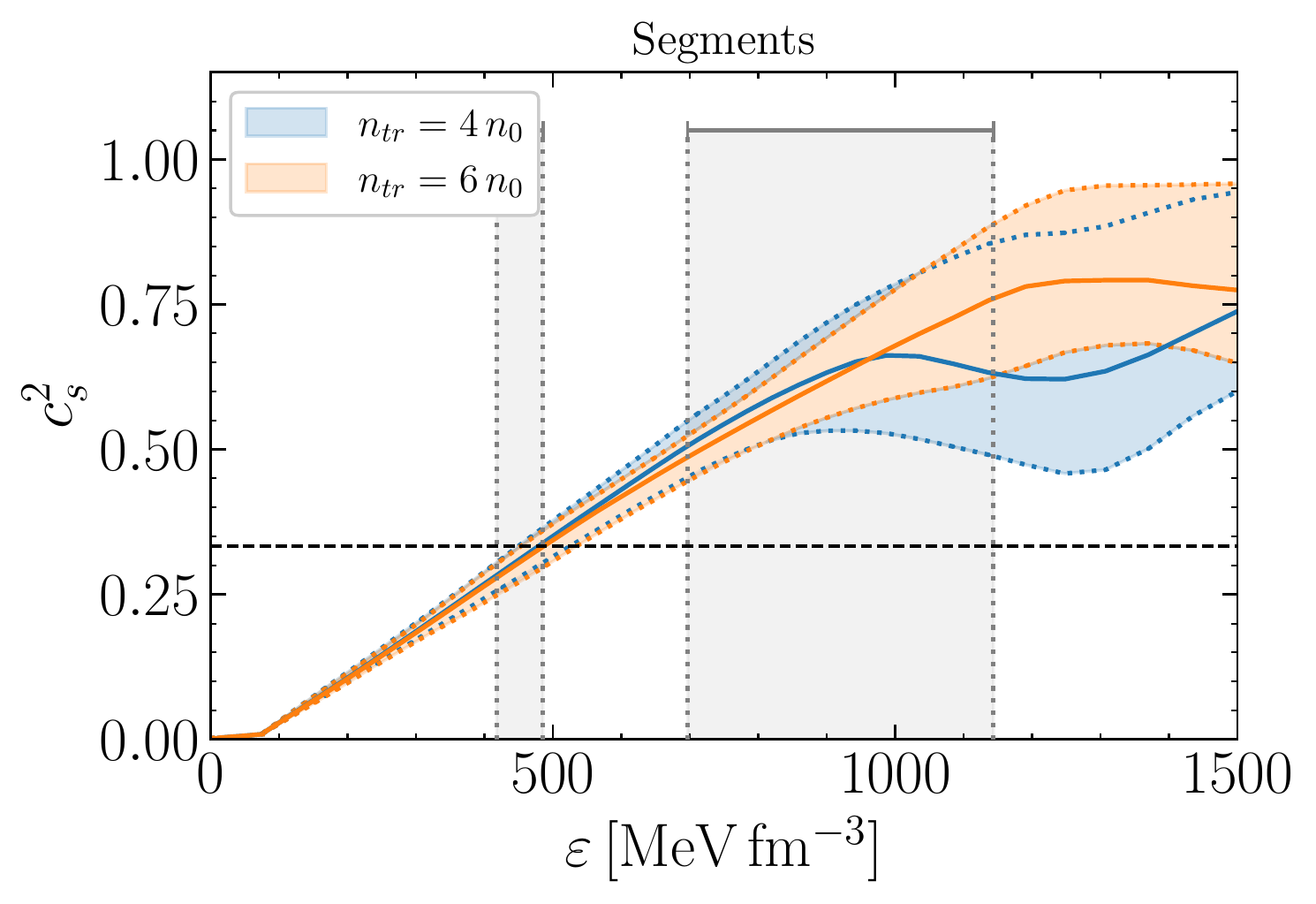} \\
		\caption{Posterior  95\% (top) and 68\% (bottom) credible bands and medians of the squared speed of sound $c_s^2$ as a function of energy density $\varepsilon$ for two different transition densities,  $n_{tr}/n_0 = 4$ and $ 6$,  up to which the speed of sound is preconditioned to rise monotonically.  In grey the 68\% intervals of the central energy densities of neutron stars with masses $M = 1.4\, M_\odot$ and $2.1\, M_\odot$ are displayed. The dashed black line indicates the value of the conformal limit.}
		\label{fig:PosteriorBands1ntr}
	\end{center}
\end{figure*}

\begin{figure*}[tp]
	\begin{center}
		\includegraphics[height=55mm,angle=-00]{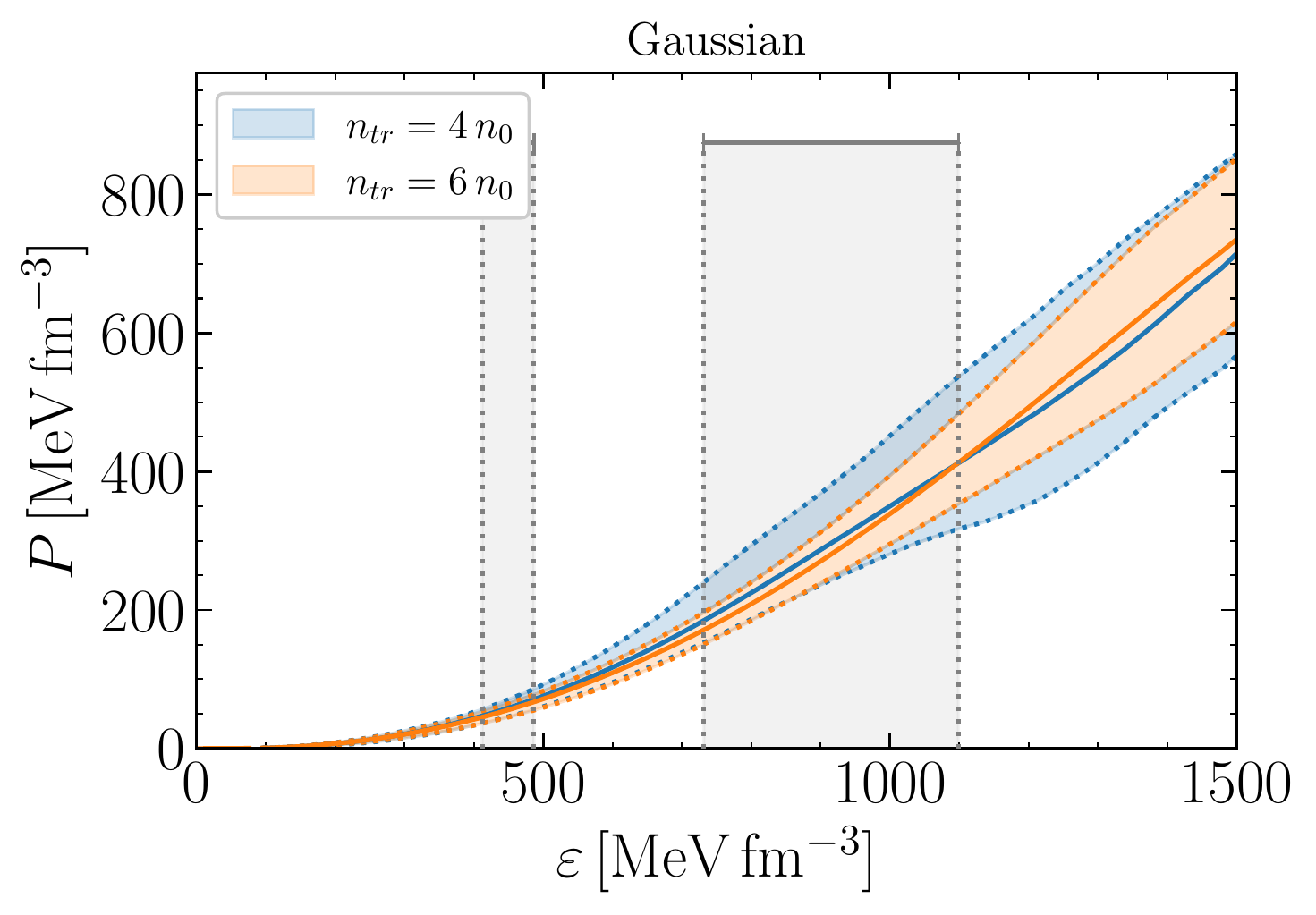} 
		\includegraphics[height=55mm,angle=-00]{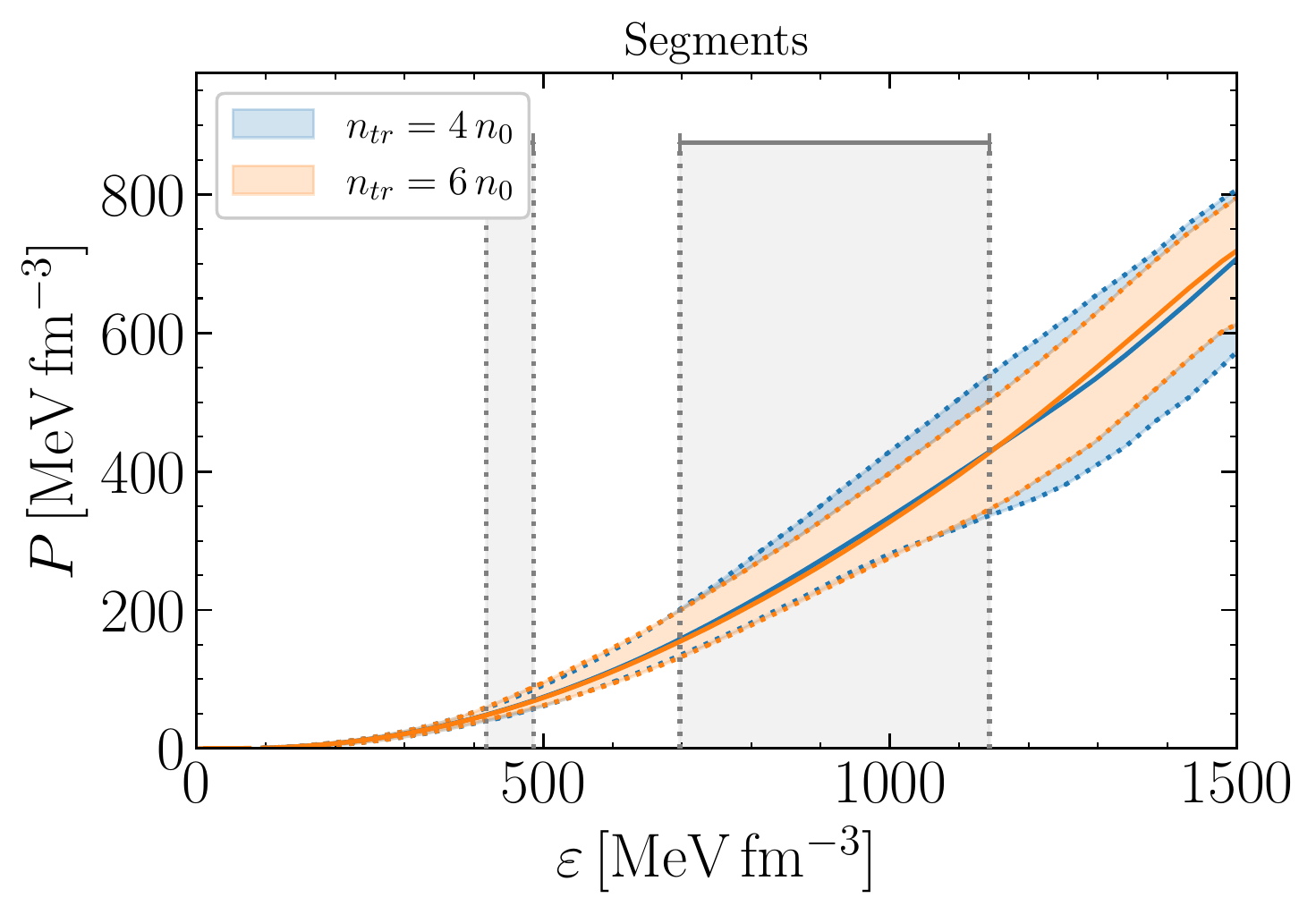} \\
		\includegraphics[height=55mm,angle=-00]{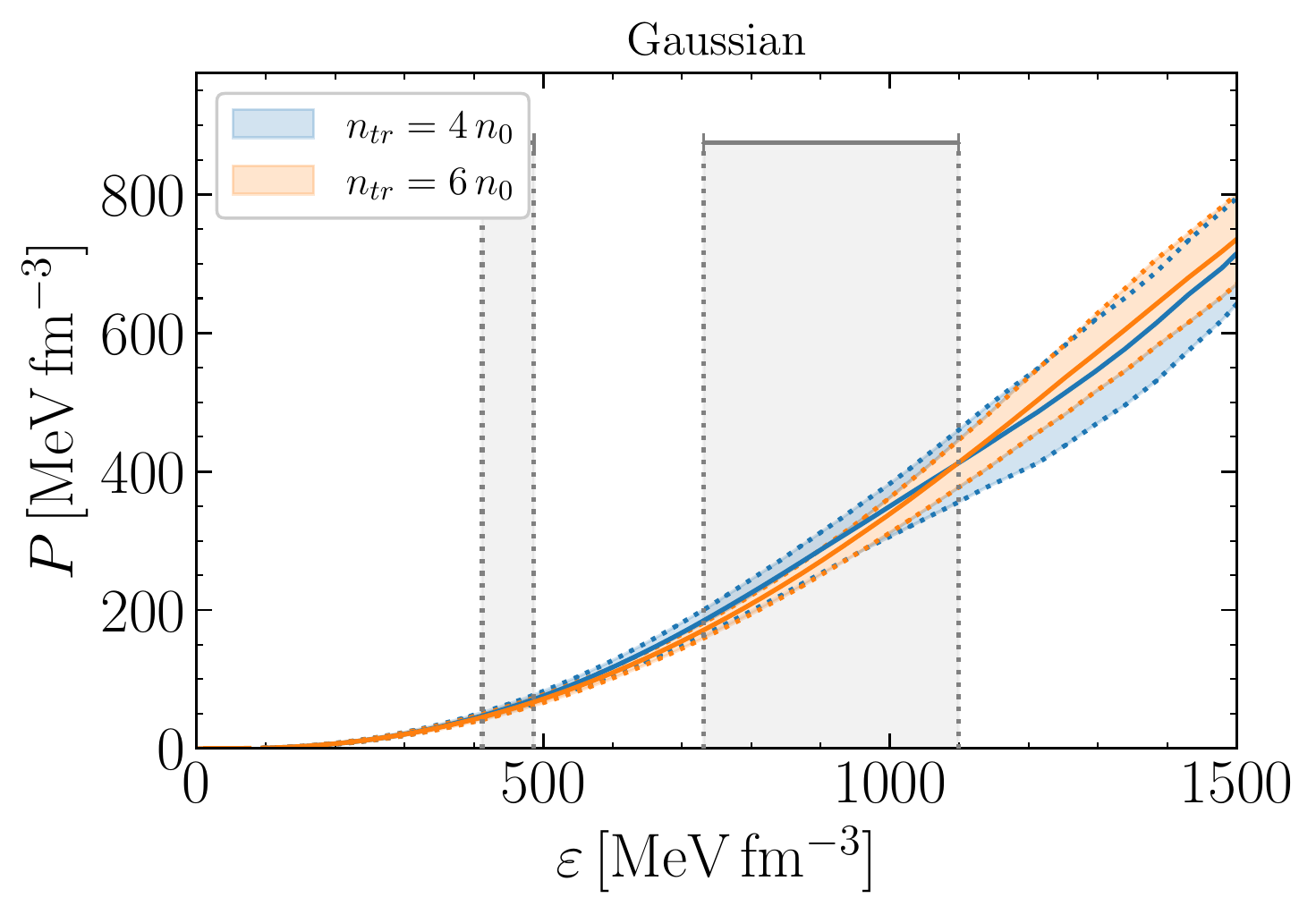} 
		\includegraphics[height=55mm,angle=-00]{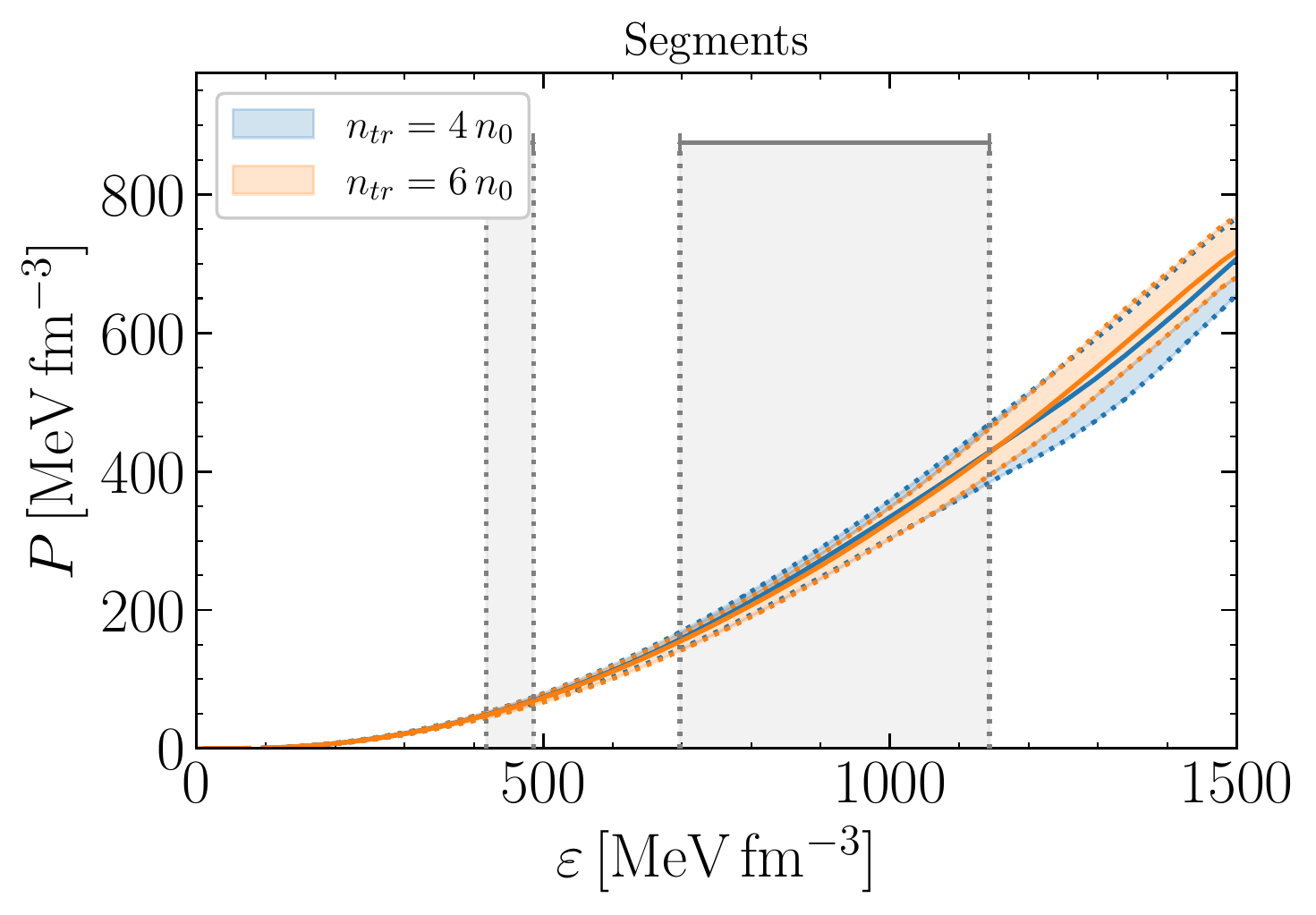} \\
		\caption{Posterior 95\% (top) and 68\% (bottom) credible bands and medians of the pressure $P$ as a function of energy density $\varepsilon$ for two different transition densities,  $n_{tr}/n_0 = 4$ and $6$,  up to which the squared speed of sound is preconditioned to rise monotonically.  In grey the 68\% intervals of the central energy densities of neutron stars with masses $M = 1.4\, M_\odot$ and $2.1\, M_\odot$ are displayed.}
		\label{fig:PosteriorBands2ntr}
	\end{center}
\end{figure*}

\begin{figure*}[tp]
	\begin{center}
		\includegraphics[height=55mm,angle=-00]{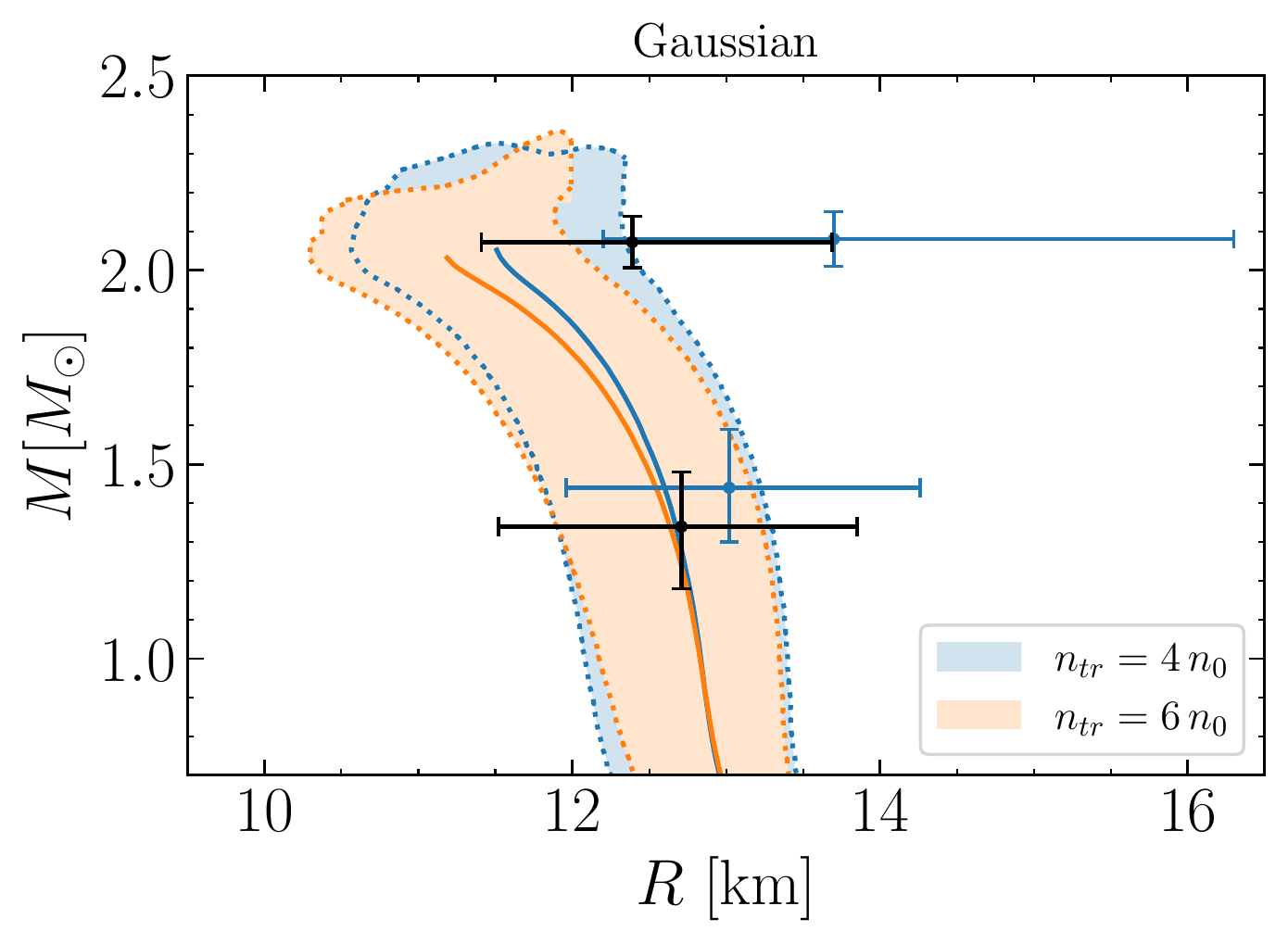} 
		\includegraphics[height=55mm,angle=-00]{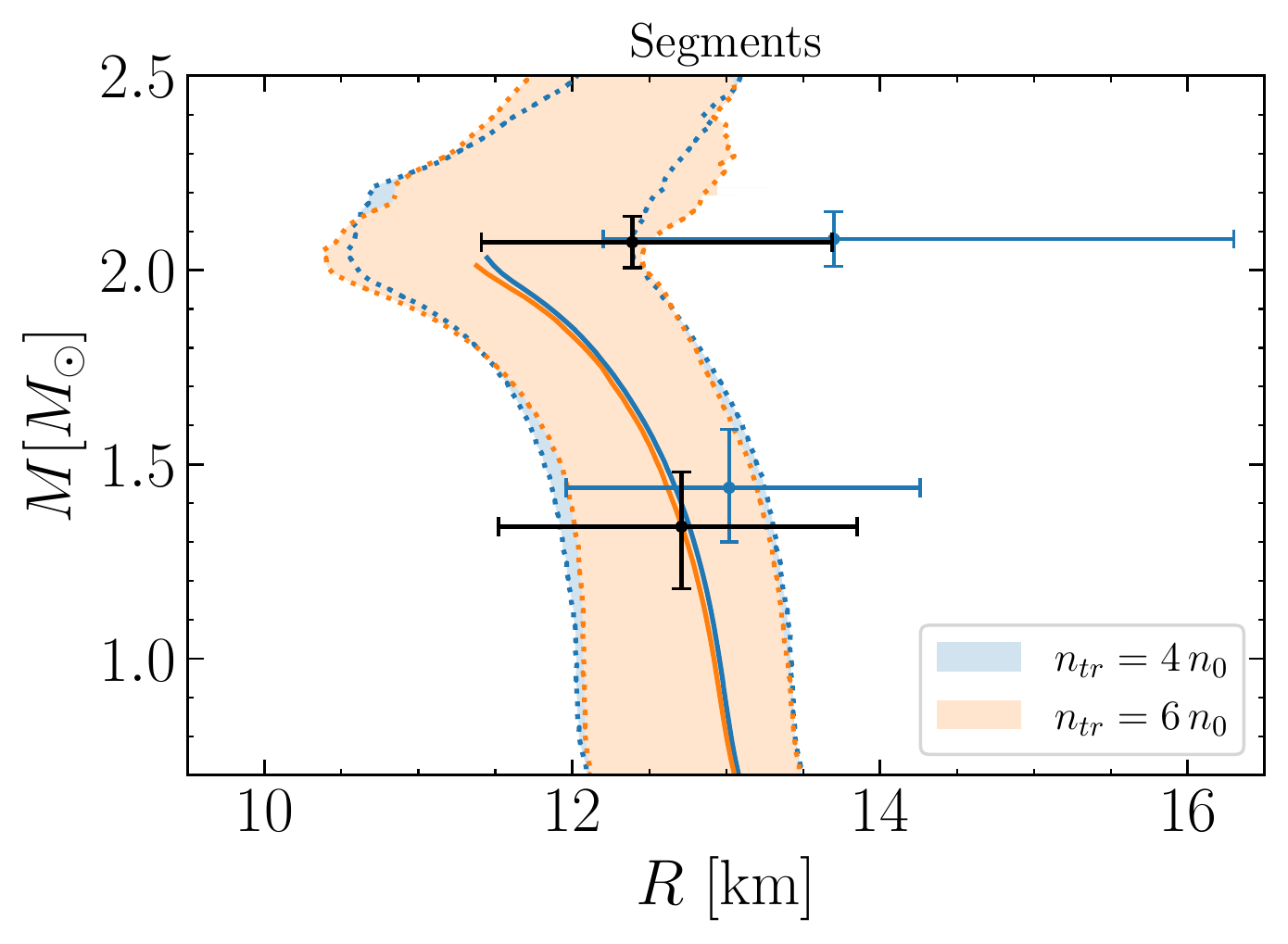} \\
		\includegraphics[height=55mm,angle=-00]{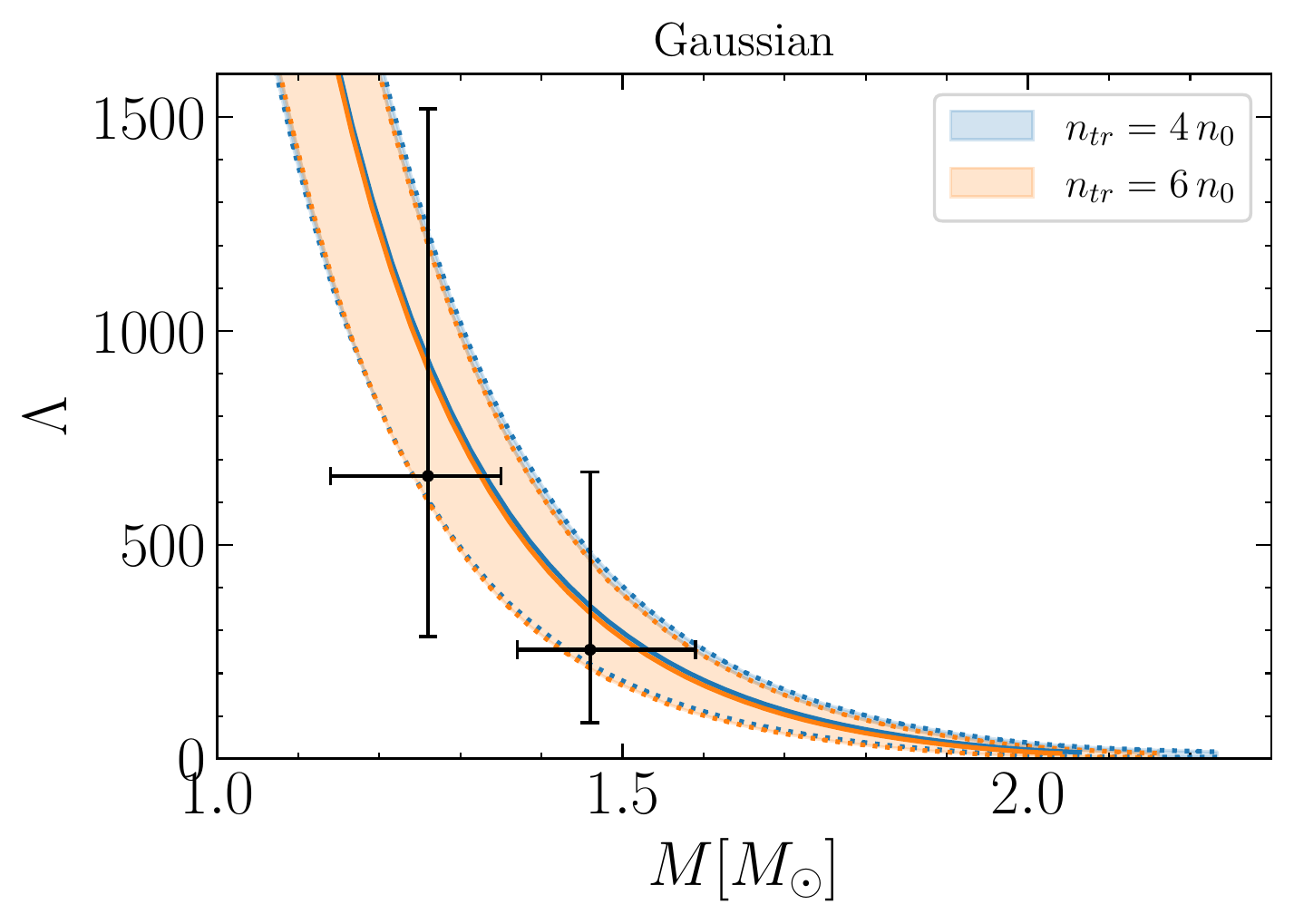} 
		\includegraphics[height=55mm,angle=-00]{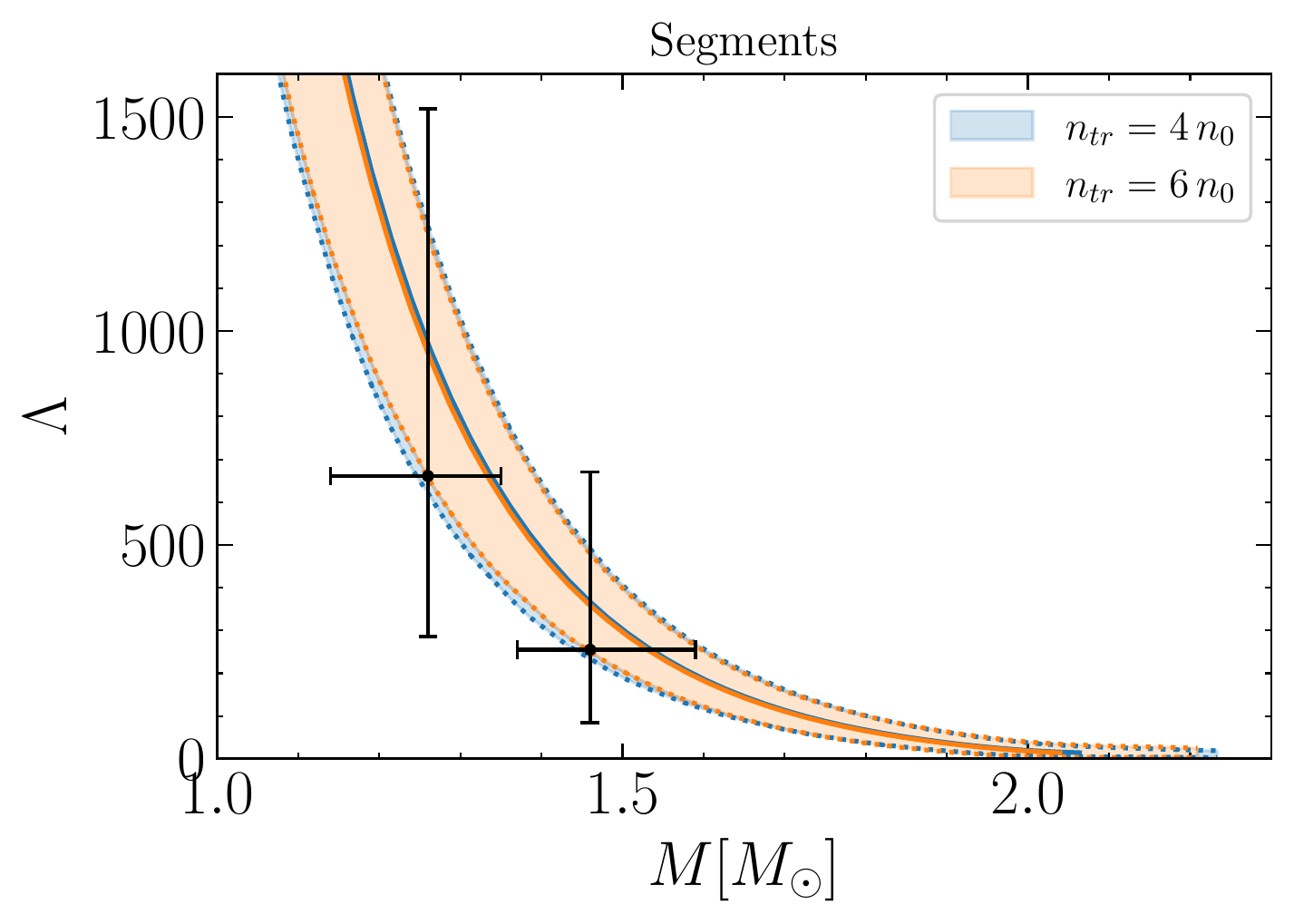} \\
		\caption{Posterior credible bands at the 95\% level and medians of the radius $R$ and the tidal deformability $\Lambda$ as a function of mass $M$ for two different transition densities, $n_{tr}/n_0 = 4$ and $6$,  up to which the speed of sound is preconditioned to rise monotonically.  $R(M)$ is compared to the marginalised intervals at the 68\% level from the analysis of the NICER measurements of PSR J0030+0451 and PSR J0740+6620 by Miller et al. (blue) \cite{Miller2019,Miller2021} and by Riley et al. \cite{Riley2019,Riley2021} (black). $\Lambda(M)$ is compared to the masses and tidal deformabilities inferred in Ref.\,\,\cite{Fasano2019} for the two neutron stars in the merger event GW170817 at the 90\% level.}
		\label{fig:PosteriorBands3ntr}
	\end{center}
\end{figure*}

In Fig.\,\,\ref{fig:PosteriorBands1ntr} the 68\% and 95\% Posterior credible bands of $c_s^2(\varepsilon)$ for $n_{tr}/n_0 = 4$ and $6$ are displayed.  The credible bands with $n_{tr} = 4\, n_0$ differ very little from those with $n_{tr} = 0$ shown in Fig.\,\,\ref{fig:PosteriorBands1},  most likely because the data indeed prefer equations of state that rise continuously with positive curvature at least up to $3\, n_0$,  as explained in the previous paragraph.  For $n_{tr} = 6\,n_0$ the differences to the $n_{tr} = 0$ case are more visible.  The speed of sound is now preconditioned to increase up to high energy densities $\varepsilon \sim 1.3\,$GeV$\,$fm$^{-3}$.  No complex phase structure appears in neutron stars with masses $M \leq 2.1 \, M_\odot$.  In the Segments parametrisation with its greater freedom,  the 95\% band is much wider for $n_{tr} = 6\, n_0$ and comparable to the $n_{tr} = 4 \, n_0$ case up to $\varepsilon \sim 550\,$MeV$\,$fm$^{-3}$. This behaviour of the sound speed is also reflected in the Posterior credible bands for $P(\varepsilon)$ shown in Fig.\,\,\ref{fig:PosteriorBands2ntr} and in the mass-radius relation and tidal deformability plotted in Fig.\,\,\ref{fig:PosteriorBands3ntr}.

\begin{table}[tp]
	\centering
	\begin{tabularx}{\linewidth}{|l|XXX|XXX|}
		\hline \hline  
		& \multicolumn{6}{l|}{$\mathcal{B}^{n_{-}\leq n_{tr}}_{n_{-} > n_{tr}}$} \\
		& \multicolumn{3}{l|}{Gaussian} & \multicolumn{3}{l|}{Segments} \\
		$M_{new} \, [M_\odot]$ & $2.2$ & $2.3$ & $2.4$ & $2.2$ & $2.3$ & $2.4$ \\ 
		$n_{tr} /n_0$ &&&&&& \\ \hline
		3 & 0.04 & 0.07 & 0.15 &   0.66 & 0.98 & 1.51\\
		4 & 0.32 & 0.54 & 1.02 &   1.88 & 3.15 & 5.40\\
		5 & 0.73 & 1.27 & 3.07 &   3.69 & 6.57 & 10.75 \\
		6 & 3.32 & 6.83 & 22.68 &   6.97 & 10.05 & 12.64 \\
		\hline \hline 
	\end{tabularx}
	%\end{ruledtabular}
	\caption{Similar to Table \,\ref{tab:BayesFactorntr},  here the Bayes factors $\mathcal{B}^{n_{-}\leq n_{tr}}_{n_{-} > n_{tr}}$ are displayed for various transition densities $n_{tr}$. The Bayes factors are computed assuming the existence of an additional hypothetical heavy neutron star with mass $M_{new} = 2.2\, M_\odot$, $2.3\, M_\odot$ or $2.4\, M_\odot$ and an uncertainty $\sigma_{M_{new}} = \pm 0.1 \, M_\odot$. The observation of a $M_{new} = 2.4(1)\, M_\odot$ neutron star would lead to strong evidence that the speed of sound does not rise monotonically up to a density of $n_{tr} = 6\, n_0$ in the Gaussian and in the Segments parametrisation.}
	\label{tab:BayesFactorHeavyNS}
\end{table} 
 
Next, we investigate the impact that the observation of a possible supermassive neutron star would have on the previous analysis.  Assume the  existence of a speculative heavy neutron star with mass $M_{new} = 2.2 \, M_\odot$, $2.3 \, M_\odot$ or $2.4 \, M_\odot$.   Its mass is included as an additional hypothetical (Gaussian) measurement such that the Likelihood can be computed according to Eq.\,\,(\ref{eq:LikelihoodShapiro}).  We choose $\sigma_{M_{new}} = \pm 0.1\, M_\odot$ assuming the uncertainty to be comparable to the measurements in Eqs.\,\,(\ref{eq:ShapiroMass1} - \ref{eq:ShapiroMass3}). 
We repeat the analysis of the Bayes factors in Tab. \ref{tab:BayesFactorntr} with the additional hypothetical data in Tab. \ref{tab:BayesFactorHeavyNS}.  A future measurement of a heavy neutron star with mass $M = 2.2\, M_\odot$ would lead to moderate evidence for the possibility that $c_s$ may drop at a density $n < n_{tr} = 6 \, n_0$.  For even higher masses, $M = 2.4\, M_\odot$, this evidence becomes still more definite. This means the speed of sound would need to drop before this transition density, hinting towards a more complex phase structure possibly with a transition to different degrees of freedom.  Therefore the observation of such a very heavy neutron star would open up the discussion of a more involved phase structure in the EoS.  In particular, in order to support such a heavy object, the sound velocity would need to rise steeply at densities beyond the ChEFT range. As seen in Fig.\,\,\ref{fig:PosteriorBands1ntr} such a behaviour has less support for $n_{tr} = 6\, n_0$ compared to lower transition densities.

The presently available empirical information including up to $\sim 2\, M_\odot$ objects may be updated by the recent observation of the fastest spinning and heaviest known galactic neutron star \cite{Romani2022}.  In this context the ongoing speculations about a very heavy neutron star based on the recent observation of gravitational wave signals from a black hole merger with a compact object of mass $2.6\, M_\odot$ \cite{Abbott2020a} are worth mentioning. Note, however, if the binary merger product produced in GW170817 was indeed a hyper-massive neutron star supported against gravitational collapse by differential rotation, as suggested by the measured electromagnetic counterparts \cite{Margalit2017,Rezzolla2018,Ruiz2018}, then a reduced upper limit is likely for the maximum possible mass of non-rotating neutron stars. Similarly, the condition that the EoS needs to be causally connected to pQCD results at very high densities makes extremely massive neutron stars more unlikely \cite{Gorda2022}. 

\subsection{Conformal limit reached from above}

\begin{figure*}[tp]
	\begin{center}
		\includegraphics[height=55mm,angle=-00]{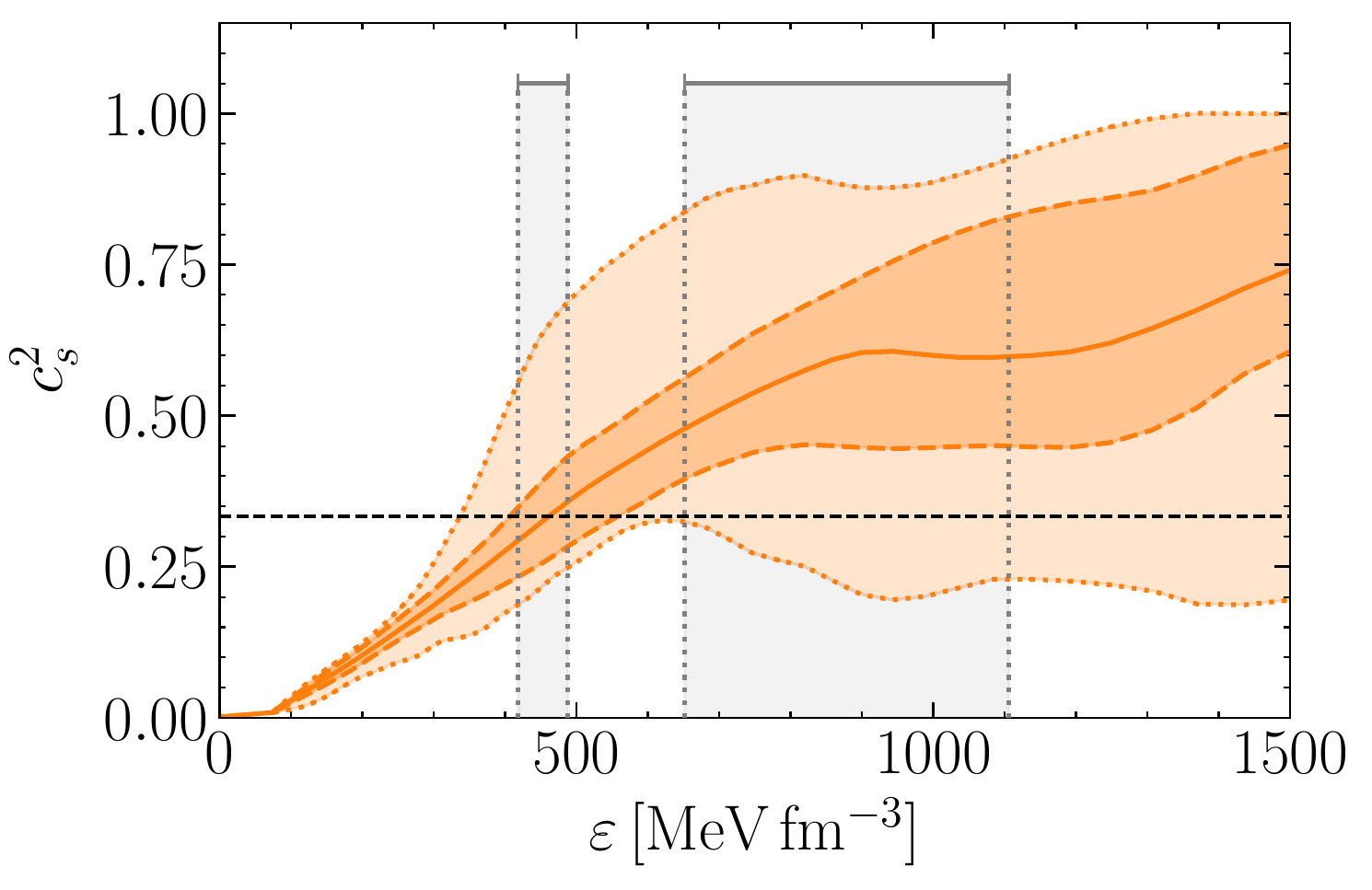} 
		\includegraphics[height=55mm,angle=-00]{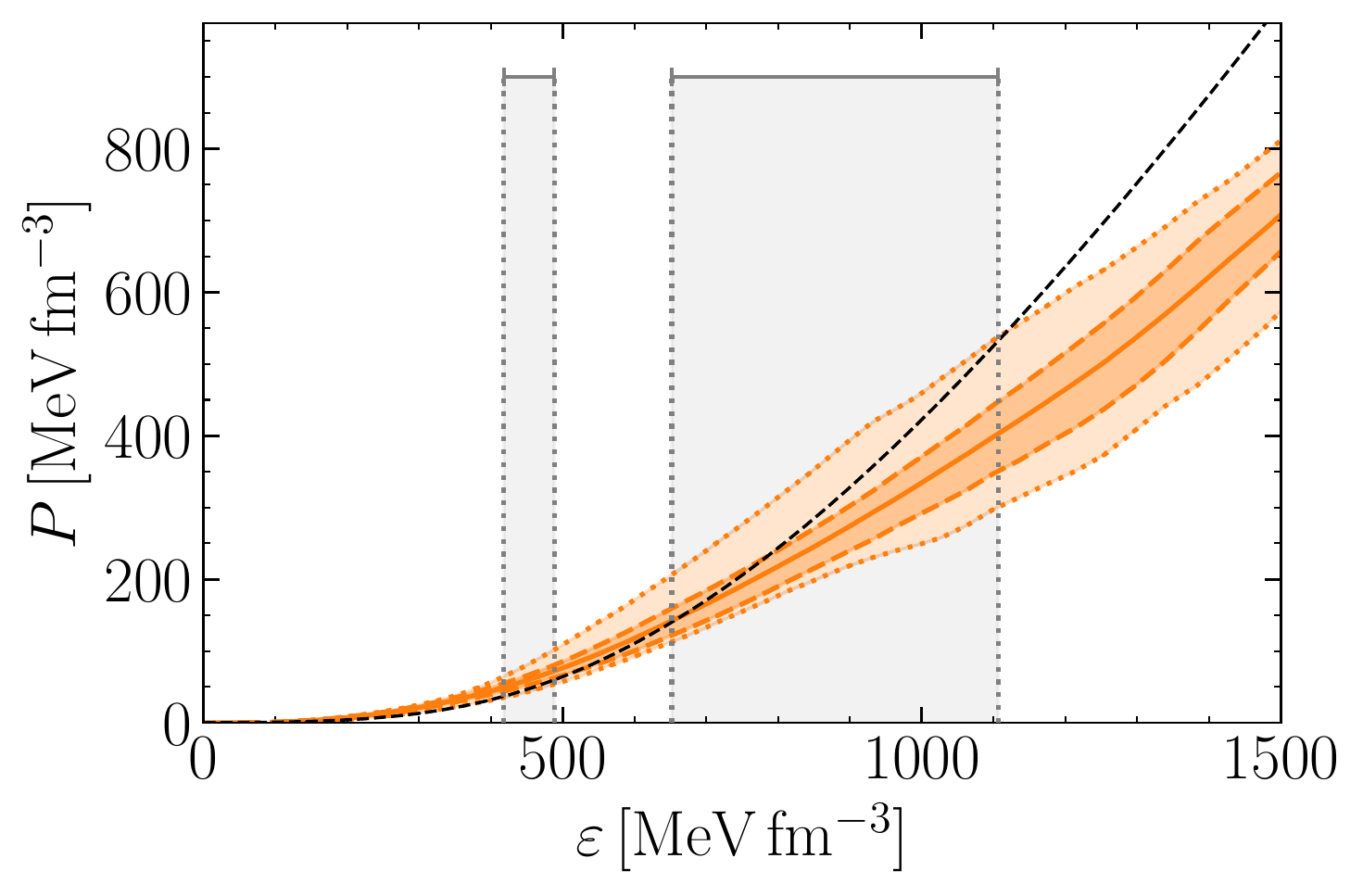} \\
		\includegraphics[height=55mm,angle=-00]{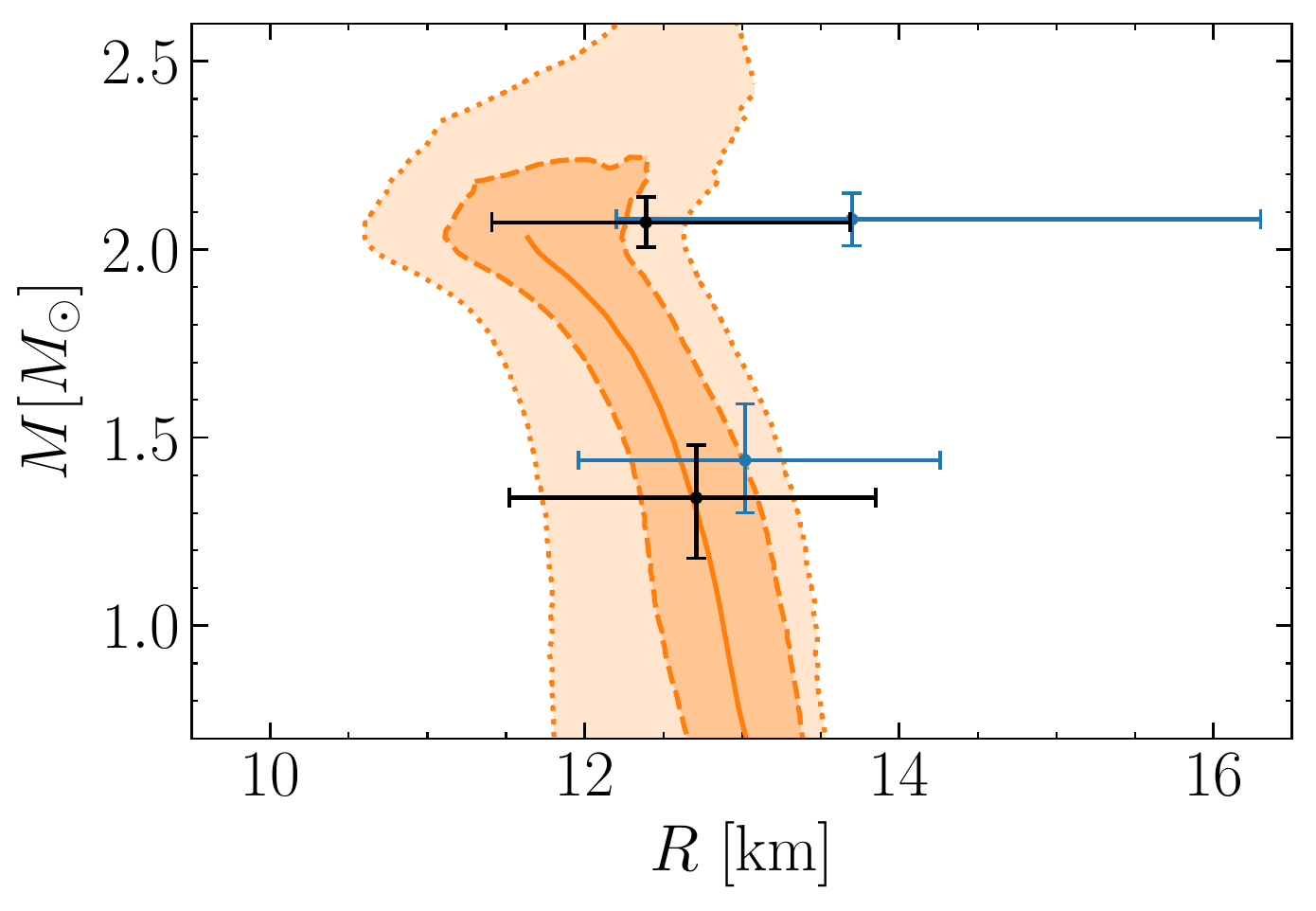}
		\includegraphics[height=55mm,angle=-00]{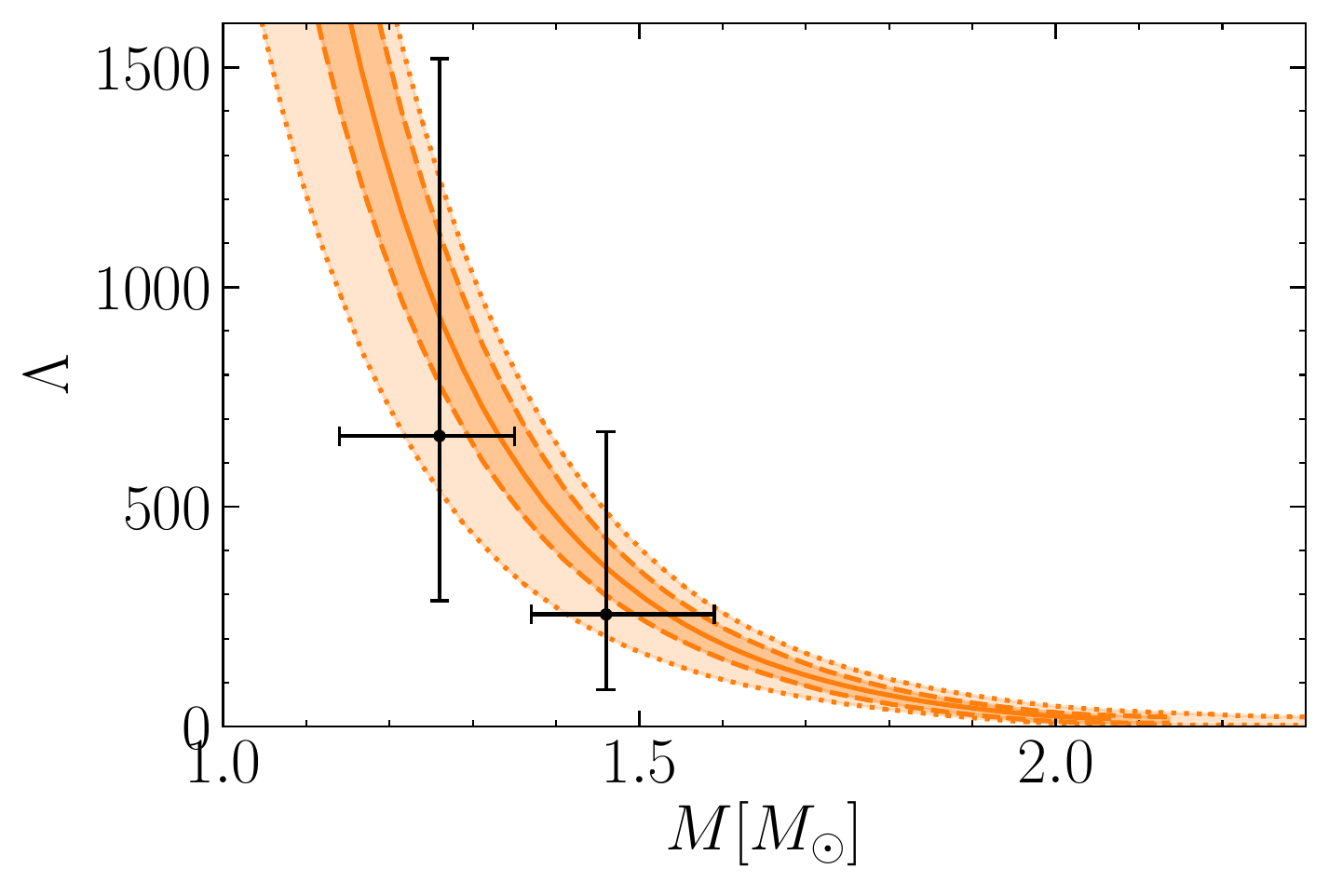} \\
		\caption{Posterior 95\% and 68\% credible bands and medians for the Segments parametrisation with the conformal limit reached asymptotically from above. 
			Shown are the squared speed of sound, $c_s^2$,  and pressure $P$ as a function of energy density $\varepsilon$,  as well as the mass-radius relation $M(R)$ and tidal deformability $\Lambda$ as a function of neutron star mass $M$.}
		\label{fig:PosteriorBandsAbove}
	\end{center}
\end{figure*}

\begin{figure*}[tp]
	\begin{center}
		\includegraphics[height=55mm,angle=-00]{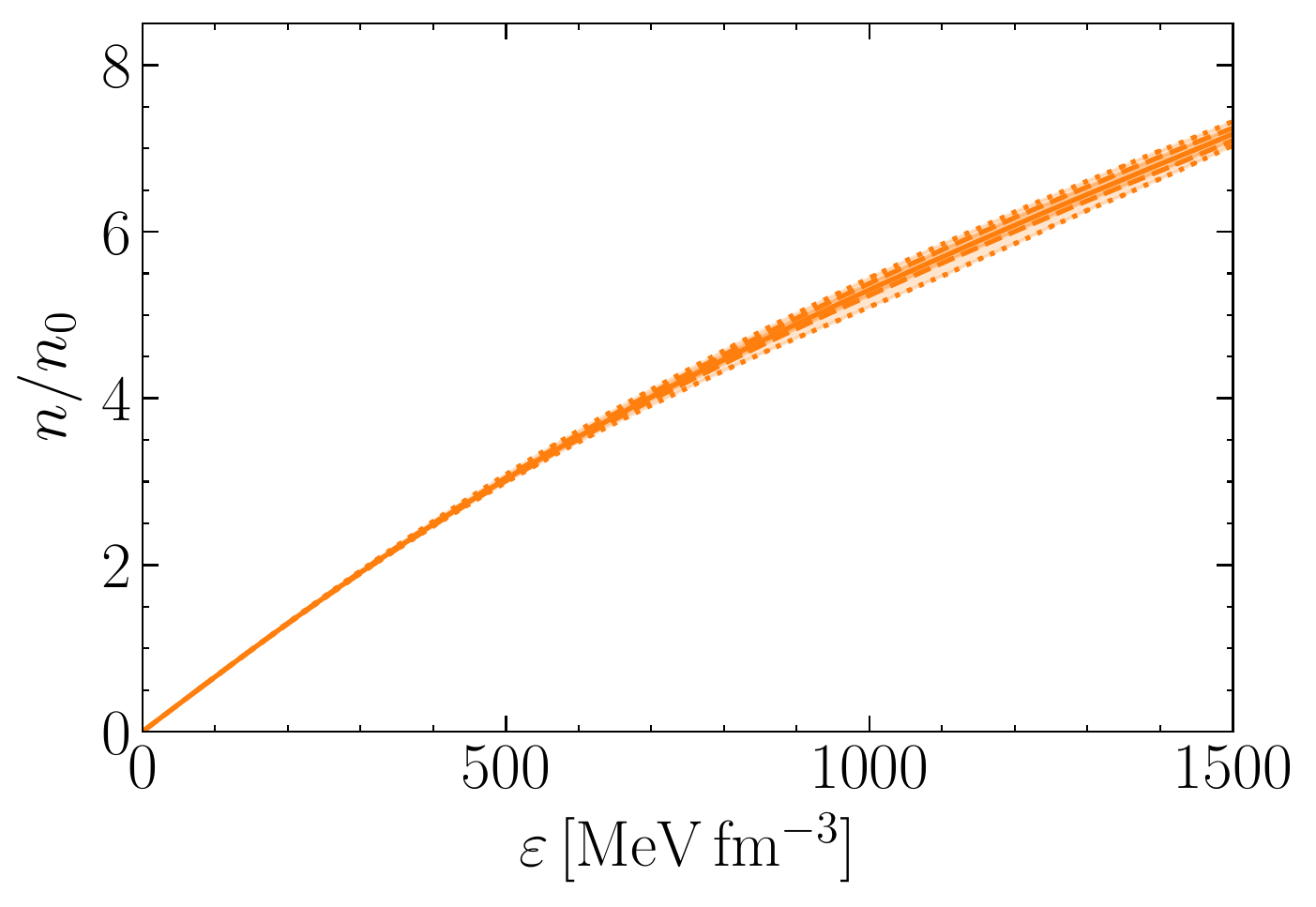} 
		\includegraphics[height=55mm,angle=-00]{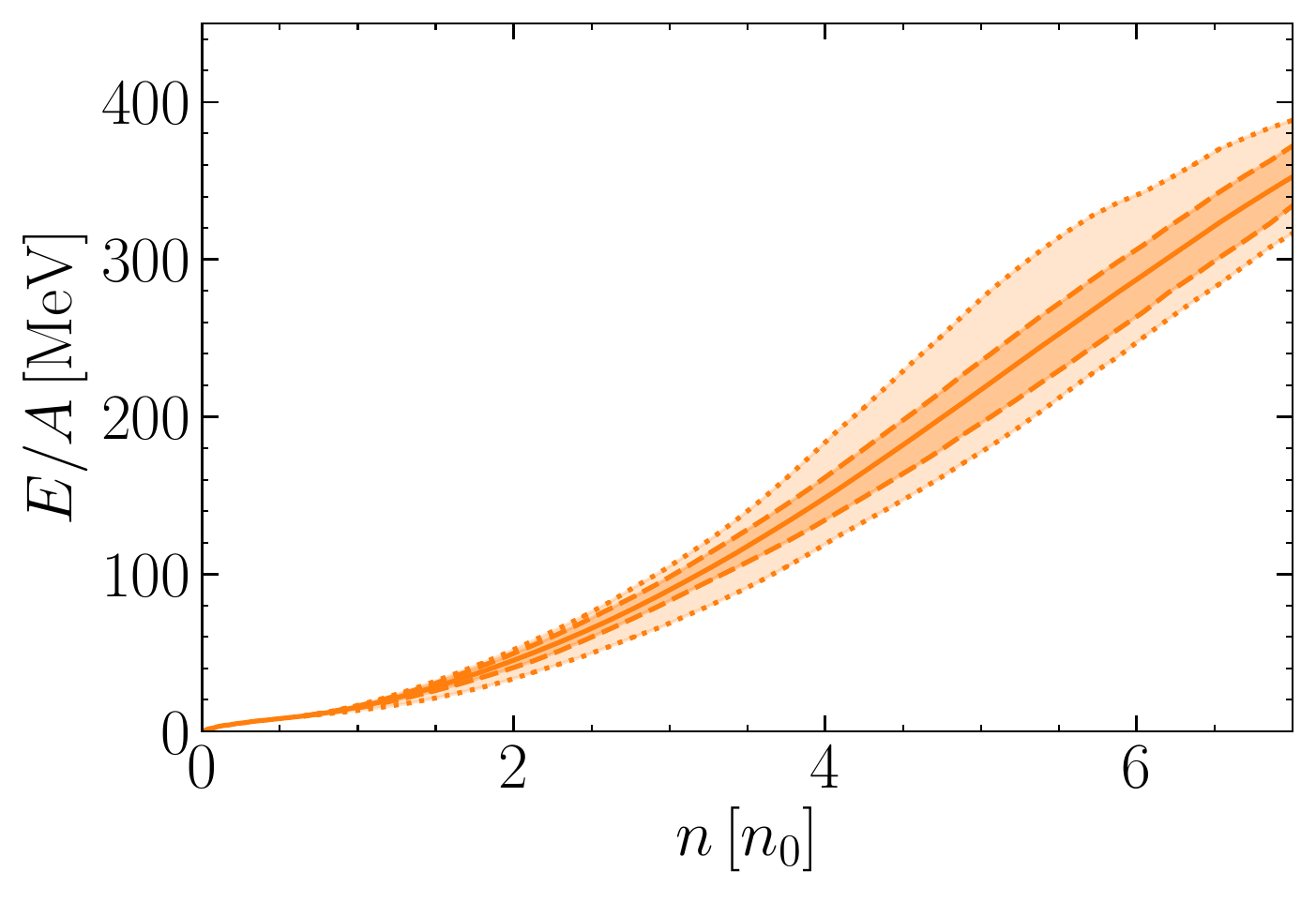} \\
		\caption{Posterior  95\% and 68\% credible bands and medians for the Segments parametrisation with the conformal limit reached asymptotically from above.  Shown are the baryon density $n$ as a function of energy density $\varepsilon$ and the energy per particle $E/A$ as a function of density $n$.}
		\label{fig:DensityAndEnergyPerParticleAbove}
	\end{center}
\end{figure*}

A recent analysis based on Hard Dense Loop resummation techniques found that,  in contrast to standard perturbative QCD results,  the speed of sound reaches the conformal limit from above at asymptotically high densities \cite{Fujimoto2022}. To analyse the impact of this assumed alternative asymptotic behaviour,  we repeat our analysis with the same steps as before,  but now modified such that the squared speed of sound reaches the conformal limit, $c_s^2 \rightarrow 1/3$,  from above.  We restrict ourselves to the Segments parametrisation. With the changed asymptotic behaviour the descriptive power of the Gaussian parametrisation is severely hindered. The resulting Posterior credible bands are displayed in Fig.\,\,\ref{fig:PosteriorBandsAbove}. Compared to the case with $c_s^2 \rightarrow 1/3$ reached from below in Figs.\,\,\ref{fig:PosteriorBands1} and \ref{fig:PosteriorBands2}, the sound velocities up to energy densities $\varepsilon \lesssim 650\,$MeV$\,$fm$^{-3}$ remain unchanged,  implying that both asymptotic behaviours lead to $M = 1.4\, M_\odot$ neutron stars with almost exactly the same properties as in Tab. \ref{tab:NS_properties1}.  Even at higher energy densities the 68\% credible bands look very similar.  However,  when the conformal limit is reached from above,  the lower limit of the 95\% credible band lies at higher sound speeds.  In contrast the standard pQCD asymptotic constraint leads to a softening of the EoS at high densities as was already concluded in the Bayesian analyses of Refs.\,\,\cite{Gorda2022, Somasundaram2022}. The similarity in the speed of sound translates into the credible bands for $P(\varepsilon)$, $M(R)$ and $\Lambda(M)$. Additionally, the 68\% credible $M(R)$ band extends to slightly larger masses. There are also only minor changes in $\varepsilon(n)$ and $E(n)/A$ as displayed in Fig.\,\,\ref{fig:DensityAndEnergyPerParticleAbove}. For a $2.1 \, M_\odot$ neutron star,  when the conformal limit is reached from above,  we find a radius $R = 11.6^{+1.0}_{-0.9}\,$km,  the central pressure  $P_c = 311^{+248}_{-189}\, $MeV$\,$fm$^{-3}$, tidal deformability $\Lambda = 15^{+17}_{-9}$ and central density $n_c = 4.9^{+1.6}_{-1.7} \, n_0$,  again very similar to the previous results in Tab. \ref{tab:NS_properties1}. This means that the description of neutron stars at all mass ranges is to large extent independent of the high density asymptotic behaviour as long as the speed of sound is causally connected to the conformal limit. However, with less support for small speeds of sound, the modified asymptotic behaviour makes strong first-order phase transitions inside neutron stars even more unlikely. 

\section{Summary and conclusions}
\label{sec:Summary}

The present work has focused on several principal questions relevant to the inference of neutron star properties  (masses, radii, tidal deformabilities) from the presently available empirical astrophysical data and their detailed analysis:

a) To what extent can the inference results be considered independent of the choices of Priors ?

b) Is it feasible to draw statistically significant conclusions about the equation of state of dense matter concerning the possible occurrence of a phase transition or crossover inside neutron stars ? 

A key quantity to address these issues is the speed of sound, $c_s = \sqrt{\partial P/ \partial\varepsilon}$,  in neutron star matter.  We have modelled $c_s$  using two generic parametrisations,  a skewed Gaussian in combination with a logistic function,  and a more general form based on piecewise linear segments.  
Using Bayesian inference methods,  multimessenger data sets from Shapiro time delay observations of selected pulsars,  NICER X-ray measurements and gravitational wave signals from binary neutron star mergers have been translated into constraints on the sound velocity inside neutron stars.  The asymptotic behaviour of the squared sound speed,  reaching the conformal limit,  $c_s^2\rightarrow 1/3$,  either from below following perturbative QCD considerations,  or from above when applying Hard Dense Loop (HDL) resummation techniques in QCD,  is implemented and discussed.  At low baryon densities $n$ around $n_0 = 0.16\,$fm$^{-3}$,  the equilibrium density of normal nuclear matter,  state-of-the-art Chiral EFT constraints are incorporated.  But unlike several approaches in the recent literature,  these nuclear physics constraints are implemented in terms of Likelihoods similar to those for the empirical data in order to warrant a statistically consistent Bayesian framework. 

The results and conclusions are summarised as follows:

i) Good agreement is found between the output Posteriors for both parametrisations up to energy densities $\varepsilon \sim 1.2\,$GeV$\,$fm$^{-3}$ which cover the central energy densities of $M \sim 2\, M_\odot$, neutron stars.  For larger energy densities differences between the parametrisations become more prominent because this high-mass region is unrestricted by observational data.  Bayes factors comparing the two hypotheses indicate that no parametrisation is statistically preferred over the other.  The implementation of a conservative upper limit at $n = 2\, n_0$ based on ChEFT results prohibits steeply rising sound speeds seen in some previous analyses,  highlighting the importance of nuclear physics constraints at low and intermediate densities.  The overall conclusion is that the Bayesian inference approach generates results that are indeed stable with respect to variations in the functional form of the Prior if the initial parametrisation is chosen sufficiently general.

ii) A quantitative Bayes factor analysis gives extreme evidence that the conformal bound $c_s^2 \leq 1/3$ is violated inside neutron stars.  If combined with the behaviour at ultra-high densities derived from standard perturbative QCD,  with the conformal limit reached asymptotically from below,  this suggests that the squared speed of sound displays a non-monotonic behaviour including at least two extrema,  a maximum $c_{s,max}^2$ followed by a minimum $c_{s,min}^2$. However,  this minimum would occur at very high baryon densities, $n(c_{s,min}^2) = 6.7^{+0.7}_{-0.5}\, n_0$ (at 68\% level) and correspondingly at neutron star masses $M \gtrsim 2.1\,M_\odot$,  i.e.  at the borderline or beyond the objects presently observed.  Accordingly,  there is extreme evidence that $c_{s,min}^2$ stays larger than 0.1 for neutron stars with mass $M \leq 1.9 M_\odot$ and still strong evidence for $M \leq 2.0 M_\odot$.  This indicates that a first-order phase transition in the core of even the heaviest observed neutron stars is unlikely,  while a continuous crossover (as realised for example in the quark-hadron continuity picture) is not ruled out. In our analysis, we assume that such a phase change does not produce twin-star solutions which are implausible given the available data. 

 If the asymptotic behaviour of the sound velocity is changed such that the conformal limit is reached from above as suggested in the HDL scenario,  the gross features of the inferred speed of sound inside neutron stars do not alter. The only difference is that at high energy densities,  the 95\% level of $c_s^2$ remains at larger sound speeds, shifting possible  phase changes to even higher densities.  Nevertheless,  the behaviour of $c_s^2$ at asymptotic densities has relatively minor influence on the properties of typical neutron stars with masses between $M = 1.4 \, M_\odot$ and $2.1 \, M_\odot$.  

iii) The characteristic baryon densities reached in the centre of a two-solar mass neutron star,  $n_c \sim (5-6)\,n_0$,  are not extreme, suggesting that a description based entirely on baryonic degrees of freedom can still be viable.  Such a picture is realised for example in Functional Renormalization Group studies based on chiral baryon-meson field theory or in a related treatment of neutron star matter as a relativistic Fermi liquid.  Its basic feature is a monotonically increasing sound velocity as a function of baryon density.  We have investigated whether such an option can be accommodated with or excluded by the existing empirical data.  With the additional assumption that the speed of sound is preconditioned to rise monotonically up to a certain transition density $n_{tr}$,  a corresponding Bayes factor analysis points out that there is strong evidence in the Gaussian and moderate evidence in the Segments parametrisation that $n_{tr} \gtrsim 3\, n_0$.  Even a monotonically rising speed of sound up to very high densities, $n_{tr} = 6\, n_0$ cannot be excluded by the current data and further supports the moderate evidence against a phase transition at any density in the core of neutron stars.  However,  an extension of the Bayes factor evaluation including a hypothetical object with a mass beyond $2.1\,M_\odot$ suggests that the observation of a superheavy neutron star with mass $M \sim 2.3 - 2.4 \, M_\odot$ would provide indication for a more complex phase structure in the deep interior of the star.  

The expected expansion of the observational data base in the future will lead to even tighter constraints on the speed of sound in neutron stars, with a chance of further clarifying the phase structure of QCD at high baryon densities and low temperatures.  In this context our studies point out that the observation of a neutron star with even larger mass than the presently heaviest existing one would be most informative.

\begin{acknowledgments}
	This work has been supported in part by Deutsche Forschungsgemeinschaft (DFG) and National Natural Science Foundation of China (NSFC) through funds provided by the Sino-German CRC110 “Symmetries and the Emergence of Structure in QCD” (DFG Grant No. TRR110 and NSFC Grant No. 11621131001), and by the DFG Excellence Cluster ORIGINS.
\end{acknowledgments}

\appendix

\section{Bayes factors}
\label{sec:Bayesfactors}

With Bayes factors one can compare two competing hypotheses $H_0$ and $H_1$ and quantify the evidence for one hypothesis over the other. Given a data set $\mathcal{D}$, the Bayes factor is defined as the quotient of the marginal likelihoods: 
\begin{eqnarray}
	\mathcal{B}_{H_0}^{H_1} = \frac{\text{Pr}(\mathcal{D}|H_1, \mathcal{M}) }{\text{Pr}(\mathcal{D}|H_0, \mathcal{M})} ~.
\end{eqnarray} 
Using Bayes' theorem in Eq.\,\,(\ref{eq:BayesTheorem}), we can rewrite the marginal Likelihoods in terms of Posterior and Prior probabilities
\begin{eqnarray}
	\mathcal{B}_{H_0}^{H_1} = \frac{\text{Pr}(H_1|\mathcal{D}, \mathcal{M}) } {\text{Pr}(H_1|\mathcal{M})}\frac{\text{Pr}(H_0| \mathcal{M})}{\text{Pr}(H_0|\mathcal{D}, \mathcal{M})} ~.
\end{eqnarray}
The probability distribution for a general hypothesis $H$ is given by the integral over all parameter sets that support this hypothesis, $\theta \in \Theta_H$:
\begin{align}
	\frac{\text{Pr}(H|\mathcal{D, \mathcal{M}}) } {\text{Pr}(H|\mathcal{M})} =& \frac{\int_{\theta\in\Theta_H} d\theta \,~ \text{Pr}(\theta|\mathcal{D}, \mathcal{M})}{\int_{\theta\in\Theta_H} d\theta \,~ \text{Pr}(\theta|\mathcal{M})} \nonumber \\
	&\approx \frac{\sum_{\theta\in\Theta_H} \text{Pr}(\theta|\mathcal{D}, \mathcal{M}) }{ N_{\theta\in\Theta_H} } ~,
\end{align} 
where $N_{\theta\in\Theta_H}$ denotes the number of parameter sets that support the hypothesis $H$. The last line holds only if the number of samples is large enough to approximate the integration over the multidimensional parameter space by a sum. To interpret the resulting Bayes factors we use the established evidence classification of Ref.\,\,\cite{Jeffreys1961,Lee2016} listed in Tab. \ref{tab:BayesFactorInterpretation}. 

\begin{table}[tp]
	\centering
	%\begin{ruledtabular}
	\begin{tabularx}{\linewidth}{|l|X|}
		\hline \hline 
		$B^{H_1}_{H_0}$ \hspace{15mm} & Interpretation \\ \hline
		$>$ 100 & Extreme evidence for $H_1$ \\
		30 - 100 & Very strong evidence for $H_1$ \\
		10 - 30 & Strong evidence for $H_1$\\
		3 - 10 & Moderate evidence for $H_1$\\
		1 - 3 & Anecdotal evidence for $H_1$\\
		1 & No evidence \\
		1/3 - 1 & Anecdotal evidence for $H_0$ \\
		1/10 - 1/3 & Moderate evidence for $H_0$ \\
		1/30 - 1/10 & Strong evidence for $H_0$ \\
		1/30 - 1/100 & Very strong evidence for $H_0$ \\ 
		$<$ 1/100 & Extreme evidence for $H_0$ \\
		\hline \hline 
	\end{tabularx}
	%\end{ruledtabular}
	\caption{Interpretation of Bayes factors for comparing the evidence for hypotheses $H_0$ and $H_1$ according to the evidence classification in Ref.\,\,\cite{Jeffreys1961} with the updated terminology of Ref.\,\,\cite{Lee2016}.}
	\label{tab:BayesFactorInterpretation}
\end{table}
	
\section{Kernel Density Estimation}	
\label{sec:KDE}
Kernel Density Estimation is a non-parametric method to determine the probability density function of a given data set. Assume a set of $N$ points, $(x_1, x_2, \dots, x_N)$, which are independent and identically distributed according to the unknown density function $f(x)$. The Kernel Density Estimator (KDE) of this underlying density function is
\begin{equation}
	\hat{f}_h(x) = \frac{1}{Nh} \sum_{i=1}^N K\left(\frac{x-x_i}{h}\right)~,
\end{equation}
where $h$ is a smoothing parameter called the bandwidth and $K$ is a Kernel function. This Kernel function must integrate to one and be symmetric and non-negative. There is a range of possible Kernel functions, e.g. uniform,  linear or exponential and here we use a normalized Gaussian Kernel:
\begin{align}
	K(z) = \frac{1}{\sqrt{2\pi}}\,\exp(-z^2/2) ~.
\end{align}
The choice of the bandwidth $h$ is done such that a proper balance is achieved between maintaining important features in the density function and smoothing over irrelevant fine structure in the estimator.  To find an appropriate value for $h$ we use the rule of thumb developed by  Silverman \cite{Silverman1986}. The above approach can be straightforwardly generalized to the case with data on a higher dimensional space.

\section{EoS Tabular}
\label{sec:ThermoDynQuantities}
For practical purposes and applications,  the median values of the baryon density $n(\varepsilon)$ as a function of energy density,  as shown in Fig.\,\,\ref{fig:DensityOfEnergDensity},  are listed in Tab. \ref{tab:ThermoDynQuantities}.  Based on these values the energy per particle can be computed using Eq.\,\,(\ref{eq:EA}). The pressure can be computed using the Gibbs-Duhem relation,  Eq.\,\,(\ref{eq:GibbsDuhem}),  and then Eq.\,\,(\ref{eq:soundspeed}) for squared speed of sound.  The asymmetry of the Posterior distribution causes small deviations between the pressure computed from the Gibbs-Duhem relation and the median of the pressure in Fig.\,\,\ref{fig:PosteriorBands1}.

\begin{table}[tp]
	\centering
		\begin{tabularx}{\linewidth}{|l|X|l|l|X|}
			\hline \hline  
			$\varepsilon\,[$GeV\,fm$^{-3}]$ & $n/n_0$ & $E/A\,$[MeV]& 
			$P\,[$MeV\,fm$^{-3}]$ & $c_s^2$ \\ \hline
			0.1 & 0.66 & 9.8 & 0.7 & 0.03 \\ 
			0.2 & 1.30 & 22.7 & 7.4 & 0.10 \\ 
			0.3 & 1.91 & 42.4 & 21.2 & 0.18 \\ 
			0.4 & 2.49 & 65.6 & 43.7 & 0.27 \\ 
			0.5 & 3.03 & 91.6 & 75.7 & 0.37 \\ 
			0.6 & 3.54 & 119.6 & 116.5 & 0.45 \\ 
			0.7 & 4.02 & 148.7 & 164.7 & 0.51 \\ 
			0.8 & 4.47 & 178.5 & 217.8 & 0.55 \\ 
			0.9 & 4.90 & 208.5 & 273.9 & 0.57 \\ 
			1.0 & 5.31 & 238.1 & 331.9 & 0.59 \\ 
			1.1 & 5.70 & 267.2 & 390.8 & 0.59 \\ 
			1.2 & 6.07 & 295.5 & 450.3 & 0.60 \\ 
			1.3 & 6.43 & 323.2 & 509.9 & 0.60 \\ 
			\hline \hline 
		\end{tabularx}
	\caption{Tabulated values of the median for the density $n$, in units of the nuclear saturation density $n_0$, as a function of energy density $\varepsilon$ as depicted in Fig.\,\,\ref{fig:DensityOfEnergDensity}. Only the values for the Segments parametrisation are listed. Based on these values the energy per particle $E/A$ is computed as well as the pressure $P$ and sound speed $c_s^2$ using the Gibbs-Duhem relation.}
	\label{tab:ThermoDynQuantities}
\end{table}

\bibliography{neutron_star_sound speed_library}% Produces the bibliography via BibTeX.

\end{document}